\documentclass[a4paper,11pt]{article}
\usepackage[hidelinks]{hyperref}

\usepackage{amsmath, bm, amsthm, amssymb, mathrsfs}
\usepackage{amsfonts}
\usepackage{xcolor}
\usepackage{graphicx}
\usepackage{booktabs, multirow}
\usepackage{enumitem}
\usepackage{array} 
\usepackage{caption} 

\usepackage[subrefformat=parens,labelformat=parens]{subfig}
\usepackage{float}
\usepackage{bm}
\usepackage{algpseudocode,algorithm}
\usepackage{xfrac}

\graphicspath{{graphics/}}

\title{Bayesian calibration and sensitivity analysis of heat transfer models 	
for fire insulation panels}

\author{P.-R. Wagner, R. Fahrni, M. Klippel, A. Frangi, B. Sudret}

\date{24.12.2019}

\usepackage{natbib}
\usepackage[left=25mm,right=25mm,top=1.5cm,bottom=1.5cm,includeheadfoot]{geometry}
\usepackage{bayesnotations}

\newcommand{\modelOrig}{\cm(\BParams_{\cm})}
\newcommand{\modelPC}{\cm^{\mathrm{PC}}(\BParams_{\cm})}
\newcommand{\trans}{^{\intercal}}

\newcommand{\meanRespEst}{\ve{\mu}_{\BModelOut}}
\newcommand{\covRespEst}{\ve{\Sigma}_{\BModelOut}}
\newcommand{\expc}[1]{\mathbb{E}\left[#1\right]}
\newcommand{\expe}[1]{\mathbb{E}\left[#1\right]}
\newcommand{\expcm}[2]{\mathbb{E}_{#2}\left[#1\right]}
\newcommand{\varc}[1]{\mathrm{Var}\left[#1\right]}
\newcommand{\varcm}[2]{\mathrm{Var}_{#2}\left[#1\right]}

\newcommand{\BparamsM}{\Bparams_{\cm}}
\newcommand{\BParamsM}{\BParams_{\cm}}
\newcommand{\BparamsE}{\Bparams_{\varepsilon}}
\newcommand{\BParamsE}{\BParams_{\varepsilon}}
\newcommand{\PolyBasis}{\Psi_{\ve{\alpha}}}
\newcommand{\PolyBasisB}{\Psi_{\ve{\beta}}}
\newcommand{\PolyBasisVec}{\ve{\Psi}}
\newcommand{\PolyCoeff}{a_{\ve{\alpha}}}

\newcommand{\PolyCoeffApproxPC}{\tilde{a}_{p,\ve{\alpha}}}

\newcommand{\norm}[1]{\left\vert\left\vert#1\right\vert\right\vert}
\newcommand{\card}[1]{\mathrm{card}(#1)}

\newcommand{\errorCovPMult}{\ve{\Sigma}(\BparamsE)}
\newcommand{\errorCovPMultS}{\ve{\Sigma}(\BparamsE)}
\newcommand{\cove}[1]{\mathrm{Cov}\left[#1\right]}
\newcommand{\degC}{\ensuremath{\,^{\circ}\mathrm{C}}}
\newcommand{\jpk}{{~\sfrac{\mathrm{J}}{\mathrm{kgK}}}}
\newcommand{\wpmk}{{~\sfrac{\mathrm{W}}{\mathrm{mK}}}}
\newcommand{\wpmmK}{{~\sfrac{\mathrm{W}}{\mathrm{m^2K}}}}

\newcommand{\modif}[1]{#1}

\begin{document}
	
\maketitle

\maketitle

\abstract{A common approach to assess the performance of fire insulation panels 
is the component additive method (CAM). The parameters of the CAM are based on 
the temperature-dependent thermal material properties of the panels. These 
material properties can be derived by calibrating finite element heat transfer 
models using experimentally measured temperature records. In the past, the 
calibration of the material properties was done manually by trial and error 
approaches, which was inefficient and prone to error.
	 In this contribution, the calibration problem is reformulated in a 
	 probabilistic setting and solved using the Bayesian model calibration 
	 framework. This not only gives a set of best-fit parameters but also 
	 confidence bounds on the latter. To make this framework feasible, the 
	 procedure is accelerated through the use of advanced surrogate modelling 
	 techniques: polynomial chaos expansions combined with principal component 
	 analysis. This surrogate modelling technique additionally allows one to 
	 conduct a variance-based sensitivity analysis at no additional cost by 
	 giving access to the Sobol' indices. The calibration is finally validated 
	 by using the calibrated material properties to predict the temperature 
	 development in different experimental setups.\\

}

\section{Introduction}
\label{sec:intro}

Knowledge about the basic behaviour of materials exposed to fire is
extremely important to successfully develop fire safety strategies.
Depending on the type and height of buildings, certain fire
requirements need to be fulfilled, e.g. requirements w.r.t. the
\emph{load-bearing function} (R) and \emph{separating function} (EI). In case 
of 
timber
buildings, the performance not only of the timber members, but also of
protective materials such as gypsum plasterboards and insulations is
of high importance for the fire design of the building structure.
These different materials are usually combined to build floor and wall
elements with different layups, so-called \emph{timber frame assemblies}. The 
separating function of timber frame 
assemblies is usually verified using the component additive method 
(CAM) \citep{Frangi:ES2010,Maeger:FSJ2017,Just2018}. This method is rather 
flexible for calculating the separating function because it handles 
arbitrary layups made of various materials and thickness. 
Producers of fire protection products (e.g. gypsum
plasterboard and insulation) need to determine input factors for the 
individual materials so that the separating function of a timber frame 
assembly with these materials can be verified using the CAM. 

Indeed, the same protective material (e.g. a gypsum plasterboard with a given
thickness) contributes differently to the fire resistance of a timber 
frame assembly in different setups. The CAM therefore considers (1) the 
material and thickness of a layer and (2) modification factors taking 
into account the neighbouring layers. This leads to a high number of 
possible combinations for timber frame assemblies, which cannot all be 
tested in fire resistance tests.
Therefore, the factors of the CAM are usually derived based on finite 
element
(FE) models and accompanying fire resistance tests.

Fire resistance tests using the standard EN/ISO
temperature time curve according to \citet{ISO8341} and \citet{EN1363}
constitute the basis for these simulations. In the fire tests, the
temperature is recorded over time inside the specimen at specific
distances to the fire exposed surface. These recordings are used as a
reference for FE simulations of the same setup.
Heat transfer models using
\emph{effective thermal material properties} that depend on the 
temperature $T$ (specific
heat capacity~$c(T)$, thermal conductivity~$\lambda(T)$ and material
density~$\rho(T)$) are usually employed to simulate the temperature 
development inside
timber frame assemblies exposed to fire. These material properties are 
then calibrated such that
the output matches the recorded temperatures. Since these properties are
not strictly physical quantities, they are called \emph{effective} 
material properties. They account for \modif{not explicitly modeled effects 
such as} 
fissures, cracks and moisture flow inside the specimen 
\citep{Frangi:ES2010}. \modif{Despite these simplifications, using 
temperature-dependent effective material properties together with a common heat 
transfer analysis is appropriate and state-of-the-art for the calibration of 
parameters in the CAM. This is especially true since the introduced 
simplifications are negligible compared to the simplifications made within the 
CAM.}

The conventional process of determining these effective material
properties is slow and inaccurate as the calibration is usually done
manually.
All temperature measurements are averaged, thus eliminating the 
variability in the material behaviour
and not accounting for it in the calibration. The derivation of thermal
material properties is conventionally done as follows 
\citep{Maeger:FSJ2017}:

\begin{enumerate}[label=\textbf{Step \arabic*}]
	\item \label{conventionalOne} Assume effective thermal material properties 
	and simulate 
	a specific layup with FE heat transfer models;
	\item \label{conventionalTwo} Compare the resulting temperatures with the 
	averaged 
	measurements;
	\item Iterate \textbf{\ref{conventionalOne}} and 
	\textbf{\ref{conventionalTwo}} until the simulation results are 
	similar to the measured temperatures.
\end{enumerate}

A more rigorous calibration of these effective material properties can
be achieved by parameterizing the thermal time-dependent material properties 
with a set of model parameters. Through this parametrization the 
problem of determining the time-dependent effective material properties is 
recast as a 
problem of determining the real-valued parameterizing model parameters. Then 
the 
calibration problem can be posed in a probabilistic
setting. This allows a proper treatment of uncertainties arising from
material fluctuations, measurement errors and model insufficiencies. One
general way to do this is the so-called \emph{Bayesian inversion}
framework \citep{Bayesian:Beck1998, Bayesian:Gelman2014:3rd, Yu2019}. 
In this framework, the model parameters are seen as
random variables. Instead of trying to determine one particular value
for these parameters, this probabilistic framework determines the full
probability distribution of the model parameters 
conditioned on the observed measurements. This
distribution contains much more information about the calibrated
properties than the single point estimate from the conventional
approach. For example, it allows computing expected values, maximum a
posteriori estimates, confidence intervals on the calibrated values and the 
full correlation structure. 
Furthermore, the calibration can be verified easily by computing the 
\emph{posterior predictive 
	distribution} 
\citep{Bayesian:Gelman1996}.

To determine the probability distribution of the material properties, 
it is necessary to repeatedly evaluate the 
FE heat transfer model. To reduce the computational burden associated 
with repeated model evaluations, it has become customary to 
replace the computational forward model with a cheap-to-evaluate 
surrogate. Therefore, the Bayesian inversion framework is here combined 
with the polynomial chaos expansions (PCE) surrogate modelling
technique \citep{Sudret2007, BlatmanThesis, Guo2018}.

When working with models with multiple input parameters, the question 
of 
the \emph{relative importance} of individual parameters with respect to 
the 
output arises naturally. Quantifying this influence is called sensitivity 
analysis. One family of approaches are 
the so-called \emph{variance decomposition techniques} \citep{Saltelli2000, 
	Arwade2010}. These methods attempt to 
apportion the variance of the probabilistic model output to the 
individual input parameters. The \emph{Sobol' indices} are one such 
variance 
decomposition technique \citep{Sobol1993}. Determining the Sobol' 
indices is
typically computationally expensive, but it has been shown by
\citet{SudretCSM2006} that they can be computed easily for a PCE 
surrogate model. Their computation allows valuable 
insights into the heat-transfer model's properties. 

In this paper, the material properties of four different
gypsum insulation boards (Products A-D, E1-E4) are calibrated based on 
fire resistance 
tests
carried out with these materials. These experimental results are presented and 
discussed in 
detail in Section~\ref{sec:expMod}. The calibration is carried out with the 
Bayesian model calibration framework accelerated by
constructing a PCE-based surrogate model as detailed in 
Section~\ref{sec:BayesCal}. Section~\ref{sec:sensitivity} outlines how the 
employed 
surrogate model can be used to conduct a global sensitivity analysis of 
the 
considered computational model. Finally, in Section~\ref{sec:results}, the 
calibration is verified using two fire tests 
that
use two of the calibrated materials (Product C and 
Product D) in different experimental setups (V1 and V2).

\section{Experiments and modelling}
\label{sec:expMod}

\subsection{Experiments}
\label{sec:expMod:experiments}

\begin{figure}
	\centering
	\subfloat[Test 1, E1-E4 
	\citep{Breu2016}]{{\includegraphics[width=7cm]{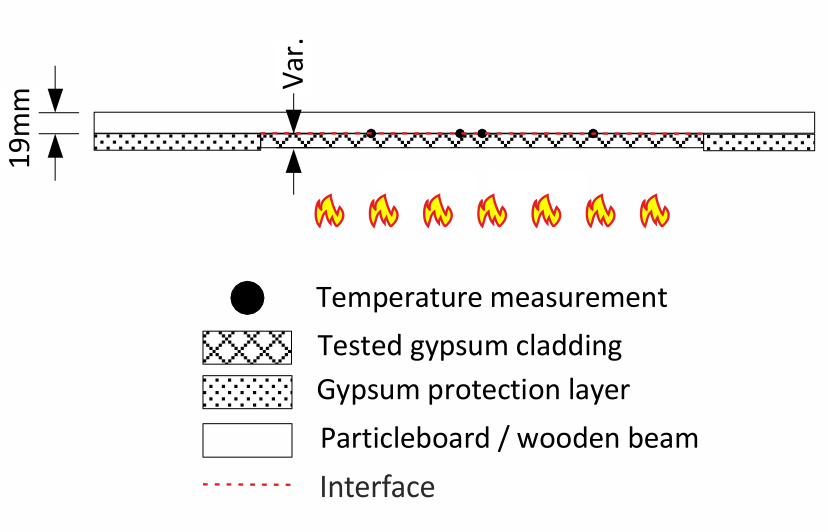}} 
		\label{fig:setupTest1}}%
	\subfloat[Test 2, V1-V2
	\citep{test:Gyproc2016}]{{\includegraphics[width=7cm]{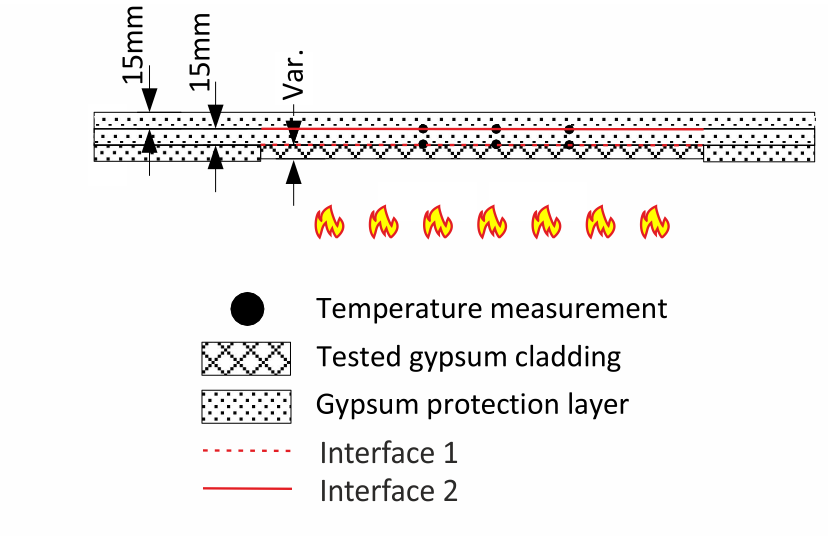}}
		\label{fig:setupTest2}}%
	\caption{Sketch of the experimental setups from Test 1 and 
		Test 2. 
		For more details refer to the respective publications 
		\citet{Breu2016} and \citet{test:Gyproc2016}.}%
	\label{fig:setupTest}%
\end{figure}

\begin{figure}
	\centering
	\includegraphics[width=12cm]{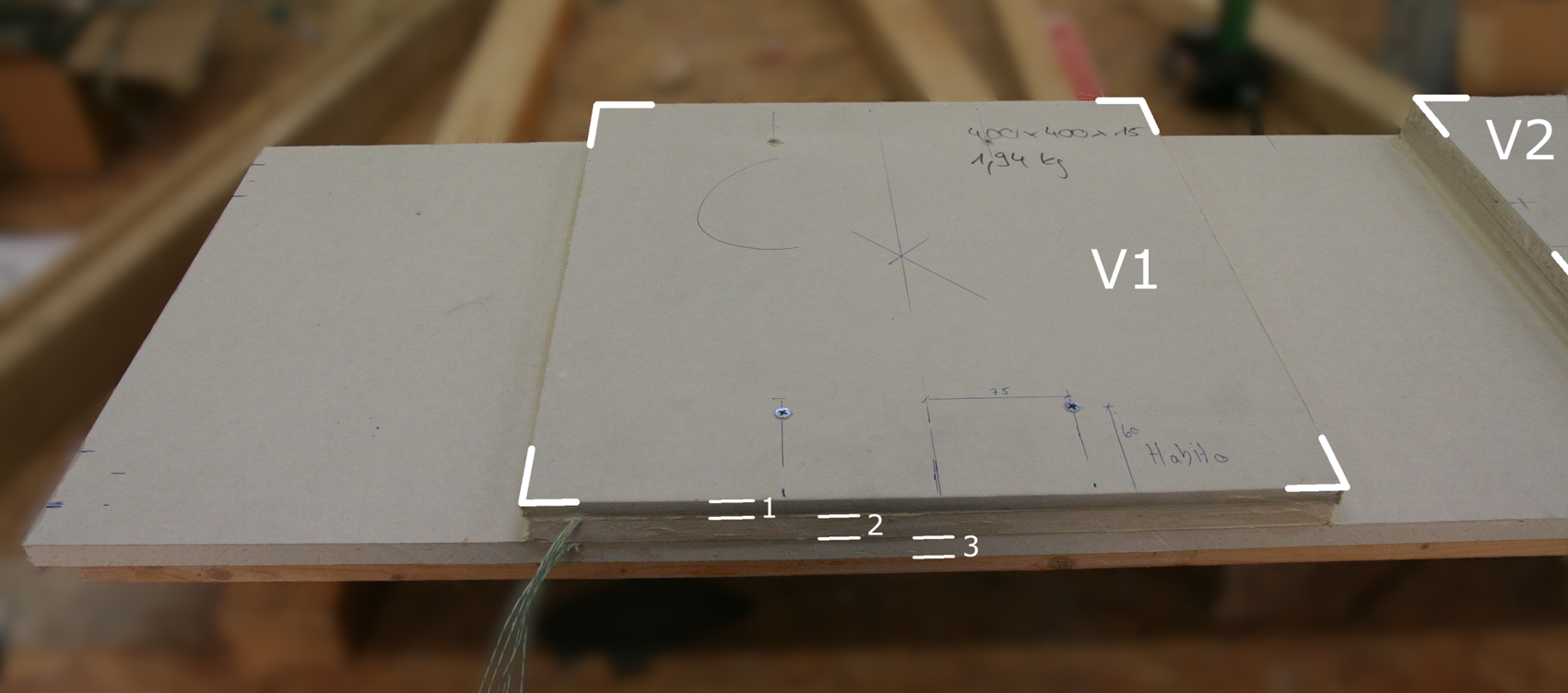}
	\caption{Specimens V1 and V2 (on the right), upside-down, exposed
		surface
		on top, before installation of the protection layers around/between the 
		specimens, three wire thermocouples between each layer; \\*
		1: exposed protection layer, $12.5~\mathrm{mm}$ Product C; 2: 
		protection layer 
		$15~\mathrm{mm}$
		Product D; 3: carrying layer $15~\mathrm{mm}$ Product D; 
		around the
		specimen other protection layers were applied to protect the carrying
		layer.}
	\label{fig:fire_test_spcimen} 
\end{figure}

Two fire resistance tests with horizontally oriented specimens constitute the 
experimental basis for the analysis in this paper 
\citep{test:Gyproc2016,
	Breu2016}. The unloaded tests were conducted on the model-scale furnace
of SP Wood Building Technology (today's Research Institute of Sweden, 
RISE) and were exposed to the EN/ISO temperature-time curve
\citep{EN1363,ISO8341}. Two (V1 and V2,
\citep{Breu2016}), respectively four (E1 to E4,
\citep{test:Gyproc2016}) different gypsum plasterboard setups with
dimensions of $0.4\times0.4~m$ were tested in each test
(Figure~\ref{fig:setupTest}). Tables \ref{tab:E1-E4} and
\ref{tab:V1-V2} show the layups of the
specimens. The space between and around the specimens was at least
100~mm and was filled with Product D boards to protect the
carrying layer, \ie the last layer. The carrying layer was a 19~mm
particle 
board with density $\rho=633$~kg/m\textsuperscript{3} in Test 1 
(specimens E1 to E4) and a Product D 15~mm in Test 2 
(specimens V1 and V2). Figure~\ref{fig:fire_test_spcimen} shows the 
specimens of Test 2 during fabrication.

\begin{figure}
	\centering
	\subfloat[E1 
	\citep{test:Gyproc2016}]{{\includegraphics[width=7cm]{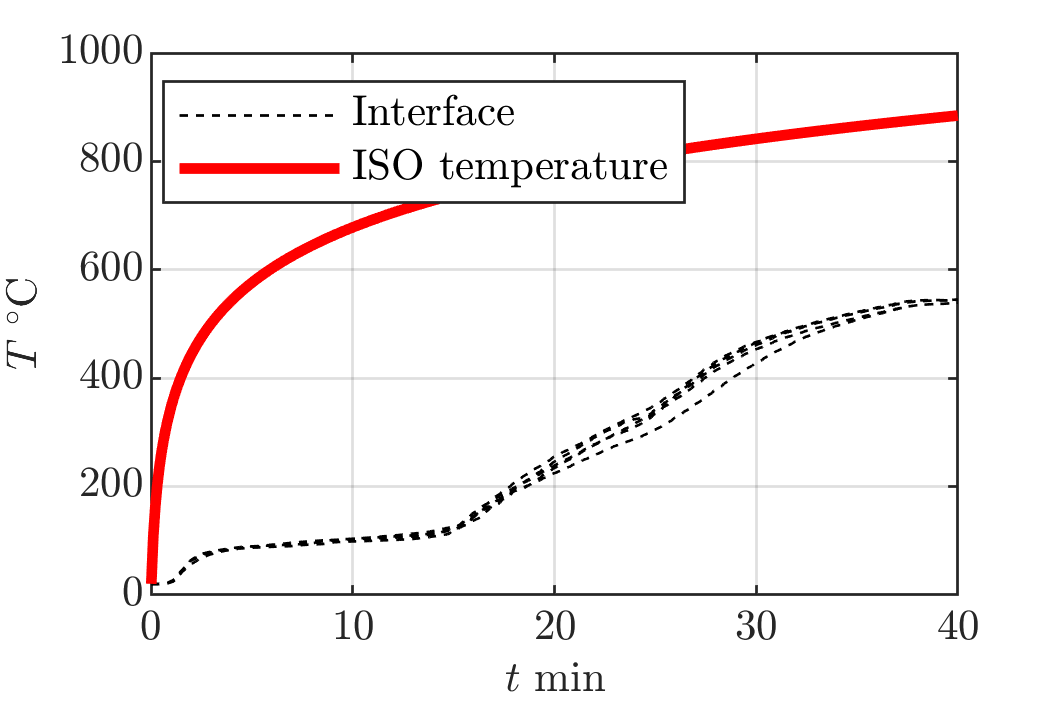}} 
		\label{fig:dataSummaryE1}}%
	\subfloat[E2 
	\citep{test:Gyproc2016}]{{\includegraphics[width=7cm]{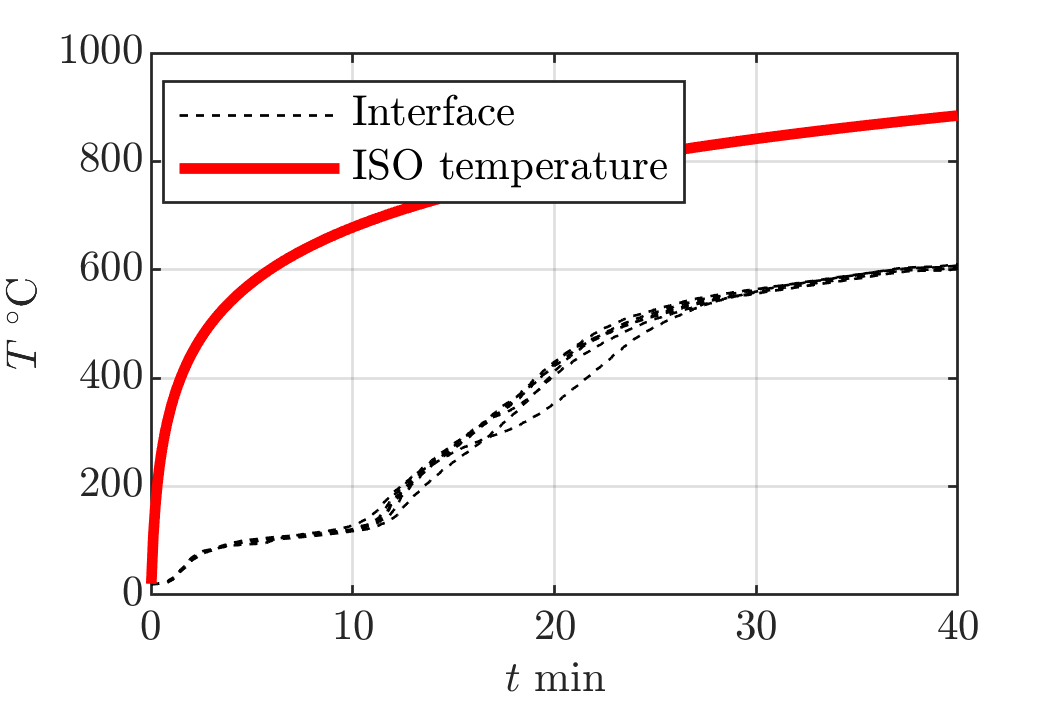}} 
		\label{fig:dataSummaryE2}}%
	\\
	\subfloat[E3 
	\citep{test:Gyproc2016}]{{\includegraphics[width=7cm]{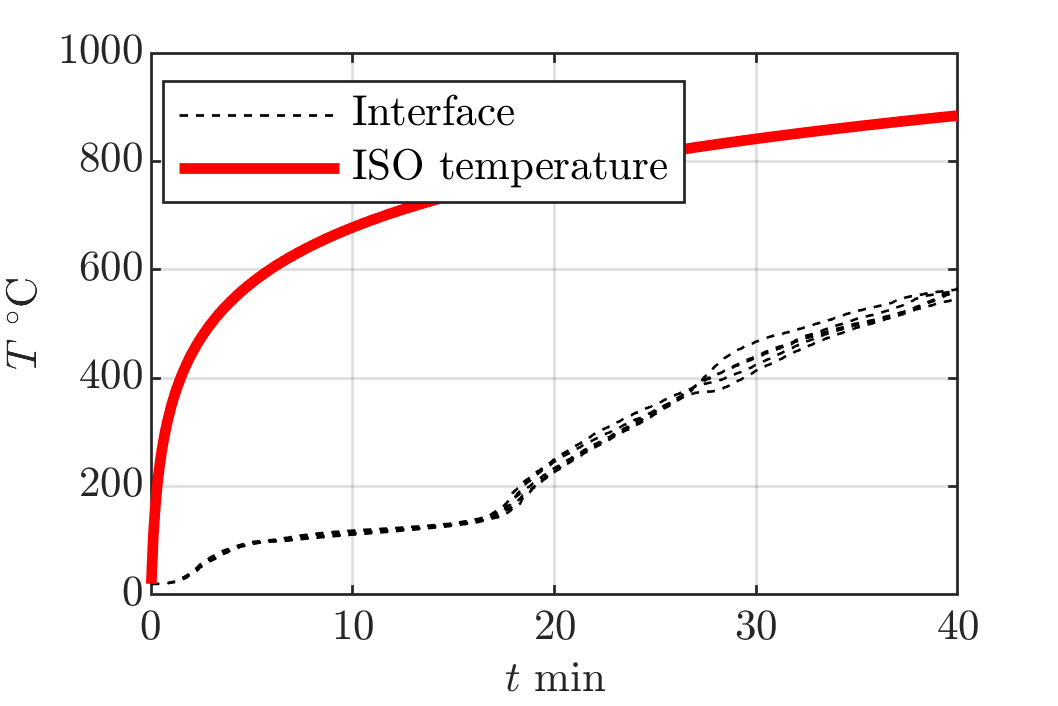}} 
		\label{fig:dataSummaryE3}}%
	\subfloat[E4 
	\citep{test:Gyproc2016}]{{\includegraphics[width=7cm]{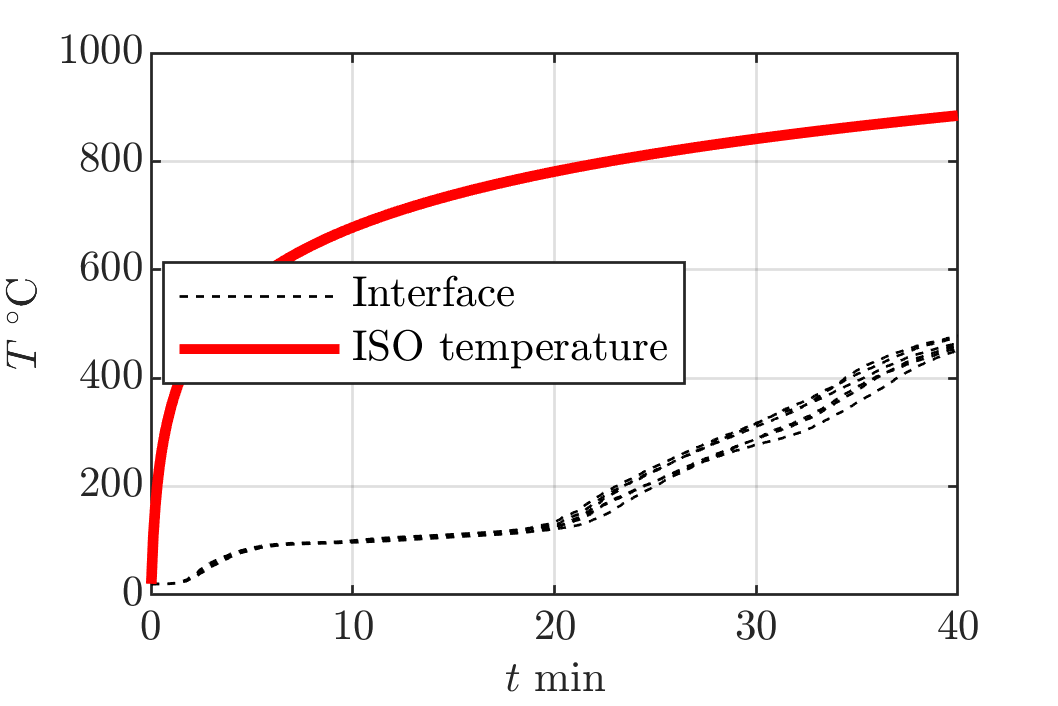}} 
		\label{fig:dataSummaryE4}}%
	\\
	\subfloat[V1 
	\citep{Breu2016}]{{\includegraphics[width=7cm]{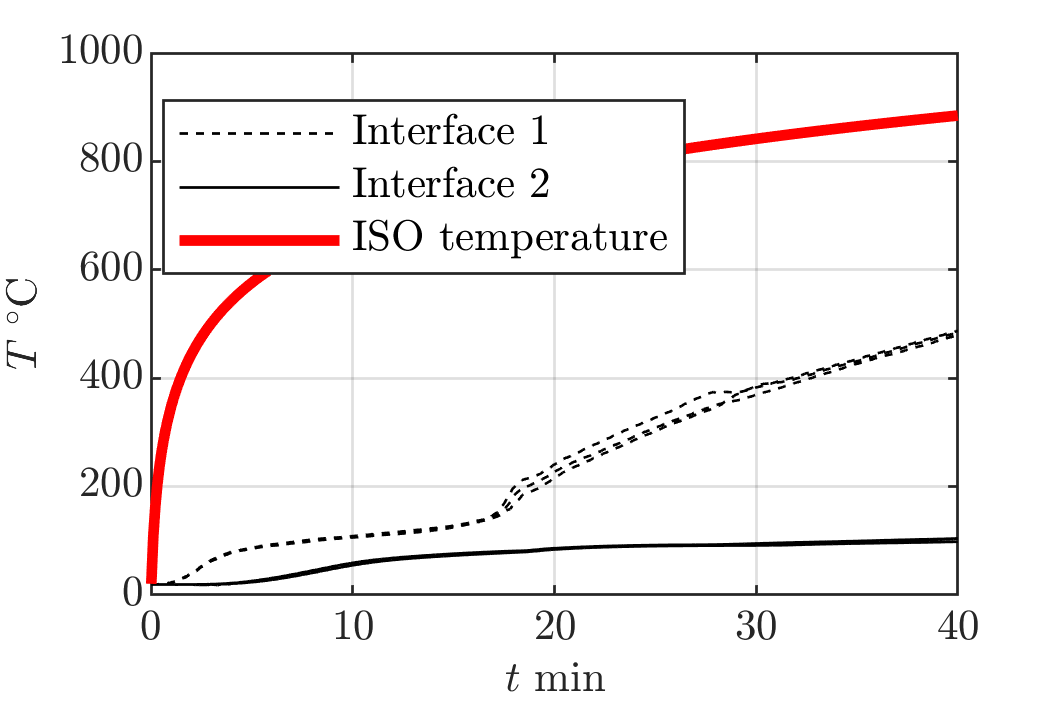}} 
		\label{fig:dataSummaryV1}}%
	\subfloat[V2 
	\citep{Breu2016}]{{\includegraphics[width=7cm]{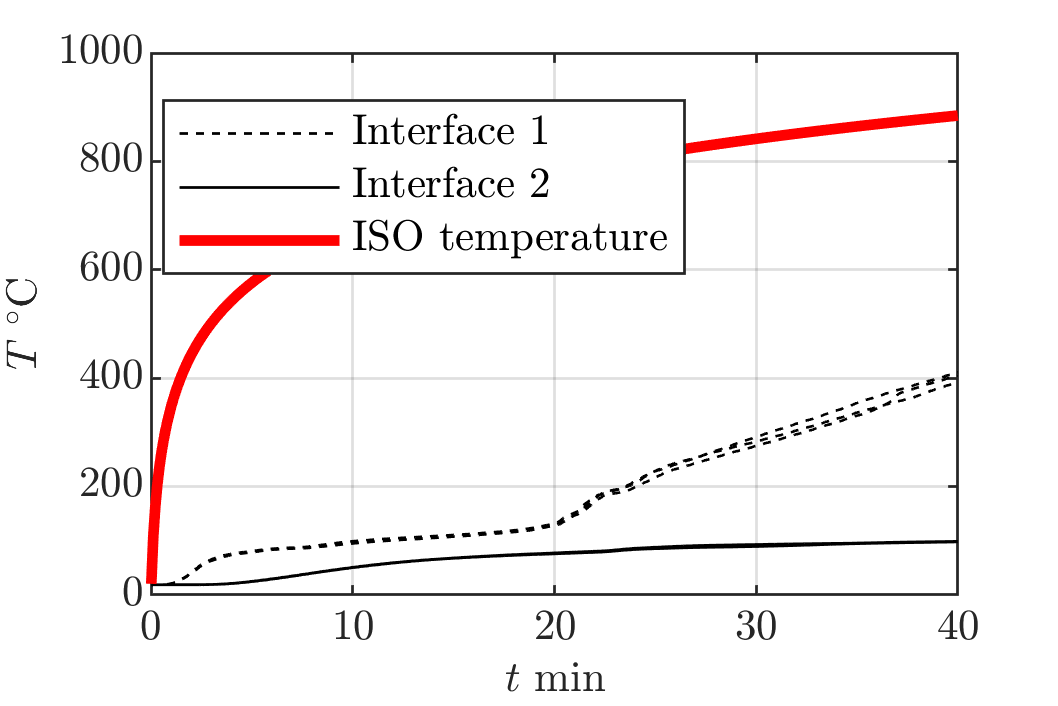}} 
		\label{fig:dataSummaryV2}}%
	\caption{Summary of the data used for calibration and the 
		underlying ISO temperature according to \citet{EN1363,ISO8341}.}%
	\label{fig:dataSummary}%
\end{figure}

In specimens E1 to E4 (Test 1), five wire thermocouples and one 
copper disc thermocouple measured the temperatures at a single 
\emph{interface} between the layers. The fluctuations of the sensor readings 
can mainly be attributed to variations in the material properties. The measured 
temperatures are 
displayed in Figures~\subref*{fig:dataSummaryE1} to 
\subref*{fig:dataSummaryE4}.

In specimens V1 and V2 (Test 2), three wire thermocouples were 
placed between
each layer. The measured temperatures at the two \emph{interfaces 1} 
and 
\emph{2} are displayed in 
Figures~\subref*{fig:dataSummaryV1} and \subref*{fig:dataSummaryV2}.

Test 1 and Test 2, with specimens E1-E4 and V1,V2
respectively, were exposed to the EN/ISO standard
temperature-time curve \citep{EN1363,ISO8341}.

The temperature measurements $y^{(s)}(t)$ at each sensor $s$ were
collected at $N$ discrete time steps $t_{i}$:
\begin{equation}
\Bdata^{(s)} = (y^{(s)}_1,\dots, y^{(s)}_N)\trans \quad \text{with} \quad 
y_i^{(s)} \eqdef y^{(s)}(t_i) \quad 
\text{for} 
\quad i = 1, \dots, N,
\end{equation}
where $t_i=i\tau$ and $\tau=10~\mathrm{s}$. For simplicity, 
the superscript $(s)$ is omitted unless required to distinguish 
between individual measurements. Therefore, in the sequel $\Bdata$ stands for a 
vector of 
measurements captured by a single sensor.

\begin{table}
	\caption{Specimens E1-E4 (Test 1)}
	\label{tab:E1-E4}
	\centering
	\begin{tabular}{lrlrl}
		\hline
		& \multicolumn{2}{c}{Layer 1} & 
		\multicolumn{2}{c}{Layer 2} \\
		\hline
		E1 &  Product A & 12.5~mm & particle board & 19~mm\\
		E2 &  Product B & 9.5~mm  & particle board & 19~mm\\
		E3 &  Product C & 12.5~mm & particle board & 19~mm\\
		E4 &  Product D & 15~mm & particle board & 19~mm\\
		\hline
	\end{tabular}
\end{table}

\begin{table}
	\caption{Specimens V1 and V2 (Test 2)}
	\label{tab:V1-V2}
	\centering
	\begin{tabular}{lrlrl}
		\hline
		& \multicolumn{2}{c}{Layer 1} & 
		\multicolumn{2}{c}{Layer 2+3} \\
		\hline
		V1 &  Product C & 12.5~mm  & Product D & 
		2$\times$15~mm\\
		V2 &  Product D & 15~mm & Product D & 
		2$\times$15~mm\\
		\hline
	\end{tabular}
\end{table}

\subsection{Forward modelling}
\label{sec:expMod:modelling}
The experiments described in the previous section were modeled using 
one-dimensional heat transfer FE-models. The reduction to a 
one-dimensional setup is justified, because it is known that the 
experimental 
heat-flux is mostly perpendicular to the exposed surface. The
simulations were conducted using the general purpose finite element software 
\emph{Abaqus} \citep{AbaqusManual}. The
energy input on the heated surface and the losses on the unexposed side
took into 
account the energy input/loss through convection and radiation. 

The radiation
temperature was assumed to be equal to the gas temperature and followed
the EN/ISO temperature-time curve \citet{EN1363,ISO8341} on the exposed
side and was constantly $19.5~^{\circ}C$ on the unexposed side (as in 
the
experiments). The emissivity was taken as $0.8$ and the convection
coefficient as $25 \wpmmK$ according to \citet{EN199112}. The element
size in the FE-mesh was $0.25~\mathrm{mm}$. The temperature-dependent 
material 
properties of the particle board 
(Test 1) were taken from \citet{Schleifer2009}.

The unknown parameters of interest are the \emph{temperature-dependent 
	effective material properties} of the insulation material: the thermal 
conductivity $\lambda(T)$, 
the heat capacity $c(T)$ and the material density $\rho(T)$ of the 
investigated insulation materials. 

With this, the computational forward model is
\begin{equation}
\BModelOut = (Y_1,\dots, Y_N)\trans =
\cm(\lambda(T),c(T),\rho(T)),
\end{equation}

For every set of 
material properties, this model returns the 
temperature evolution at locations and at times where measurements 
are available 
(see Section~\ref{sec:expMod:experiments}).

Since each of those parameters is a function of the temperature, they 
cannot be 
directly calibrated. Instead, these functions have to be parameterized with a 
set of scalar parameters, as described 
next.

\subsection{Parametrization of material properties}
\label{sec:expMod:parametrization}

The parametrization of the three temperature-dependent material
properties ($\lambda(T),c(T),\rho(T)$) is a crucial step of the
calibration procedure. It consists of specifying a set of parameters
$\BParamsM$ that define the shape of the temperature-dependent function 
that
describes each material property. The choice of these parameters is
delicate, as it imposes a certain temperature-dependent behaviour on the
material properties. A priori, there are no physical constraints on this
thermal behaviour besides positivity, so generally, the
properties are defined as 
$\lambda: [0,1200\!\degC]\to\mathbb{R}^+$, $c:
[0,1200\!\degC]\to\mathbb{R}^+$ and $\rho: 
[0,1200\!\degC]\to\mathbb{R}^+$. 

One further complication lies in the fact that these properties are
mere \emph{effective} properties and cannot generally be
measured. To find constraints on these parameters, it is thus necessary 
to rely
on previous calibration attempts of gypsum insulation boards
\citep{Breu2016,Schleifer2009} in conjunction with measurements of
certain properties, where available. 

By gathering information from such previous attempts, the 
thermal properties are parameterized by six parameters 
$\BParamsM = 
(X_1,\dots,X_6)\trans$. This parametrization is flexible enough to enable 
inference on 
$\BParamsM$ and follows physical and
empirical reasoning as described next.

We propose to distinguish two \emph{key processes} during which the 
temperature-dependent material
properties change significantly:

\begin{description}
	\item[First key process] When the free water
	content in the gypsum insulation boards evaporates, the latent 
	water 
	content of 
	gypsum, which is composed of
	sulfate dihydrate CaSO\textsubscript{4} $\cdot$ 2 
	H\textsubscript{2}O,
	evaporates. In this process, evaporation first forms calcium
	sulfate hemihydrate CaSO\textsubscript{4} $\cdot$ 1/2
	H\textsubscript{2}O (also called bassanite) and then anhydrate III
	CaSO\textsubscript{4}. The evaporation consumes heat, which is 
	modelled 
	as
	a local increase of the specific heat capacity, a reduction in the 
	conductivity and a 
	reduction in the material density.
	\item[Second key process] Thermogravimetric analyses have shown a 
	second peak in the specific
	heat due to chemical metamorphosis at elevated temperatures of 
	secondary components found in the gypsum insulation 
	boards \citep{Schleifer2009}. This second key process is modelled as 
	an increase in the material conductivity, a peak in the specific 
	heat and a further reduction of the material density.
\end{description} 

The temperatures at which these two key processes occur cannot be equally
well prescribed a priori. While the temperature of the water evaporation is well
known to occur at approximately $100\!\degC$ with its main effect taking
place at $140\!\degC$ until it tails off at $180\!\degC$, the second key 
process
cannot be characterized this precisely. It is assumed that the second 
key process starts at the unknown temperature $X_1$ and ends at
$850\!\degC$. Additionally, the relative location of its main effect 
between $X_1$ and $850\!\degC$ is parameterized with $X_2$. These two 
temperatures  heavily influence the evolution of the
thermal properties and are thus used as temperatures of change in all
effective material properties. 

In the present setting of heated gypsum boards, the
initial conductivity at ambient temperature is assumed to be
$\lambda(20\!\degC) = 0.4\!\wpmk$ \citep{Breu2016}. During the first key
process, the conductivity is assumed to linearly decrease to a second
value that is parameterized by $X_3$. Starting with the second key
process the
conductivity starts to increase linearly to another value that is
parameterized by 
$X_4$, reached at the highest simulation 
temperature of $1200\!\degC$ .

Phase changes require a significant amount of energy. To model the 
energy 
requirement associated with the evaporation of water trapped inside the 
insulation material, the specific heat $c(T)$ is modelled with two
piecewise linear spikes during both key processes, while being constant 
at $c=960\jpk$ 
\citep{Schleifer2009} for the other temperatures. The specific heat at 
the peaks is
parameterized by $X_5$ for the first process and $X_6$ for the second one.

\begin{figure}
	\centering
	\subfloat[$\lambda(T,\BParams)$]{{\includegraphics[width=14cm]{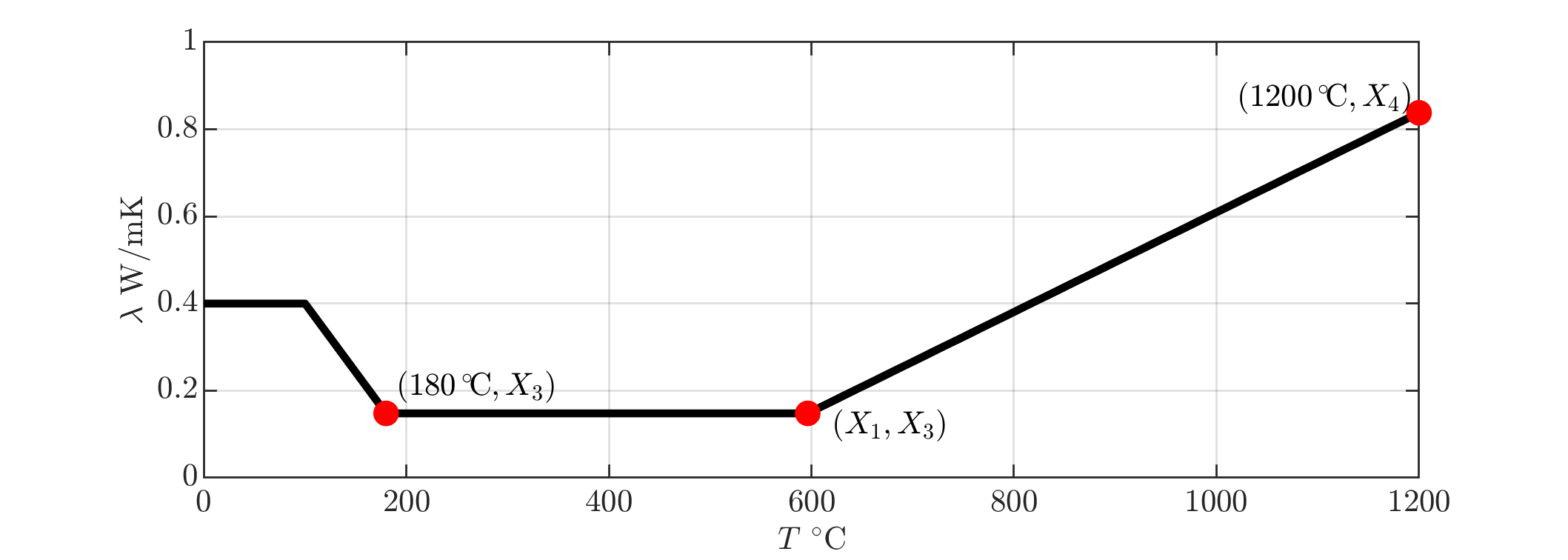}
	}}%
	
	\subfloat[$c(T,\BParams)$]{{\includegraphics[width=14cm]{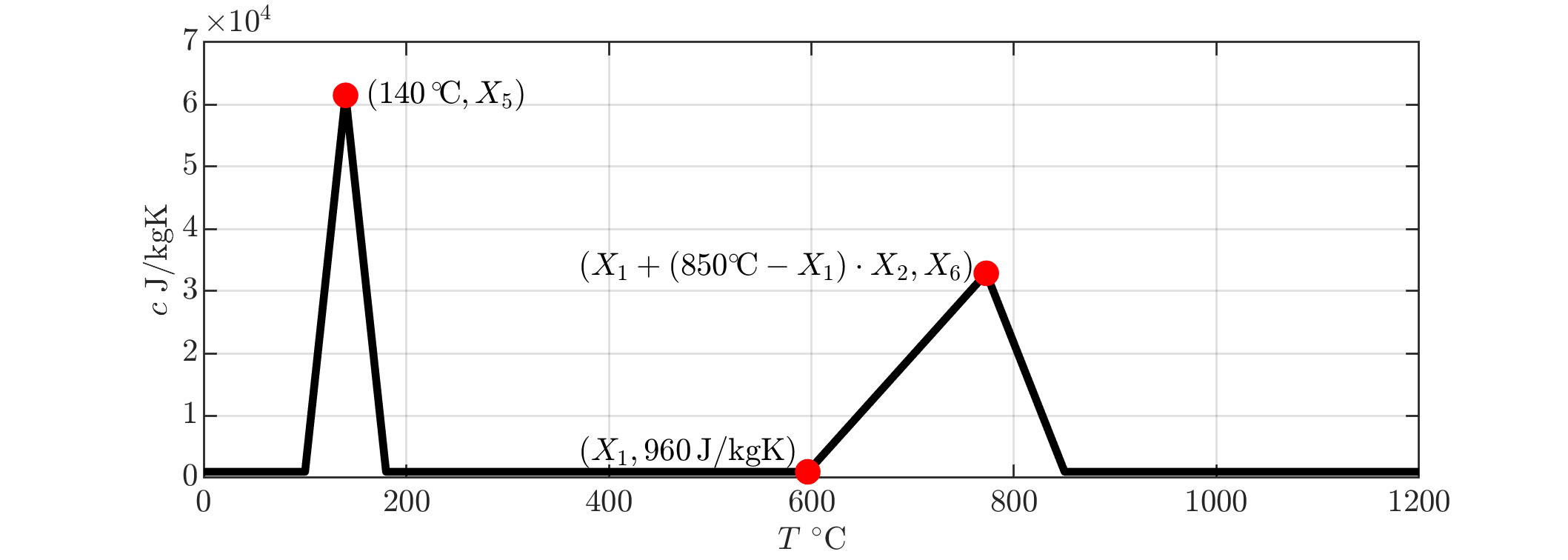}
	}}%
	\\
	
	\subfloat[$\rho(T,\BParams)$]{{\includegraphics[width=14cm]{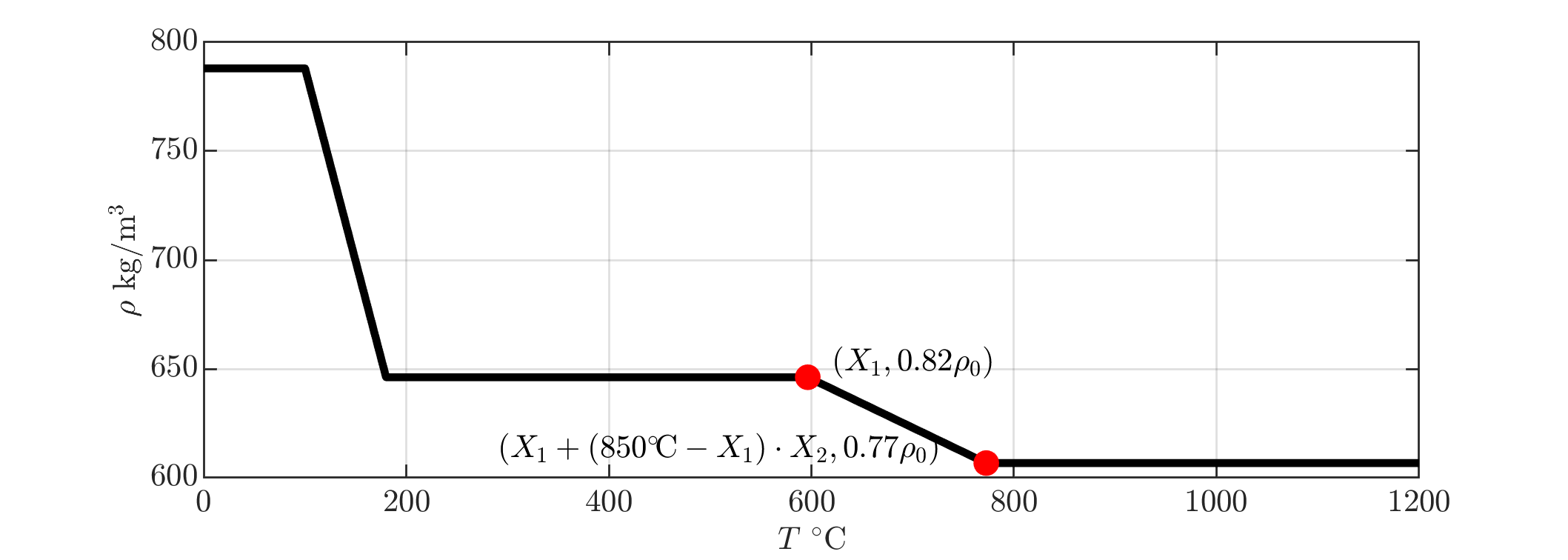}
	}}%
	\caption{Parametrization of temperature-dependent effective 	
		material properties as defined in 
		Table~\ref{tab:parametrization}.}%
	\label{fig:parametrization}%
\end{figure}

During the first and second key process, gaseous products are emitted 
(water
and carbon dioxide respectively) and thus the density of the gypsum $\rho(T)$
reduces. This density reduction
was studied in \citet{Schleifer2009} and the results are
applied here directly. Starting from the density measured at room
temperature $\rho_0$, $\rho(T)$ linearly reduces during the first key process to
$82\%$. It then remains constant and linearly reduces further to $77\%$ 
from the start of the second key process $X_1$ to the main effect of 
the second key 
process. The parametrization of the material properties is visualized in 
Figure~\ref{fig:parametrization}.

\begin{figure}
	\centering
	
	\subfloat[$\lambda(T,\BParams)$]{{\includegraphics[width=14cm]{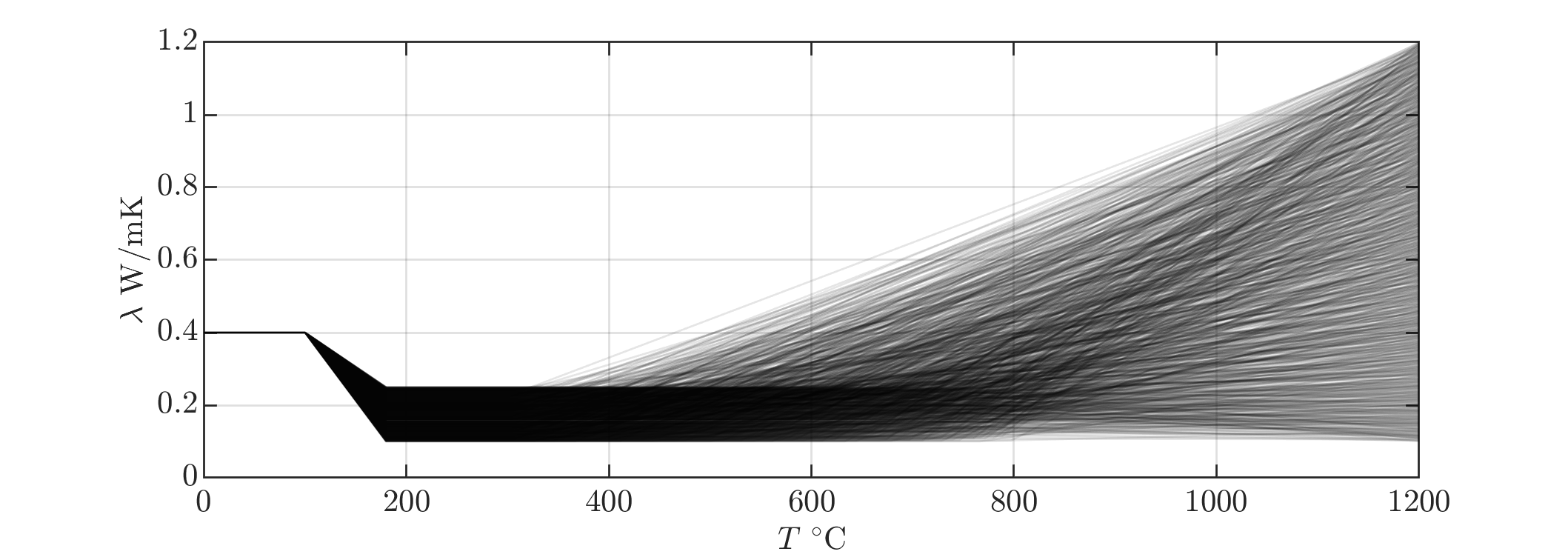}
	}}%
	
	\subfloat[$c(T,\BParams)$]{{\includegraphics[width=14cm]{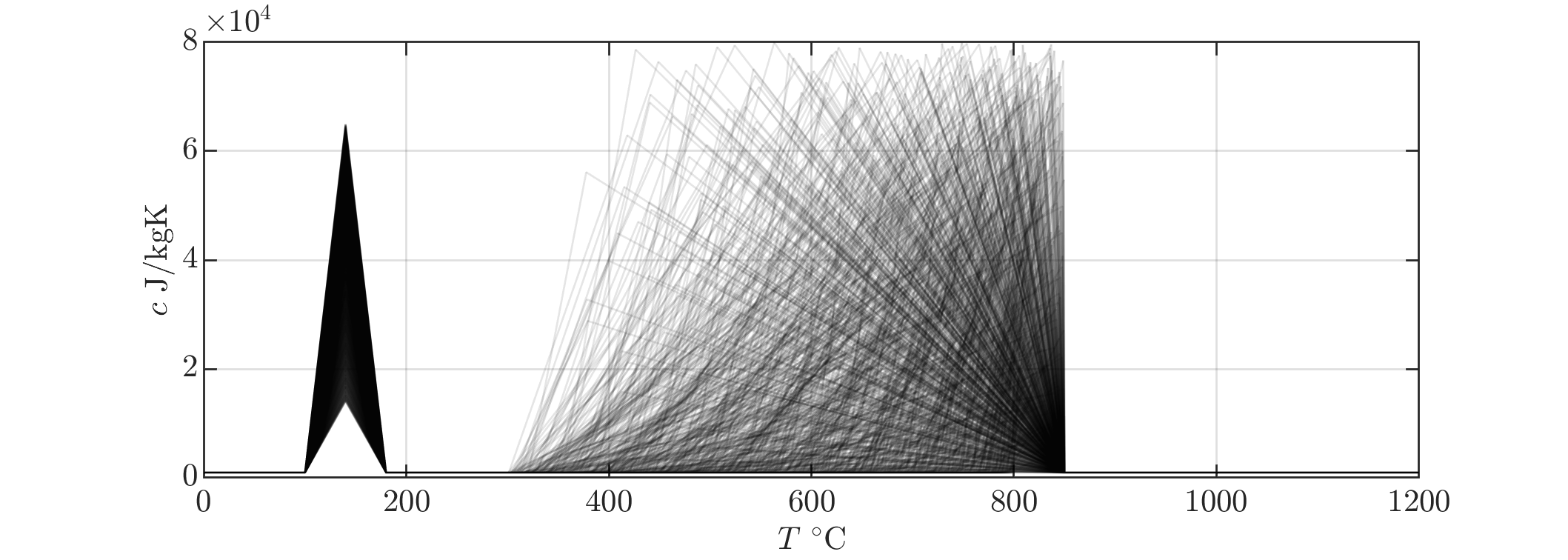}
	}}%
	\\
	
	\subfloat[$\rho(T,\BParams)$]{{\includegraphics[width=14cm]{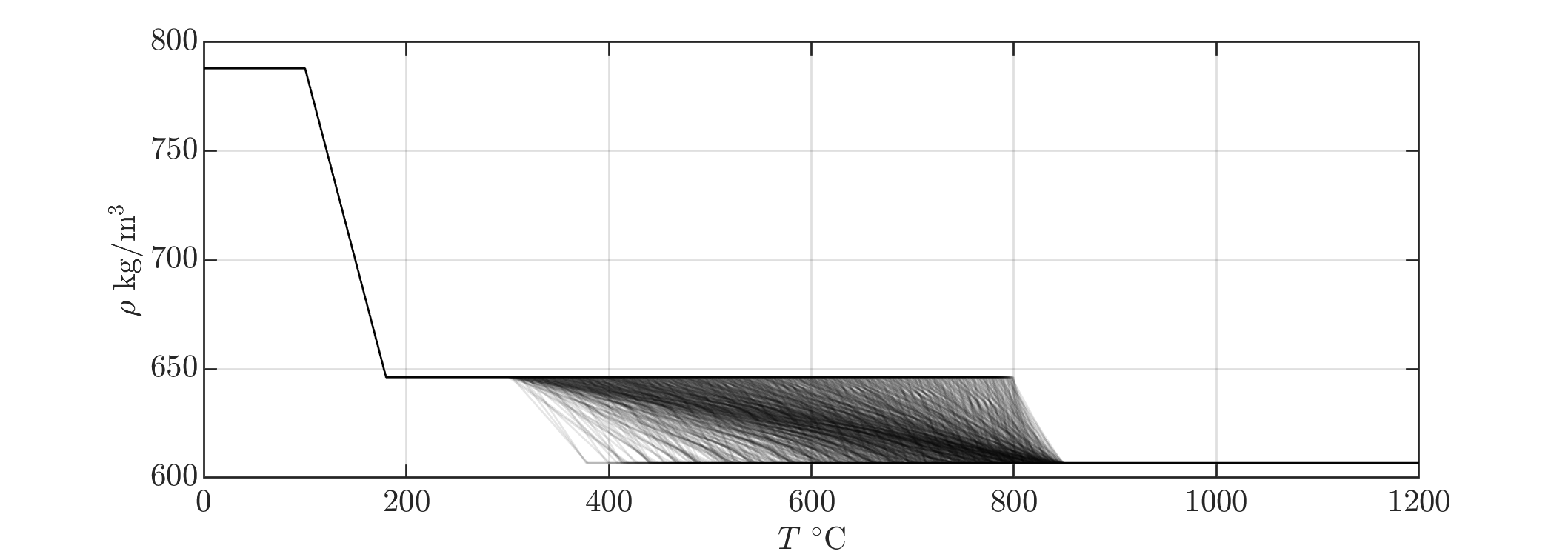}
	}}%
	\caption{Realizations of temperature-dependent effective 	
		material properties in their respective ranges as defined in 
		Table~\ref{tab:parametrization}.}%
	\label{fig:priorMaterialProp}%
\end{figure}

To finalize the parametrization, reasonable ranges are defined for all 
parameters.
These ranges correspond to bounds on the parameters that are the 
results of prior calibration attempts along with expert judgement. 
These ranges are given along with a summary of the parameters in 
Table~\ref{tab:parametrization}
with plots of the resulting temperature-dependent material properties
in Figure~\ref{fig:priorMaterialProp}. 

These 
six parameters are gathered into a vector 
$\BParamsM=(X_1,\dots,X_6)\trans$,
which fully characterizes the temperature-dependent behaviour of the
gypsum insulation boards.

\begin{table}[!ht]
	\caption{Summary of the parameters $(X_1,\dots,X_6)\trans$ that describe 
		the material
		properties with their respective ranges.}
	\label{tab:parametrization}
	\centering
	\begin{tabular}{llcc}
		\hline
		Parameter & Physical Meaning & Range & Unit\\
		\hline
		$X_1$ & Start of second key process & $[300, 800]$ & 
		$\degC$ \\
		$X_2$ & Main effect of second key process & $[0.1,1]$ & - 
		\\
		& (relative between 
		$X_{1}$ and $850\!\degC$) & &\\
		$X_3$ & $\lambda(180\!\degC)$ and $\lambda(X_1)$ & 
		$[0.1, 0.25]$ & $\wpmk$\\
		$X_4$ & $\lambda(1200\!\degC)$ & $[0.1, 1.2]$ & $\wpmk$\\
		$X_5$ & $c(140\!\degC)$ & $[1.4\cdot 10^{4},6.5\cdot 10^{4}]$ & 
		$\jpk$\\
		$X_6$ & $c(X_1+(850\!\degC-X_1)\cdot X_{2})$ & $[1\cdot 10^{3},8\cdot 
		10^{4}]$ & $\jpk$\\
		\hline
	\end{tabular}
\end{table}

\subsection{Finite element model}
\label{sec:expMod:FEModel}

The FE model $\cm(\BParams_{\cm})$ is considered as a verified 
simulator for the transient 
heat propagation in gypsum insulation panels under fire 
exposure. This means that the model is assumed to accurately solve the 
underlying differential equations posed by the mathematical heat 
transfer model. For a review of techniques for rigorous model 
verification see \citet{VVUQ:Oberkampf2004, VVUQ:Oberkampf2010}. 

The FE model yields a discretized time-dependent temperature curve
$\BParams_{\cm}\mapsto\BModelOut 
=  
(Y_1,\dots, Y_N)\trans$ for
each 
realization of the parameter vector 
$\BParams_{\cm}=(X_1,\dots,X_6)\trans$ that parameterizes the effective
thermal 
properties $\lambda(T,\BParams_{\cm})$, $c(T,\BParams_{\cm})$ and 
$\rho(T,\BParams_{\cm})$. The 
discretization of the time steps is identical to the available 
measurements, so that $t_i=i\tau$ with $\tau=10~\mathrm{s}$.

\section{\modif{Model calibration}}
\label{sec:BayesCal}
\modif{The process of finding model parameters so that the model 
	evaluation using this parameter vector agrees with some observations 
	is called calibration. A general probabilistic framework for calibration 
	is presented next. For simplicity, it is assumed that only one
	measurement series $\Bdata=\Bdata^{(s)}$ is available for now. This 
	restriction is
	lifted in Section~\ref{sec:results}.}

%
%

\subsection{\modif{The Bayesian calibration approach}}
\label{sec:BayesCal:approach}

\modif{All models are simplifications of reality and all 
	observations made in the real world contain measurement 
	errors. To explicitly account for this combined mismatch between model 
	output and observations, one option is to model the discrepancy as an 
	\emph{additive} mismatch between the model predictions and the 
	observations:
	\begin{equation}
	\label{eq:additiveDiscrepancy}
	\Bdata = \cm(\BParamsM) + \ve{E}.
	\end{equation}}

\modif{One way to address the calibration 
	problem of determining $\BParamsM$ is to formulate it in a probabilistic 
	setting.}
The unknown model 
discrepancy $\ve{E}$ from Eq.~\eqref{eq:additiveDiscrepancy} is then seen 
as a 
random vector. Commonly and in this work, $\ve{E}$ is 
assumed to follow a zero 
mean 
normal distribution with a covariance matrix parameterized by a set of 
parameters $\BParamsE$:
\begin{equation}
\label{eq:additiveDiscrepancyRV}
\ve{E}\sim\cn(\ve{\varepsilon}\vert\ve{0},\errorCovPMult).
\end{equation}

Additionally, in the probabilistic setting, the combined 
parameter vector $\BParams 
= (\BParamsM,\BParamsE)\in\cd_{\BParams}$ is assumed to be
distributed according to a so-called prior distribution (to be further 
specified)
\begin{equation}
\BParams\sim\Bprior.
\end{equation}

Then the discrepancy distribution from 
Eq.~\eqref{eq:additiveDiscrepancyRV} can be used together with 
Eq.~\eqref{eq:additiveDiscrepancy} to construct a model giving the probability 
of the observations given a realization of the parameter vector. Denoting by  
$\Bparams 
= (\BparamsM,\BparamsE)$ a realization of $\BParams$, this probability as a 
function of the parameters is 
the so-called \emph{likelihood 
	function}
\begin{equation}
\label{eq:likelihoodDef}
\cl(\BparamsM, \BparamsE; 
\Bdata)=\cn(\Bdata\vert\cm(\BparamsM), 
\errorCovPMult),
\end{equation}
which reads more explicitly:
\begin{equation}
\label{eq:likelihoodExplicit}
\cl(\BparamsM, \BparamsE; 
\Bdata)=\frac{1}{(2\pi)^{N/2}\det{\errorCovPMult}}\exp{\left[ 
	(\cm(\BparamsM)-\Bdata)\trans\errorCovPMult^{-1}(\cm(\BparamsM)-\Bdata) 
	\right]}.
\end{equation}

With these definitions, it becomes possible to apply the  
Bayes' theorem for conditional probabilities \citep{Bayesian:Gelman2014:3rd}: 
\begin{equation}
\label{eq:BayesTheorem}
\Bpost = \frac{\Blikelifun\Bprior}{Z}, \quad \text{with} \quad Z = 
\int_{\cd_{\BParams}}\Blikelifun\Bprior\,\di{\Bparams},
\end{equation}
where $\Bprior$ is the \emph{prior distribution} of the input parameters 
$\BParams$, $\Bpost$ is the \emph{posterior distribution} and $Z$ is a 
normalizing factor called \emph{evidence}.

The probability distributions $\pi(\cdot)$
in 
this expression can be interpreted as degrees of belief about the 
parameters $\BParams$ \citep{Bayesian:Beck1998}. 
Low values of the distribution at a realization $\BParams$ indicate 
low confidence in this particular value, whereas high values indicate 
high confidence. With this interpretation of probabilities, 
Eq.~\eqref{eq:BayesTheorem} encodes the shift of belief about the
parameter vector from \emph{before} $\BParams\sim\Bprior$ to \emph{after} the 
observation of experiments $\BParams\vert\Bdata\sim\Bpost$. 
This process is called \emph{Bayesian updating}, \emph{Bayesian 
	inference} or
\emph{Bayesian inversion}.

As mentioned above, the probability distributions in Bayes' theorem are named 
according to 
their 
information content about the parameters $\BParams$ in the setting of 
the updating procedure:

\begin{description}
	\item[Prior distribution $\Bprior$:] this distribution captures the 
	belief about the parameters before (\ie \emph{prior} to) observing 
	data. In the setting of Bayesian updating for model calibration, it 
	is chosen according to expert opinion and possible prior
	calibration attempts. A typical choice is to select a reasonable, although 
	sufficiently large range (lower/upper bounds) for each parameter.
	\item[Posterior distribution $\Bpost$:] the posterior distribution 
	is the conditional distribution of the parameters given the 
	observations. It can be regarded as the state of information about 
	the parameters $\BParams\vert\Bdata$ after (\ie $posterior$ to) making 
	observations.
\end{description}

Thus, the computation of the posterior distribution $\Bpost$ can be 
considered as the solution of the calibration problem. Since it is a 
probability distribution rather than a single value, it encompasses 
all information specified by the prior distribution and the newly 
observed data. Already conceptually it is thus a much broader way of 
defining calibration than single value estimators.

Another probability distribution of interest in the Bayesian inference 
setting is the \emph{posterior predictive distribution} $\Bpostpred$. 
It is defined as
\begin{equation}
\label{eq:postPred}
\Bpostpred = 
\int\cl(\Bparams;\Bdata^*)\Bpost~\mathrm{d}\Bparams.
\end{equation}

This distribution expresses beliefs about future (\ie predictive) 
observations $\Bdata^*$ given the already observed ones 
$\Bdata$. If it is possible to sample from the posterior distribution, \ie 
$(\BparamsM,\BparamsE)\sim\Bpost$, 
a sample from the posterior predictive distribution is obtained by 
drawing
\begin{equation}
\BData^{\text{postpred}} 
\sim\cn(\Bdata^*\vert\cm(\BparamsM),\mat{\Sigma}(\BparamsE)),
\quad 
\text{where} \quad 
(\BparamsM,\BparamsE)\sim\Bpost.
\end{equation}

The posterior predictive distribution allows to assess the predictive 
capabilities of the model following calibration. It contains the 
uncertainty about the model parameters $\BParamsM$ and the mismatch 
parameters 
$\BParamsE$. Because this distribution is defined 
in 
the 
space where the data $\Bdata$ are collected, it can be used to visually
check the calibration results.

\subsection{Sampling from the posterior distribution}
\label{sec:BayesCal:sampling}
The analytical computation of the posterior distribution is typically 
not possible. This is mostly due to difficulties in evaluating the 
normalizing constant $Z$ defined in Eq.~\eqref{eq:BayesTheorem}. Its 
computation typically relies on 
estimating an integral in the parameter space, which is in most cases 
intractable.

A breakthrough technique called \emph{Markov chain Monte Carlo} (MCMC) 
sampling, originally developed by 
\citet{MCMC:Metropolis1953} and \citet{MCMC:Hastings1970}, completely avoids 
the 
need to evaluate this high-dimensional integral. It is a type of 
stochastic simulation technique that
constructs Markov chains that are guaranteed to produce samples 
distributed according to the posterior distribution. 
Posterior characteristics (\eg quantities of interest, expected 
values, 
marginal distributions etc.) can then be estimated using this sample. 

Following the initial development of the MH algorithm 
\citep{MCMC:Metropolis1953,MCMC:Hastings1970}, recent developments to 
improve the algorithm's efficiency strived towards adaptive proposal 
distributions 
\citep{MCMC:Haario2001,MCMC:Roberts2009} and the utilization of 
gradient information \citep{MCMC:Rossky1978,Bayesian:MacKay2003}. One 
common flaw of these algorithms, however, is the requirement to tune 
them using a set of tuning parameters. This is a particularly tedious 
task that is a major source of error in practical applications. 

The \emph{affine-invariant ensemble sampler} (AIES, \citep{MCMC:Goodman2010}) 
is a fairly recent MCMC algorithm that performs particulary well in 
this respect. This algorithm 
requires only a single tuning parameter and its performance is invariant to 
affine transformations of the target distribution. This property makes 
it particularly useful for real-world applications, where strong 
correlations between individual parameters often hinder conventional 
MCMC algorithms. 

The AIES algorithm relies on a set of parallel chains where proposal 
samples are obtained by moving in the direction of a randomly chosen 
conjugate sample from a different chain. The pseudo-code for the 
implementation used in this paper is given in Algorithm~\ref{ag:AIES}.

A property that makes MCMC algorithms especially suitable for Bayesian 
computations is that they do not require the explicit computation of 
the normalization constant
$Z$ (from Eq.~\eqref{eq:BayesTheorem}), as only 
a posterior ratio, called \emph{acceptance ratio}, is required 
(Step~\ref{ag:stepAccept} of 
Algorithm~\ref{ag:AIES}). In this ratio, 
$Z$ cancels out. However, the computationally expensive 
forward model must be evaluated each time this acceptance ratio is computed. 
This 
necessity of many runs of computationally expensive models has spurred the 
idea of constructing a \emph{surrogate model} that, after 
successful construction, can be used in the MCMC algorithms in lieu of the 
original model, whereby  
the overall computational burden is reduced to feasible levels.

\begin{algorithm}
	\caption{Affine-invariant ensemble sampler \citep{MCMC:Goodman2010}}
	\label{ag:AIES}
	\begin{algorithmic}[1]
		\Procedure{AIES}{$\Bparams^{(1)}_0,\dots,\Bparams^{(L)}_0,a$, 
			NSteps}
		\For{$i\gets 1,$ NSteps}
		\For{$l\gets 1, L$}
		\State $\tilde{\Bparams}\gets\Bparams^{(l)}_{i-1}$ 
		\State $\Bparams^{*}\gets\Bparams^{(k)}_{i-1}$ with 
		$k \in \{1,\dots,L\} \backslash\{l\}$ chosen randomly
		\State Sample $z\gets\ve{Z}\sim g(z)\propto 
		\begin{cases}
		\frac{1}{\sqrt{z}} 
		&\text{if}~z\in\left[\frac{1}{a},a\right]\\
		0 &\text{otherwise}
		\end{cases}$
		\State $\hat{\Bparams}\gets\tilde{\Bparams}+ 
		z(\Bparams^*-\tilde{\Bparams})$
		\State Sample $u\gets U\sim\cu(0,1)$
		
		\If{$z^{M-1}\frac{\pi(\hat{\Bparams}\vert\Bdata)\label{ag:stepAccept}
			}{\pi(\tilde{\Bparams}\vert\Bdata)}>u$}
		\Comment{$M$ is dimension of $\Bparams\in\mathbb{R}^{M}$}
		\State $\Bparams^{(l)}_{i}\gets\hat{\Bparams}$
		\Else
		\State $\Bparams^{(l)}_{i}\gets\tilde{\Bparams}$
		\EndIf
		\EndFor
		\EndFor
		\EndProcedure
	\end{algorithmic}
\end{algorithm}

\subsection{Surrogate modelling of the temperature time series}
\label{sec:BayesCal:surrogate}
Often, sampling based techniques (\eg MCMC algorithms) are considered 
infeasible because of the high number of computationally expensive 
model runs $\cm(\BParamsM)$ required. Surrogate modelling techniques try to 
solve this problem by 
constructing a computationally cheap emulator that can be used instead of 
the original model.

Non-intrusive approaches to construct surrogate models are solely based on 
realizations of the 
model 
parameters and 
corresponding model outputs
\citep{Choi2004a}. The 
set of parameters used for constructing the surrogate is referred to as 
\emph{experimental design}, which means here a \emph{set of computer 
	experiments} and shall no be confused with physical experiments. Following 
	the 
assembly 
of the experimental design, the constructed surrogate 
model aims to approximate the original model predictions denoted by 
$\cm^{\mathrm{PC}}$,
\begin{equation}
\cm^{\mathrm{PC}}(\BParamsM) \approx \cm(\BParamsM).
\end{equation}

This section details the construction of a surrogate model combining 
\emph{polynomial chaos expansions} (PCE) with the \emph{principal 
	component analysis}
(PCA).

\subsubsection{Polynomial Chaos Expansions}
\label{sec:BayesCal:surrogate:PCE}

\emph{Polynomial chaos expansions} (PCE) are a surrogate modelling 
technique 
that has been used extensively in the engineering disciplines 
\citep{Xiu2002,Soize2004,Guo2018} to 
construct surrogate models of scalar-valued functions of random 
variables. A brief 
introduction to the method is presented next.

Assume a random vector 
$\BParams=(X_1,\dots,X_M)$ with mutually independent components 
$X_i\sim\pi_i(x_i)$. 
Its joint probability density function is thus given by
\begin{equation}
\label{eq:inputIndependence}
\Bprior = \prod_{i=1}^M\pi_i(x_i).
\end{equation}

The functional inner product of two polynomials 
$\psi_{k}^i,\psi_{l}^i:x_i\in\cd_{X_i}\mapsto\mathbb{R}$ of degree $k$ and $l$ 
respectively, is then defined by 
\begin{equation}
\label{eq:orthonormality}
\left\langle\psi_{k}^i,\psi_{l}^i\right\rangle_{\pi_i} 
\eqdef 
\int_{\cd_{X_i}}\psi_{k}^i(x_i)\psi_{l}^i(x_i)\pi_i(x_i)\,\di{x_i}.
\end{equation}

By choosing these polynomials to fulfil 
$\left\langle\psi_{k}^i,\psi_{l}^i\right\rangle_{\pi_i} = \delta_{k,l}$, \ie
$\delta_{k,l}=1$ if $k=l$ and 0 otherwise, these polynomials form a family of 
\emph{univariate orthonormal polynomials} $\{\psi_k^i\}_{k=0}^{\infty}$. There 
exist well-known families of 
polynomial 
functions that fulfil the fundamental condition of 
Eq.~\eqref{eq:orthonormality} w.r.t.\ standard parametric probability
distributions $\pi_i$ \citep{Askey1985}. 

These univariate polynomials can be used to build \emph{multivariate} 
polynomials by tensor product. Introducing the multi-indices 
$\ve{\alpha}=(\alpha_i,\dots,\alpha_M)\in\mathbb{N}^{M}$ the latter are defined 
by:
\begin{equation}
\Psi_{\ve{\alpha}}(\Bparams)\eqdef\prod_{i=1}^M\psi_{\alpha_i}^i(x_i).
\end{equation}

It can be shown that the univariate orthonormality property of 
Eq.~\eqref{eq:orthonormality} extends to the multivariate case and that the 
following holds:
\begin{equation}
\label{eq:orthonormalityMulti}
\left\langle\Psi_{\ve{\alpha}},\Psi_{\ve{\beta}}\right\rangle_{\pi} 
\eqdef 
\int_{\cd_{\BParams}}\Psi_{\ve{\alpha}}(\Bparams)\Psi_{\ve{\beta}}(\Bparams)\Bprior\,\di{\Bparams}
= \delta_{\ve{\alpha},\ve{\beta}}.
\end{equation}

These polynomials $\{\Psi_{\ve{\alpha}}\}_{\ve{\alpha}\in\mathbb{N}^M}$ form a
so-called \emph{orthonormal basis} of the space of \emph{square 
	integrable 
	functions} with respect to the probability distribution $\Bprior$. Any such 
function can be represented by:
\begin{equation}
\label{eq:PCEfull}
f(\Bparams) = \sum_{\ve{\alpha}\in\mathbb{N}^M} 
a_{\ve{\alpha}}\Psi_{\ve{\alpha}}(\Bparams),
\end{equation}
where $a_{\ve{\alpha}}\in\mathbb{R}$ are the \emph{coefficients} of the 
expansion.

In practical applications it is not feasible to compute the infinite number of 
coefficients $a_{\ve{\alpha}}\in\mathbb{N}^{M}$. Instead, a \emph{truncation 
	scheme} is 
typically proposed that reduces the number of considered polynomials to a 
finite set. This truncated set denoted by $\ca\subset\mathbb{N}^M$ transforms 
the equality of Eq.~\eqref{eq:PCEfull} to an approximation
\begin{equation}
\label{eq:PCEapprox}
f(\Bparams) \approx f^{\mathrm{PCE}}(\Bparams) \eqdef \sum_{\ve{\alpha}\in\ca} 
a_{\ve{\alpha}}\Psi_{\ve{\alpha}}(\Bparams).
\end{equation}

In regression-based approaches, the coefficient vector 
$\ve{a}\in\mathbb{R}^{\card{\ca}}$ is 
typically estimated by 
least-squares analysis, as originally proposed in 
\citet{Berveiller2006}. This 
corresponds to selecting a truncation set $\ca$ \citep{BlatmanJCP2011} and 
using an \emph{experimental 
	design} $\cx\eqdef\{\Bparams^{(i)},i=1,\dots,K\}$ to minimize the 
expression
\begin{equation}
\label{eq:leastSquares}
\tilde{\ve{a}} = 
\arg\min_{\ve{a}\in\mathbb{R}^{\card{\ca}}} 
\frac{1}{K}\sum_{i=1}^K\left(f(\Bparams^{(i)})- \sum_{\ve{\alpha}\in\ca} 
a_{\ve{\alpha}}\Psi_{\ve{\alpha}}(\Bparams^{(i)})\right)^2.
\end{equation}

By storing the function evaluations at $\cx$ in a vector 
$\cy\eqdef\{f(\Bparams^{(1)}),\dots,f(\Bparams^{(K)})\}$ the solution of 
Eq.~\eqref{eq:leastSquares} reads:
\begin{equation}
\label{eq:leastSquaresSolution}
\tilde{\ve{a}} = (\mat{B}\trans\mat{B})^{-1}\mat{B}\trans \cy,
\end{equation}
where $\mat{B}=\{B_{ij}\eqdef\Psi_{j}(\Bparams^{(i)}), i=1,\dots,K, 
j=1,\dots,\card{\ca}\}$ are the evaluations of the basis polynomials 
$\Psi_{\ve{\alpha}}$ on the experimental design $\cx$. 

To assess the accuracy of the obtained polynomial chaos expansion, the 
so-called \emph{generalization error} 
$\expc{(f(\BParams)-f^{\mathrm{PCE}(\BParams)})^2}$ shall be evaluated. A 
robust error measure  can be obtained by using the \emph{leave-one-out} 
(LOO) cross validation technique. This estimator is obtained by
\begin{equation}
\label{eq:LOO}
\varepsilon_{\mathrm{LOO}} = 
\frac{1}{K}\sum_{i=1}^{K}\left( 
f(\Bparams^{(i)}) - 
f^{\mathrm{PCE}}_{\sim i}(\Bparams^{(i)}) \right)^{2},
\end{equation}
where $f^{\mathrm{PCE}}_{\sim i}$ is constructed by leaving out the $i$-th 
point from 
the experimental design. After some algebraic manipulation, it can be shown 
that the LOO error can be computed as a mere post-processing of the PCE 
expansion as follows
\begin{equation}
\varepsilon_{LOO} = \frac{1}{K}\sum\limits_{i = 1}^K \left( 
\frac{f(\Bparams^{(i)}) - 
	f^{PCE}(\Bparams^{(i)})}{1-h_i}\right)^2,
\end{equation}
where $h_i$ is the $i^{th}$ component of the vector given by: 
\begin{equation}
\Ve{h} = \text{diag}\left(\mat{B}(\mat{B}\trans\mat{B})^{-1} 
\mat{B}\trans\right),
\end{equation}
for more details refer to \citet{BlatmanPEM2010}.

This section outlined the approach to use PCE for approximating scalar 
quantities. Since the heat transfer model $\BModelOut = \cm(\BParamsM)$ 
considered in 
this paper returns 
a vector of interface temperatures at $601$ time steps, a pure PCE 
approach would require the construction of $N=601$ independent 
\emph{polynomial chaos expansions}. Instead, 
a dimensionality reduction technique on the output is applied before 
using the PCE technique.

\subsubsection{Principal Component Analysis}
\label{sec:BayesCal:surrogate:PCA}

Because the discretized temperature evolution $\BModelOut$ is expected to be 
rather 
smooth (see Figure~\ref{fig:conv}), considerable 
correlation between the 
individual time steps is expected. 
This correlation can be exploited to reduce the dimensionality of the 
output in the context of surrogate modelling. 

There exist numerous so-called \emph{dimensionality reduction techniques} 
\citep{Maaten09}, one of which is \emph{principal 
	component analysis} 
(PCA, \citet{Statistics:Jolliffe2002}). The latter utilizes an 
orthogonal transformation to express $\BModelOut$ in a new basis of 
uncorrelated
\emph{principal 
	components} $\ve{Z}$. 

In practice, PCA is carried out by 
computing estimators of the 
expectation $\meanRespEst \approx \expe{\BModelOut}$ 
and 
the covariance matrix 
$\covRespEst\approx\cove{\BModelOut}$. The $N$ 
eigenvectors of 
this covariance matrix are denoted by $\ve{\phi}_{p}$ for 
$p=1,\dots,N$. The 
associated eigenvalue $\lambda_p$ corresponds to the variance of 
$\BModelOut$ 
in 
direction of the $p$-th principal component. Thereby the 
random vector $\BModelOut$ can be expressed through its $N$ principal 
components $z_p(\BParamsM)$ as $\BModelOut  =  \meanRespEst + 
\sum_{p=1}^{N}z_p(\BParamsM)\ve{\phi}_{p}$.

The model output can then be compressed to a lower dimensional subspace 
by 
retaining only those $N'$ principal components with the highest 
variance:
\begin{equation}
\label{eq:PCA}
\BModelOut \approx \BModelOut^{\mathrm{PCA}} =  \meanRespEst + 
\sum_{p=1}^{N'}z_p(\BParamsM)\ve{\phi}_{p}.
\end{equation}

The number of terms $N'$ is selected such that 
$\sum_{p=1}^{N'}\lambda_p=(1-\varepsilon_0)\sum_{p=1}^{N}\lambda_p$, with 
$\varepsilon_0$ typically chosen as $0.01$. This way, the model output 
$\BModelOut\in\mathbb{R}^{N}$ can be approximated by a linear 
transformation of the principal 
component vector $\ve{Z} = 
(z_1(\BParamsM),\dots,z_{N'}(\BParamsM))\trans$ thereby reducing the 
problem 
dimensionality from $N$ to $N'\ll N$.

\subsubsection{Combining PCA with PCE}
\label{sec:BayesCal:surrogate:PCAPCE}

The combination of PCA with PCE gives rise to an efficient surrogate 
modelling technique as shown originally in \citet{BlatmanICASP2011}. 
Constructing $N'$ 
polynomial chaos expansions of each retained principal component 
$z_p(\BParamsM)\approx 
z^{\mathrm{PCE}}_p(\BParamsM)= 
\sum_{\ve{\alpha}\in\ca}\PolyCoeffApproxPC\PolyBasis(\BParamsM)$, 
together with the PCA formulation from Eq.~\eqref{eq:PCA} yields a 
surrogate model relating the model parameters to the vector valued time 
series output of the transient heat transfer problem:
\begin{equation}
\label{eq:PCAPCE}
\BModelOut \approx \modelPC\eqdef\BModelOut^{\mathrm{PCA+PCE}} = 
\ve{\mu}_{\BModelOut} + 
\sum_{p=1}^{N'}\left(
\sum_{\ve{\alpha}\in\ca_p}\PolyCoeffApproxPC\PolyBasis(\BParamsM)
\right)\ve{\phi}_{p}, 
\end{equation}
which can be rewritten by introducing the union set 
$\ca^{\star}\eqdef\bigcup_{p=1}^{N'}\ca_p$:
\begin{equation}
\label{eq:PCAPCEComb}
\BModelOut^{\mathrm{PCA+PCE}} = 
\ve{\mu}_{\BModelOut} + 
\sum_{\ve{\alpha}\in\ca^{\star}}\sum_{p=1}^{N'}
\PolyCoeffApproxPC\PolyBasis(\BParamsM)
\ve{\phi}_{p}.
\end{equation}

For compactness, this equation can also be expressed in matrix form by 
letting $\ve{\Phi}=(\ve{\phi}_1,\dots,\ve{\phi}_{N'})$ be a $N\times 
N'$ matrix containing the retained eigenvectors
$\ve{\phi}_p=(\phi_{p1},\dots,\phi_{pN})\trans$.
For the PCE part of the equation the vector 
$\PolyBasisVec(\BParamsM)= 
\{\Psi_{\ve{\alpha}}(\BParamsM),\ve{\alpha}\in\ca^{\star}\}$ is 
introduced 
that holds the individual multivariate orthogonal polynomials. Let 
$\mat{A}$ be a $\card{\ca^{\star}}\times N'$ matrix that stores the 
corresponding PCE coefficients, then 
Eq.~\eqref{eq:PCAPCE} can be written as
\begin{equation}
\label{eq:surrogate}
\BModelOut \approx \ve{\mu}_{\BModelOut} + \ve{\Phi}
\left(\mat{A}\trans\PolyBasisVec(\BParamsM)\right).
\end{equation}

For completeness, the response can also be expressed for each random variable 
$Y_t$ individually. For this, the
row vector $\ve{\phi}^{\mathrm{row}}_{t}=(\phi_{1t},\dots,\phi_{N't})$,
taken from the $t$-th row of the matrix of eigenvectors
$\ve{\Phi}$, is introduced:
\begin{equation}
\label{eq:surrogateIndiv}
Y_t \approx \mu_{Y_t} + \ve{\phi}^{\mathrm{row}}_{t}
\mat{A}\trans\PolyBasisVec(\BParamsM).
\end{equation}

This surrogate model can then be used in lieu of the 
original computationally expensive forward model. The evaluation of 
the surrogate model is orders of magnitude faster than the original 
finite element model. For comparison, in our application example a single FE 
run takes about $1~\mathrm{min}$ 
on a conventional computer, while in the same time $10^7$ 
evaluations of the surrogate model can be made.

This reduction in computational time is a promising feature of the 
presented surrogate modelling technique. It does, however, come at the 
cost of a series of approximations that are introduced during the PCA 
and PCE computation. To ensure confidence in the produced surrogate 
model, a general error measure has to be devised. It includes the approximation 
error due to the PCA truncation and the truncated polynomial chaos expansion. 
Such an error
measure $\tilde{\eta}$ was derived in \citet{BlatmanIcossar2013}. For the sake 
of completeness 
the details are given in \ref{sec:SurrogateError}.

\subsection{Summary of the proposed method}
\label{sec:BayesCal:summary}

In this section, a procedure to efficiently conduct Bayesian 
inference with expensive vector valued models was presented. It is 
assumed that the 
parametrization of the temperature-dependent effective material properties 
(see Section~\ref{sec:expMod:parametrization}) is known. Bayesian 
inference then aims at determining the distribution of the parameters 
$\BParams\vert\Bdata\sim\Bpost$ after observations (\ie experimental 
measurements) have been made. A brief step-by-step 
account of this procedure is given below for reference:

\begin{enumerate}[label=\textbf{Step \arabic*}]
	\item \label{enum:start}Choose a prior distribution $\Bprior$ on 
	$\BParams$ and 
	construct an experimental design $\cx$ using $K$ 
	samples from 
	this 
	prior. Evaluate the forward model at $\cx$ and store the evaluations in 
	$\cy$.
	\item Approximate $\cm(\BParamsM)$ using the surrogate model 
	$\cm^{\mathrm{PC}}(\BParamsM)$ from 
	Eq.~\eqref{eq:surrogate}. This requires the combination of the 
	dimensionality reduction technique PCA with the PCE uncertainty 
	propagation technique.
	\item Compute the error estimate $\tilde{\eta}$ from 
	Eq.~\eqref{eq:errorEst}. This error is only valid over the prior 
	domain. If it is too large, enrich the 
	experimental design by increasing the number of samples $K$ and 
	restart from \textbf{\ref{enum:start}}. \modif{The size of the admissible 
		error depends on the application but should typically not exceed $5\%$.}
	\item Define a likelihood function $\Blikelifun$ from 
	Eq.~\eqref{eq:likelihoodExplicit} that captures the discrepancy between 
	a 
	model run and the observations.
	\item Run the AIES defined in Algorithm~\ref{ag:AIES} where the 
	likelihood function uses the surrogate model $\modelPC$ instead of 
	the original model $\modelOrig$ to obtain a sample from the 
	posterior distribution $\Bpost$. \label{sum:MCMC}
	\item Verify the fit of the calibrated model using a sample from 
	the posterior predictive distribution from 
	Eq.~\eqref{eq:postPred}. 
	Samples from the posterior predictive distribution can be obtained 
	by reusing parameter samples distributed according to the posterior 
	distribution from \ref{sum:MCMC}.
\end{enumerate}

This method works if the support domain of the prior distribution contains that 
of the 
posterior distribution. In this respect, sufficiently large prior ranges shall 
be selected based on the expert's judgment.

The successful calibration of the parameters through the 
Bayesian inference approach gives insight into the model mismatch and 
correlation structure between individual parameters. The 
distribution of the parameters can further be used in probabilistic 
analysis using these models and, given new observations, can be updated 
to reflect beliefs incorporating the newly acquired information.

A fundamental ingredient of the presented approach is the necessity to define a 
parametrization of the thermal effective material 
properties as described in Section~\ref{sec:expMod:parametrization}. 
To judge the quality of the parametrization, it can be helpful to 
assess the relative 
importance of a single model parameter with respect to the output. For 
this, it is necessary to resort to the field of 
sensitivity analysis.

\section{Sensitivity analysis}
\label{sec:sensitivity}

\subsection{PCE based Sobol' indices}
\label{sec:sensitivity:Sobol}

Global sensitivity analysis aims at finding which input parameters of a 
computer model (or combination thereof) explain at best the uncertainties in 
the model predictions. In this respect, variance decomposition techniques rely 
on 
assigning fractions of the model output variance 
$\varc{Y}=\varc{\BModel}$ to the individual input parameters 
$X_i$. For 
simplicity, in this section the subscript $(\cdot)_{\cm}$ from the 
parameter vector $\BParams = \BParams_{\cm}$ is dropped. 

Consider a scalar-valued computational model $\cm 
:\BParams\in\left[0,1\right]^M\mapsto\cm(\BParams)\in\mathbb{R}$, which maps a 
vector of input parameters in the unit hypercube to the
real numbers. This computational model can be 
decomposed into a sum of terms that only depend on a subset of 
the input parameters, \ie a constant $\cm_0$, univariate functions 
$\{\cm_i(X_i),i=1,\dots,M\}$, bivariate functions \etc
\begin{equation}
\label{eq:hoeffSobol}
\cm(\BParams) = \cm_0 + \sum_{i=1}^M\cm_i(X_i) + \sum_{1\le i < 
	j 
	\le M} \cm_{ij}(X_i,X_j) + \cdots +  
\cm_{1,2,\dots,M}(X_1,\dots,X_M).
\end{equation}

This decomposition is called the \emph{Hoeffding-Sobol' decomposition} and is 
unique for any function $\cm$ that is 
square-integrable over the unit hypercube \citep{Sobol1993}.

Denoting by $\iu\eqdef\{i_1,\dots,i_s\}\subset\{1,\dots,M\}$ a subset 
of indices, Eq.~\eqref{eq:hoeffSobol} can be written in short:
\begin{equation}
\cm(\BParams) = \cm_0 + \sum_{\iu\subset\{1,\dots,M\}}\cm_{\iu}(\BParams_{\iu}).
\end{equation}

It can be shown that the terms of this equation called \emph{summands}, are 
orthogonal \citep{Sobol1993}. The variance of each term 
$\cm_{\iu}(\BParams_{\iu})$, called 
\emph{partial variance}, is obtained by:
\begin{equation}
\label{eq:sobolInt}
D_{\iu}\eqdef 
\int_{\left[0,1\right]^{\card{\iu}}}\cm_{\iu}^2(\Bparams_{\iu})\,\di{\Bparams_{\iu}}.
\end{equation}

Due to the orthogonality of the terms in this equation, the total variance of 
the model output $D=\varc{\BModel}$ is finally obtained as the sum of the 
partial variances
\begin{equation}
D = \sum_{\iu\subset\{1,\dots,M\}}D_{\iu}.
\end{equation}

Each partial variance describes the amount of the output variance that can be 
attributed to the interaction of the input variables 
$\BParams_{\iu}$. In particular, $D_i$ describes the fraction of the 
variance that can be attributed to one input variable $X_i$ taken separately.

Moreover, the total contribution to the variance attributable to a single 
input parameter $X_i$ is captured in the sum of the partial variances 
$D_{\iu}$ that contain the $i$-th input variable. The sum of these 
partial variances normalized by the 
total variance is called the $i$-th \emph{total Sobol' 
	index} and is defined as
\begin{equation}
S^{T}_{i} \eqdef \frac{1}{D}\sum_{\iu\supset\{i\}}D_{\iu}.
\end{equation}

\modif{It is noted here that the sum of all total Sobol' indices, {\em i.e.}, 
	$\sum_{i\in\{1,\dots,M\}}S_i^T$, is larger than one because the same 
	interaction effect contributes to multiple total Sobol' indices.}
Usually the integral in Eq.~\eqref{eq:sobolInt} can be 
computed through Monte Carlo integration. However, if the model $\cm$ is 
expressed in an orthogonal basis (as is the case for 
PCE, Eq.~\eqref{eq:PCEfull}), the Sobol' indices can be computed 
analytically 
by post-processing the PCE coefficients $\PolyCoeff$ 
\citep{SudretCSM2006, SudretRESS2008b}:
\begin{equation}
S^{T}_{i} = \frac{1}{D}\sum_{\ve{\alpha}\in\ca_{i>0}} \PolyCoeff^2, \quad 
\text{with} \quad 
\ca_{i>0}=\{\ve{\alpha}\in\ca:\alpha_i>0\},
\end{equation}
\ie $\ca_{i>0}$ is the set of multivariate polynomials that are 
non-constant in the $i$-th input parameter $X_i$. For scalar-valued models, 
this yields a measure of the variance fraction that can 
be attributed to a certain input parameter. In the following section, 
this concept is extended to models with multiple outputs.

\subsection{PCA-based Sobol' indices}
\label{sec:sensitivity:vectorSobol}

In the present paper models with multiple outputs (\ie time-series of computed 
temperatues) are considered. By 
using a 
surrogate model that combines PCE with PCA, as discussed in 
Section.~\ref{sec:BayesCal:surrogate:PCAPCE}, the total Sobol' 
indices for each output vector component (\ie time step) can also be 
computed 
analytically \citep{MarelliICASP2015, Nagel2019}. 

For this, the partial variances of the model response components $Y_t$ are 
computed by using the expression from Eq.~\eqref{eq:surrogateIndiv}. The total 
Sobol' index for the $t$-th component of the output
random vector then reads
\begin{equation}
S_{i,t}^{T} = 1 -
\frac{\sum_{\ve{\alpha}\in\ca^{\star}_{i=0}}\left(\sum_{p=1}^{N'}\phi_{pt} 
	\PolyCoeffApproxPC\right)^2}{\sum_{\ve{\alpha}\in\ca^{\star}} 
	\left(\sum_{p=1}^{N'}\phi_{pt}\PolyCoeffApproxPC\right)^2}.
\end{equation}
where $\ca_{i=0}^{\star}$ is the subset of $\ca^{\star}$ for 
which $\alpha_i=0$. The interested 
reader 
is referred to \ref{sec:PCASobolDeriv} for the derivations.

\section{Results}
\label{sec:results}

In this section, the procedure presented in 
Sections~\ref{sec:BayesCal} and~\ref{sec:sensitivity} is 
applied to calibrate the temperature-dependent material properties of gypsum 
based insulation boards. The 
experimental data stems from experiments conducted by \citet{Breu2016}
and \citet{test:Gyproc2016}
that were presented in Section~\ref{sec:expMod:experiments}. 
As 
explained in Section~\ref{sec:expMod:parametrization}, the material 
properties are 
parameterized with a set of $6$ parameters. In the Bayesian 
inference framework introduced in Section~\ref{sec:BayesCal:approach}, 
determining the posterior distribution of these parameters constitutes 
the calibration of the temperature-dependent material properties. 

To further investigate the effects of the introduced parametrization, 
the surrogate models $\modelPC$ used for 
calibration
are reused to conduct time-dependent sensitivity analyses 
(Section~\ref{sec:sensitivity}). These
analyses show the influence each model parameter $X_i$ has on the
simulation output. They deliver valuable insights and can be used to
further refine the model parametrization.

Finally, the calibrated time-dependent material properties are validated
by simulating insulation panels in a different measurement setup and
comparing these simulation results with actual measurements.

\subsection{Calibration of material properties for gypsum boards}
\label{sec:results:calibration}
In this section the general calibration 
procedure from 
Section~\ref{sec:BayesCal:summary} is applied to the specific problem of 
calibrating heat transfer models describing the experiments of 
specimens 
E1-E4 (Test 1) presented
in Section~\ref{sec:expMod:experiments}.

The model parameters $\BParamsM$ are assumed to be priorly independent 
and uniformly distributed with the 
lower and upper bounds ($\underline{x}_i$ and $\overline{x}_i$ 
respectively) defined by the ranges 
given in Table~\ref{tab:parametrization}. The prior distribution of 
the model parameters is 
thus given by
\begin{equation}
\pi(\BparamsM) = \prod_{i=1}^6\cu (x_i;\underline{x}_i,\overline{x}_i).
\end{equation} 

Since multiple 
measurements
$\Bdata^{(s)}$ are available for each experiment, the formulation for
the likelihood Eq.~\eqref{eq:likelihoodDef} has to be slightly adapted.
Under the assumption of independence between the individual
measurement locations, it can be written as the product
\begin{equation}
\label{eq:likeliGyps}
\cl(\BparamsM,\BparamsE;\Bdata) = 
\prod_{s=1}^S\cn\left(\Bdata^{(s)};\cm^{\mathrm{PC}}(\BparamsM),\errorCovPMultS\right),
\end{equation}
which generalizes Eq.~\eqref{eq:likelihoodDef} where only a single time series 
of
measurements was considered. Consequently, the posterior distribution
obtained from Bayes' theorem should strictly be written as
$\pi(\Bparams\vert\Bdata^{(1)},\dots,\Bdata^{(S)})$, but for notational
simplicity the superscript $^{(s)}$ is again dropped. 

The covariance matrix $\errorCovPMultS$ is parametrized by
\begin{equation}
\label{eq:covMatrixParam}
\errorCovPMult = \{\Sigma(\BparamsE)_{ij} \eqdef \sigma_i\sigma_j 
R(t_i,t_j,\theta),\quad i,j = 1,\dots,N\},
\end{equation}
where we choose a so-called \emph{Mat\'ern $5/2$} autocorrelation function 
($h=t_i-t_j$):
\begin{equation}
R(h,\theta) = \left(1+\frac{\sqrt{5}\vert h 
	\vert}{\theta}+\frac{5h^2}{3\theta^2}\right)\exp\left(-\frac{\sqrt{5}\vert
	h \vert}{\theta}\right).
\end{equation}

In this autocorrelation function, $\theta$ is the correlation length and 
$\sigma_i$ is the standard 
deviation at the $i$-th time step. To reduce the 
number of calibration parameters, it is assumed that the standard 
deviation $\sigma(t)$ follows a degree-$6$ polynomial function
\begin{equation}
\label{eq:Results:standardDev}
\ve{\sigma} = (\sigma_1,\dots,\sigma_N), \quad \text{with} \quad \sigma_i 
\eqdef \sigma(t_i) = 
\sum_{k=0}^6 \varpi_k\psi_k\left(2\frac{t_i-t_1}{t_N}-1\right),
\end{equation}
where $\psi_k$ is the $k$-th Legendre polynomial (defined over $[-1,1]$) and 
$\varpi_k$ are the coefficients to be calibrated. As a summary, there are $8$ 
parameters to define our discrepancy term, namely $\{\varpi_0,\dots,\varpi_6\}$ 
which define the non-stationary variance of the model discrepancy, and the 
autocorrelation length $\theta$. Using the notation from 
Section~\ref{sec:BayesCal:approach}, we pose $\BParamsE 
=(\varpi_0,\dots,\varpi_6,\theta)$. To 
complete the prior information, a uniform distribution is assumed for 
the discrepancy parameters with non-restrictive bounds. A summary of 
the full prior distribution is given in Table~\ref{tab:priorDist}.

\begin{table}
	\caption{Summary of the prior distribution 
		$\Bprior=\prod_{i=1}^{14}=\pi_{i}(x_{i})$ for the parameter 
		vector $\BParams = 
		(X_{1},\dots,X_{14})\trans$}
	\label{tab:priorDist}
	\centering
	\resizebox{\textwidth}{!}{
		\begin{tabular}{rlllllc}
			\hline
			&& $\pi_{i}(x_{i})$ & $\mu$ & $\sigma$ & c.o.v. & 
			units \\ 
			\hline
			$\BParamsM:$&$X_{1}$ & $\cu(300, 800)$ & $5.50 \cdot 10^{2}$ & 
			$1.20 
			\cdot 10^{1}$ & $2.18 \cdot 10^{-2}$ & $\degC$\\
			& $X_{2}$ & $\cu(0.1, 1)$ &$5.50 \cdot 10^{-1}$ & $5.10 \cdot 
			10^{-1}$ 
			& $9.27 \cdot 10^{-1}$ & -\\
			& $X_{3}$ & $\cu(0.1, 0.25)$ &$0.175$ & 
			$6.58$ & $3.76 \cdot 10^{1}$ & $\wpmk$\\
			& $X_{4}$ & $\cu(0.1, 1.2)$ &$0.65$ & 
			$17.8$ & $2.74 \cdot 10^{1}$ & $\wpmk$\\
			&$X_{5}$ & $\cu(1.4\cdot 10^{4}, 6.5\cdot 10^{4})$ &$3.95 \cdot 
			10^{4}$ 
			& $1.21 \cdot 10^{2}$ & $3.07 \cdot 10^{-3}$ & $\jpk$\\
			& $X_{6}$ & $\cu(10^{3}, 8\cdot 10^{4})$ &$4.05 \cdot 10^{4}$ & 
			$1.51 
			\cdot 10^{2}$ & $3.73 \cdot 10^{-3}$ & $\jpk$\\
			$\BParamsE:$ & $X_{7}$ & $\cu(0,20)$ &$1.00 \cdot 10^{1}$ & $2.40$ 
			& 
			$2.40 \cdot 10^{-1}$ & $\degC$\\
			& $X_{8}$ & $\cu(-20,20)$ &$0$ & $3.40 $ 
			& - & $\degC$\\
			& $\vdots$ & $\vdots$ & $\vdots$ & $\vdots$ & $\vdots$ & $\vdots$\\
			& $X_{13}$ & $\cu(-20,20)$ &$0$ & $3.40 $ 
			& - & $\degC$\\
			& $X_{14}$ & $\cu(0,50)$ &$2.50 \cdot 10^{1}$ & $3.80 $ & 
			$1.52 \cdot 10^{-1}$ & s\\
			\hline
	\end{tabular}}
\end{table}

\begin{figure}
	\centering
	\subfloat[$\pi(\BparamsM\vert\Bdata)$]{{\includegraphics[width=9cm]{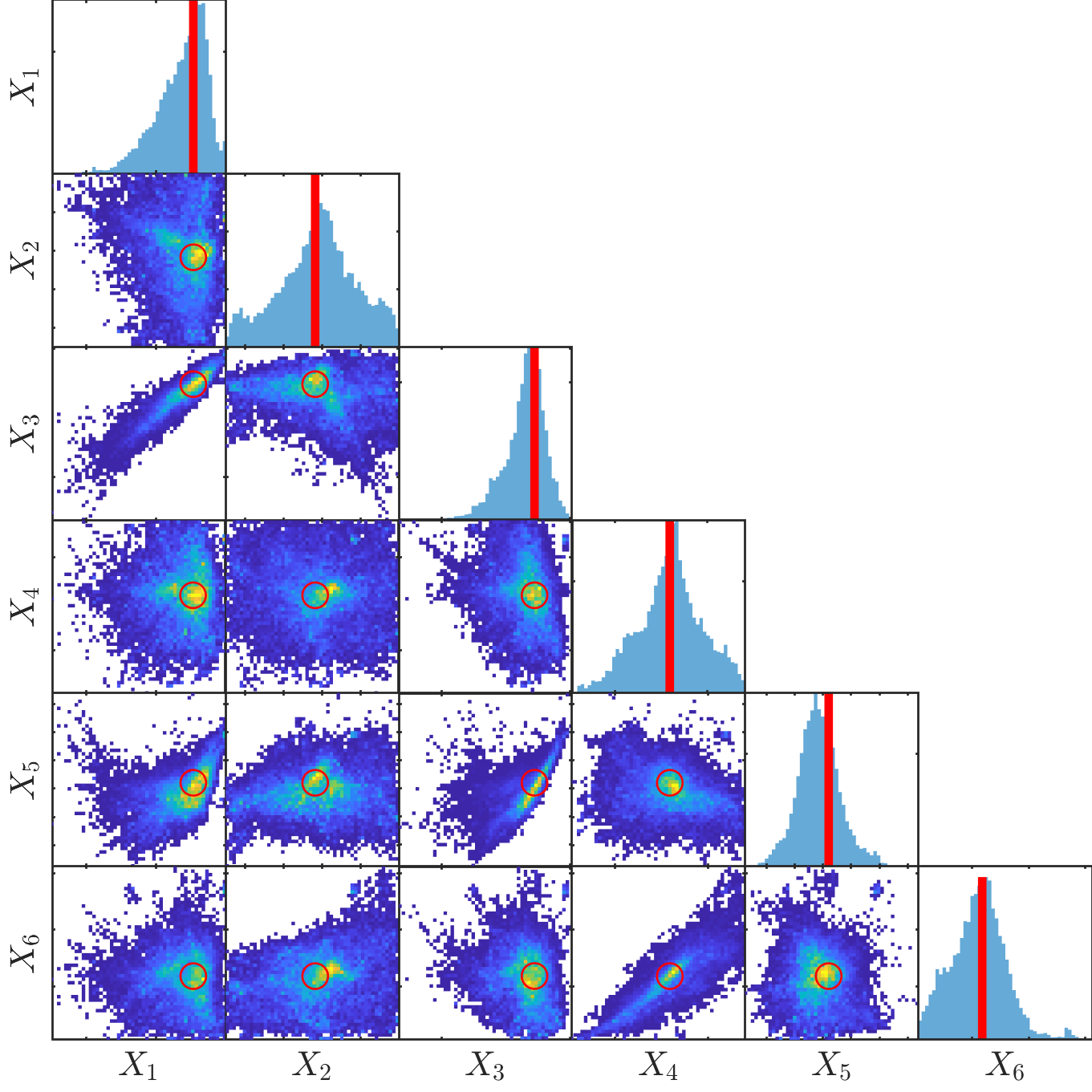}
	}}
	
	\subfloat[$\pi(\BparamsE\vert\Bdata)$]{{\includegraphics[width=9cm]{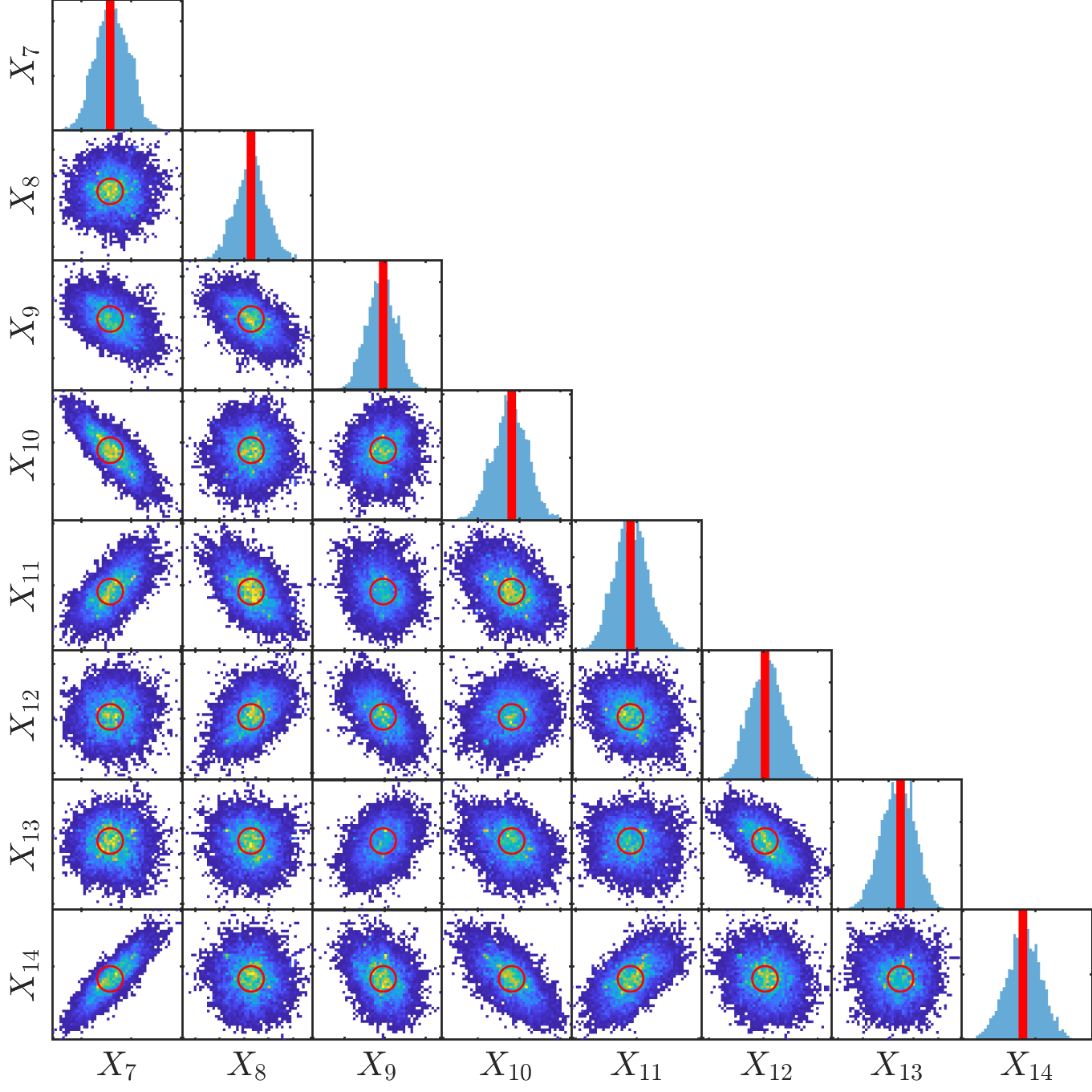}
	}}%
	\caption{Univariate and bivariate marginals from the posterior 	
		distribution of the model parameters $\pi(\BparamsM\vert\Bdata)$ 	
		and discrepancy parameters $\pi(\BparamsE\vert\Bdata)$ calibrated 	
		using the data from Product A (E1). The \modif{vertical} line (dot) 
		indicates the MAP parameter $\Bparams^{\mathrm{MAP}}$ defined 
		in Eq.~\eqref{eq:MAP}.}%
	\label{fig:posteriorSamples}%
\end{figure}

Figure~\ref{fig:posteriorSamples} shows the resulting sample points of 
the posterior distribution obtained 
for an
exemplary calibration run for the E1 setup. This sample was produced
using the previously presented AIES algorithm
(Algorithm~\ref{ag:AIES}). 

Despite the broad information contained in the full posterior plot, one
is often interested in the set of parameters that \emph{best} describe
the observations. In accordance with the Bayesian interpretation of
probabilities, this parameter set is located at the maximum value of
the posterior distribution (\emph{maximum a posteriori}, MAP). It can 
be found by solving the optimization
problem
\begin{equation}
\label{eq:MAP}
\Bparams^{\mathrm{MAP}} =
\underset{\Bparams}{\mathrm{arg}\,\mathrm{max}}\,\Bpost.
\end{equation}

This problem can be approximately solved by picking the parameter point from 
the available posterior sample that maximizes the unnormalized posterior 
distribution $\tilde{\pi}(\BParams\vert\Bdata)=\Blikelifun\Bprior 
\propto\Bpost$. The resulting
\emph{maximum a posteriori} estimator is also shown in
Figure~\ref{fig:posteriorSamples}. 

The calibrated posterior parameters for Product A (E1) are summarized in 
Table~\ref{tab:postStatE2}. It gives an overview of the calibrated parameters, 
including a set of summary statistics.

A major 
advantage full samples have compared to point estimators
is that they 
allow investigating
characteristics of the posterior distribution. This provides a fuller 
picture of
the calibrated parameter vector,\eg by showing dependence between individual
parameters $X_i$, allowing the computation of confidence intervals or revealing 
problems with identifiability.

Additionally, the full parameter distribution explains why it can be
hard to calibrate with the conventional approach. When strong
correlations exist, such as for $X_3$
and $X_5$ in Figure~\ref{fig:posteriorSamples}, it is
hard to move to a better guess by changing just one parameter.

In conclusion, the reduction of the standard deviation in all posterior 
parameters in
conjunction with the unimodal posterior distribution can be seen as an 
indicator of a successful calibration.

\begin{table}
	\caption{Posterior statistics for the calibration with Product A 
		(E1). The 
		values are computed from the available posterior sample and include the 
		MAP 
		estimate, the 
		empirical mean $\hat{\mu}$, the empirical
		$95\%$ confidence interval, the 
		empirical standard deviation $\hat{\sigma}$, and the empirical 
		coefficient 
		of 
		variation $\mathrm{c.o.v.}\eqdef \hat{\sigma}/\hat{\mu}$. The prior 
		statistics are shown in 
		Table~\ref{tab:priorDist}.}
	\label{tab:postStatE2}
	\centering
	\resizebox{\textwidth}{!}{
		\begin{tabular}{rccccc}
			\hline
			& MAP & $\hat{\mu}$ & $95\%$ conf. interval & $\hat{\sigma}$ & 
			c.o.v. \\
			\hline
			$X_{1}$ & $6.91 \cdot 10^{2}$ & $6.81 \cdot 10^{2}$ & $[5.29 \cdot 
			10^{2}, 7.88 \cdot 10^{2}]$ & $6.60 \cdot 10^{1}$ & $9.70 \cdot 
			10^{-2}$\\
			$X_{2}$ & $5.08 \cdot 10^{-1}$ & $5.65 \cdot 10^{-1}$ & $[1.48 
			\cdot 
			10^{-1}, 9.60 \cdot 10^{-1}]$ & $2.27 \cdot 10^{-1}$ & $4.01 \cdot 
			10^{-1}$\\
			$X_{3}$ & $0.185$ & $0.183$ & $[0.167, 0.196]$ & $6.99 \cdot 
			10^{-3}$ & 
			$3.82 \cdot 
			10^{-2}$\\
			$X_{4}$ & $0.799$ & $0.798$ & $[0.410, 1.15]$ & $0.199$ & $2.49 
			\cdot 
			10^{-1}$\\
			$X_{5}$ & $3.77 \cdot 10^{4}$ & $3.75 \cdot 10^{4}$ & $[3.46 \cdot 
			10^{4}, 4.14 \cdot 10^{4}]$ & $1.58 \cdot 10^{3}$ & $4.21 \cdot 
			10^{-2}$\\
			$X_{6}$ & $2.25 \cdot 10^{4}$ & $2.17 \cdot 10^{4}$ & $[3.55 \cdot 
			10^{3}, 4.65 \cdot 10^{4}]$ & $1.10 \cdot 10^{4}$ & $5.09 \cdot 
			10^{-1}$\\
			$X_{7}$ & $8.56 $ & $8.62 $ & $[7.99, 9.25 ]$ & $3.36 \cdot 
			10^{-1}$ & 
			$3.89 \cdot 
			10^{-2}$\\
			$X_{8}$ & $4.53 \cdot 10^{-1}$ & $4.58 \cdot 10^{-1}$ & $[1.99 
			\cdot 
			10^{-1}, 7.25 \cdot 10^{-1}]$ & $1.33 \cdot 10^{-1}$ & $2.90 \cdot 
			10^{-1}$\\
			$X_{9}$ & $1.35 $ & $1.36 $ & $[1.59, 1.13 ]$ & $1.79 \cdot 
			10^{-1}$ & 
			$5.87 \cdot 
			10^{-2}$\\
			$X_{10}$ & $2.39 $ & $2.41 $ & $[2.69, 2.14 ]$ & $2.11 \cdot 
			10^{-1}$ & 
			$3.92 \cdot 
			10^{-2}$\\
			$X_{11}$ & $2.46 $ & $2.46 $ & $[2.19, 2.75 ]$ & $1.43 \cdot 
			10^{-1}$ & 
			$5.84 \cdot 
			10^{-2}$\\
			$X_{12}$ & $5.06 \cdot 10^{-1}$ & $5.29 \cdot 10^{-1}$ & $[2.34 
			\cdot 
			10^{-1}, 8.29 \cdot 10^{-1}]$ & $1.56 \cdot 10^{-1}$ & $2.95 \cdot 
			10^{-1}$\\
			$X_{13}$ & $2.07 \cdot 10^{-1}$ & $2.16 \cdot 10^{-1}$ & $[3.93 
			\cdot 
			10^{-1}, 4.96 \cdot 10^{-2}]$ & $1.33 \cdot 10^{-1}$ & $2.75 \cdot 
			10^{-1}$\\
			$X_{14}$ & $2.76 \cdot 10^{1}$ & $2.77 \cdot 10^{1}$ & $[2.67 \cdot 
			10^{1}, 2.86 \cdot 10^{1}]$ & $4.87 \cdot 10^{-1}$ & $1.76 \cdot 
			10^{-2}$\\
			\hline
		\end{tabular}
	}
\end{table}

\subsection{Discussion of the calibration results}
\begin{figure}
	\centering
	\subfloat[Realizations of the conductivity 
	$\lambda(\BParams)$]{{\includegraphics[width=7cm]{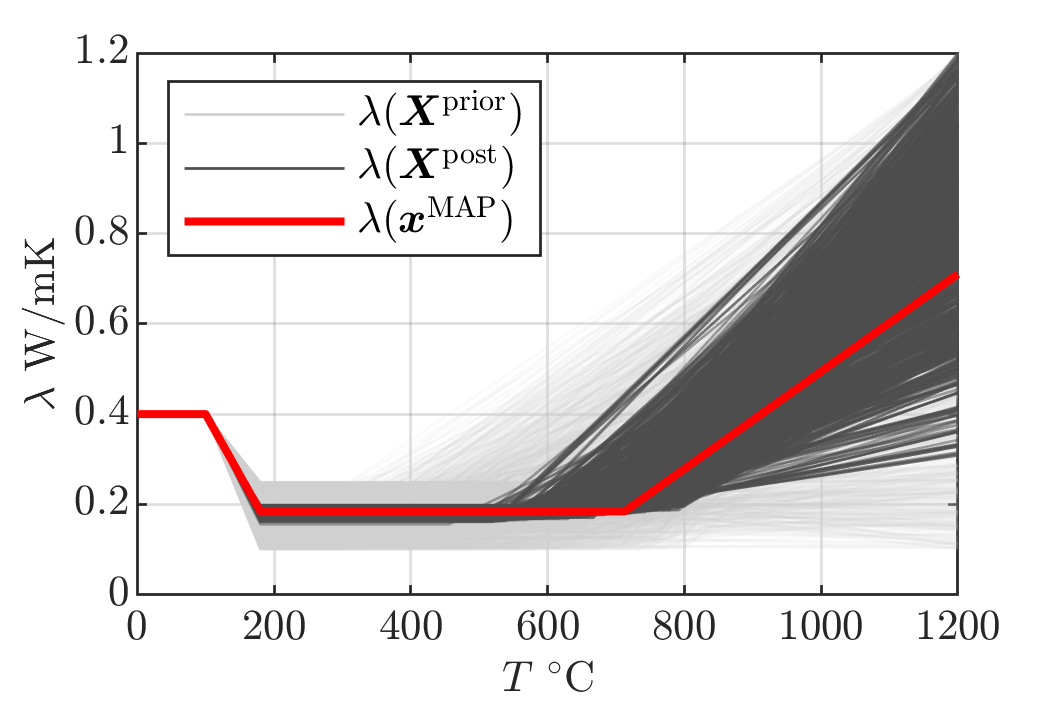}
	}}%
	\subfloat[Realizations of the heat capacity 
	$c(\BParams)$]{{\includegraphics[width=7cm]{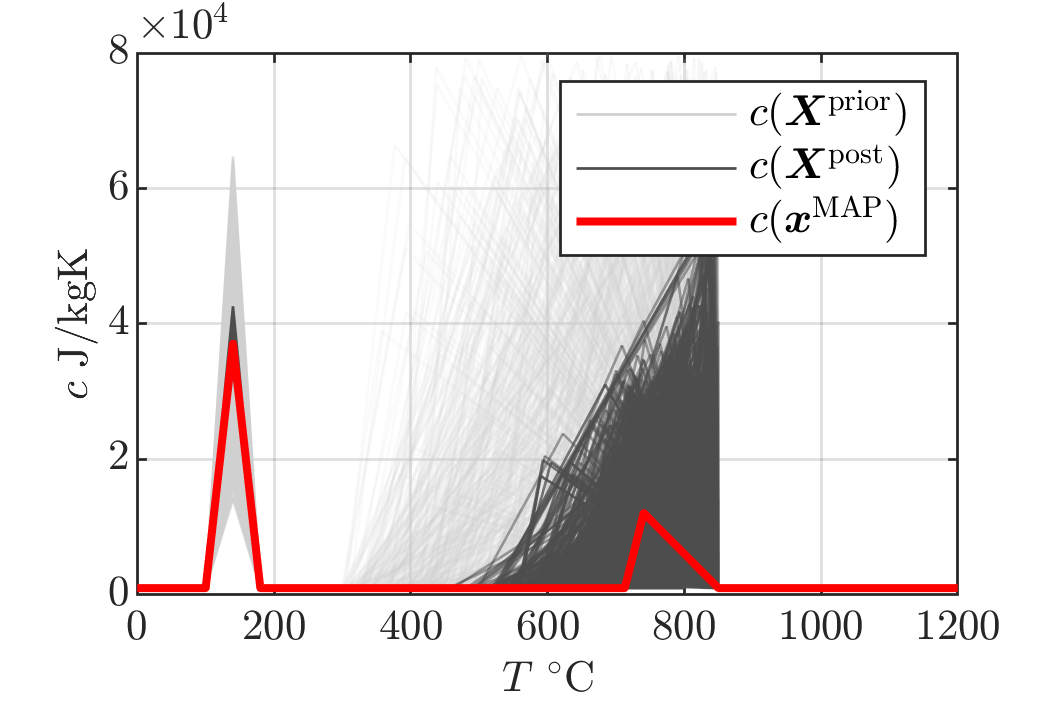}
	}}%
	\\
	\subfloat[Model predictions and calibrated discrepancy standard deviation]{{
			\begin{minipage}{\linewidth}
				\includegraphics[width=14cm,trim={0 0.5cm 0 
					0},clip]{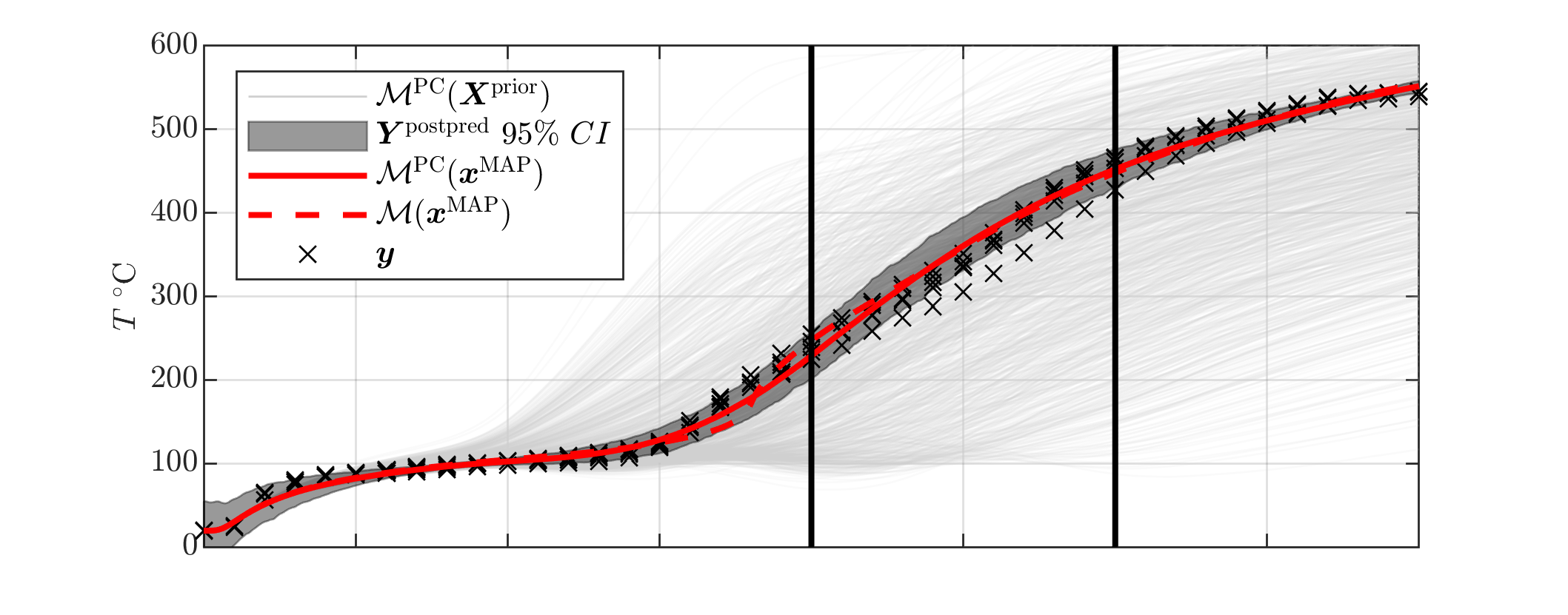}\\
				\includegraphics[width=14cm,trim={0 0 0 
					0.1},clip]{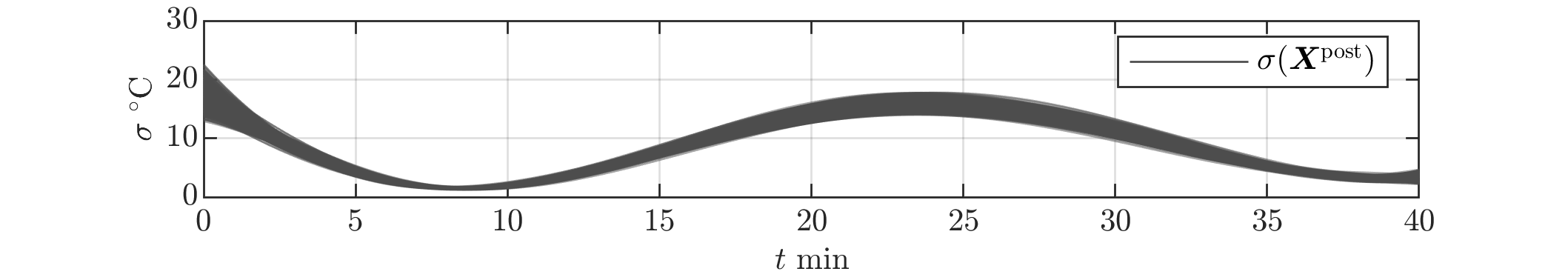} 
			\end{minipage}
	}}%
	\\
	\subfloat[Model predictions at 
	$t=20~\mathrm{min}$]{{\includegraphics[width=7cm]{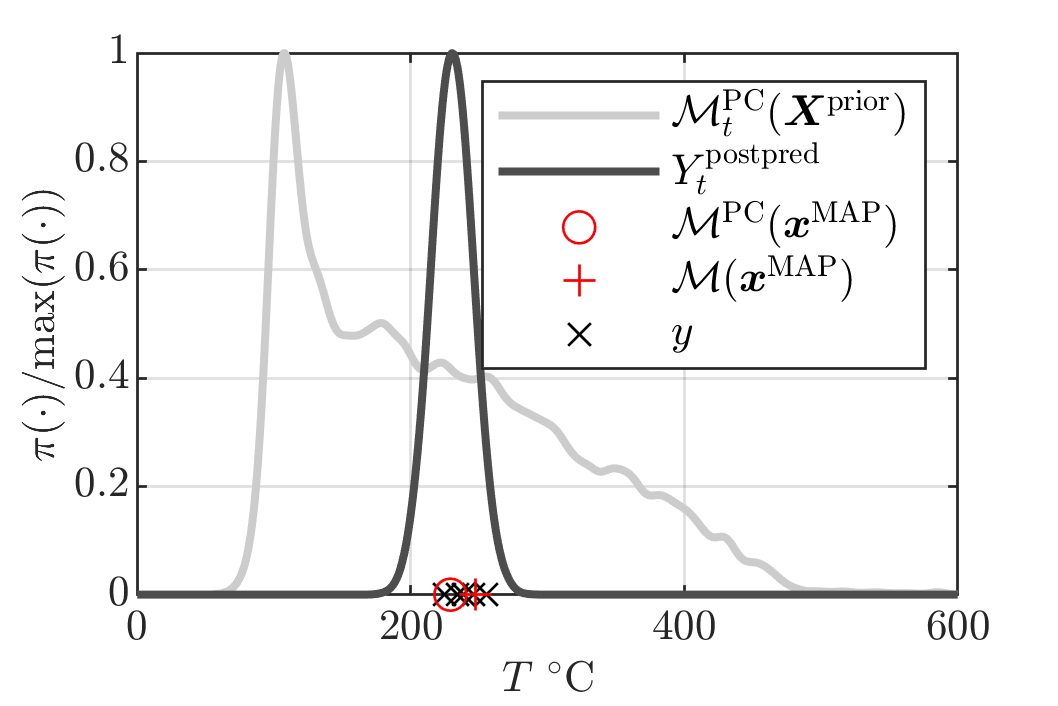}
	}}%
	\subfloat[Model predictions at $t=30~\mathrm{min}$]{{
			\includegraphics[width=7cm]{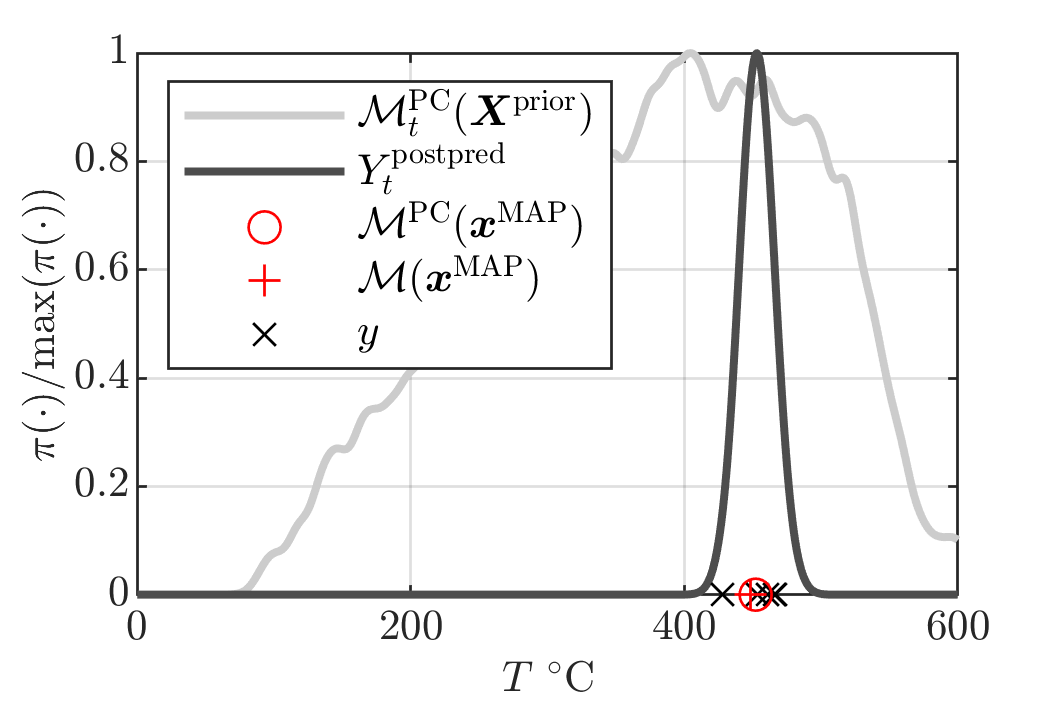}
	}}%
	\caption{Calibration results for Product A 12.5~mm insulation
		(E1), 
		experiments conducted by \citet{test:Gyproc2016}.}%
	\label{fig:E16results}%
\end{figure}

The results of the proposed calibration procedure for
four different insulation materials are discussed next. The materials 
are all
characterized by the temperature-dependent material properties
$(\lambda(T,\BParamsM),c(T,\BParamsM),\rho(t,\BParamsM))$. The 
calibration results for Product A are summarized in Figure~\ref{fig:E16results} 
and Table~\ref{tab:postStatE2}. In 
\ref{sec:additionalResults} the results of the remaining products are 
presented. The discussion in this section refers to them at times. In this 
section we use the notation 
$\BParams^{\text{prior}}\eqdef\BParams$ and 
$\BParams^{\text{post}}\eqdef\BParams\vert\Bdata$ to more clearly distinguish 
between the prior and posterior random variables.

For each material, Figures~\subref{fig:E15results:a} and 
\subref{fig:E15results:b} show the temperature-dependent conductivity 
$\lambda(T)$ and heat capacity $c(T)$.  Since the 
full 
posterior distribution of the parameters is inferred, the plots do not 
only show one line for prior 
and posterior, but $1{,}000$ samples each. The shown curves result 
from prior parameter draws $\BParams^{\text{prior}}\sim\Bprior$ (\ie 
before calibration), 
posterior parameter draws $\BParams^{\text{post}}\sim\Bpost$ (\ie after 
calibration) and the 
MAP parameter $\Bparams^{MAP}$ (see Eq.~\eqref{eq:MAP}). The
calibrated density $\rho(T)$ is not shown, since its two governing
parameters $X_1$ and $X_2$ can be seen also in the plots of the
calibrated $\lambda(T)$ and $c(T)$. It 
is obvious that the posterior samples have a smaller variance 
than the prior samples since the bandwith of the $1{,}000$ curves is much 
smaller. 

A plot of the model predictions  
together with 
the measured data $\Bdata$ is presented in 
Figures~\subref{fig:E15results:c}. These plots show runs of 
the 
surrogate model
before calibration ($\cm^{PC}(\BParams^{\text{prior}})$) and confidence 
intervals $CI$ of the predictions $\ve{Y}^{\text{postpred}}$
of the surrogate model following
calibration (see Eq.~\eqref{eq:postPred}). The MAP 
parameter $\Bparams^{\mathrm{MAP}}$ is propagated through the surrogate and the 
response of the original finite element model for $\Bparams^{\mathrm{MAP}}$ is 
also plotted. Additionally, samples from the 
posterior are used to show the calibrated discrepancy standard 
deviation $\sigma(\BParams^{\text{post}})$ (see 
Eq.~\eqref{eq:Results:standardDev}).  

To emphasize the reduction in uncertainty by the presented
calibration procedure, additionally posterior distributions of the predicted 
temperature at \emph{snapshot} times 
$t=20~\mathrm{min}$ and 
$t=30~\mathrm{min}$ are shown in Figures~\subref{fig:E15results:d} and 
\subref{fig:E15results:e}. These plots show kernel density estimates of 
the output probability density function of the surrogate model before 
calibration 
$\cm_t^{PC}(\BParams^{\text{prior}})$ and after
calibration $Y_t^{\text{postpred}}$ 
at the respective times $t$. Additionally, the measured data $y$ and the 
MAP predictions are displayed. These plots are now discussed in more 
detail.

The distributions of $X_3$ (conductivity between 
$180\!\degC$ and the start $X_1$ of the second key process) and 
$X_5$ (distribution of the heat capacity at $140\!\degC$), 
show a large variance reduction. This means that the information gained 
about these parameters through the conducted calibration is high. 

Generally, it can be expected that the heat capacity and the 
conductivity correlate, since a higher heat flow due to a higher 
conductivity can be \emph{compensated} by a higher heat capacity, 
resulting in the same temperature profile. This is indeed the case for 
the parameters $X_3$ and $X_5$ as well as for $X_4$ and $X_6$ (which describe 
the conductivity 
and the heat capacity at higher temperatures), as 
seen in Figure \ref{fig:posteriorSamples}. The effect of this 
correlation and its \emph{compensation effect} can be seen by comparing 
Figure \ref{fig:E15results}~(E2) to Figures 
\ref{fig:E16results}~(E1), \ref{fig:E17results}~(E3) and 
\ref{fig:E18results}~(E4).
Despite the fact that parameters $X_4$ and $X_6$ show a higher posterior 
variance for specimens E1, E3 and E4 than for E2, the variance of the resulting 
model predictions (Figures~\subref{fig:E15results:c}) is comparable.

Despite the fact that the shown prior thermal material 
properties in Figures~\subref{fig:E15results:d} and 
\subref{fig:E15results:e} are 
based on the same prior assumptions for the parameters 
$\{X_i,i=1,\dots,6\}$, the modes of the prior distributions estimated from 
$\cm_t^{PC}(\BParams^{\text{prior}})$ are significantly different for the 
individual specimens. This is 
caused by two factors: (1) the variable layer thickness and (2) the 
parametrization. The layer thickness influences the measured 
temperature simply by the fact that the further inside the specimen the 
measurement is placed, the colder it is at a certain time. This can 
easily be seen by comparing the different 
Figures~\subref{fig:E15results:c}, from $9.5$ mm 
(Figure~\ref{fig:E15results}) to $12.5$ mm 
(Figures~\ref{fig:E16results} 
and \ref{fig:E17results}) to $15$ mm (Figure~\ref{fig:E18results}). 
Additionally, parameter $X_1$, the start time of the second key 
process, is responsible for the large variance of the start time of the 
temperature increase after the $100\!\degC$ plateau. For higher values of 
$X_1$, the plateau is longer and the temperature increase starts later. 
Thus if the snapshots (Figures~\subref{fig:E15results:d} 
and \subref{fig:E15results:e}) are taken where 
most 
prior temperatures are still at the plateau, the probability density 
functions have positive (right) skew. The later the 
snapshot is drawn, the more negatively skewed is the probability 
density function and the larger its variance gets. 

Product B (E2) is the only product 
where the conductivity decreases after 400°C and where there is hardly 
a second key process according to the posterior curves. This shows that
this product is made of a significantly different 
material.

Generally, all posterior predictive distributions $\Bpostpred$ agree 
well with 
the measurements. This suggests that the used heat transfer model with 
the chosen parametrization is sufficiently accurate to reproduce the 
observations.

The calibrated temperature-dependent material properties show a higher 
variance reduction at lower temperatures. This is mostly due to the 
maximum 
system temperature at $40~\mathrm{min}$ being only $885\!\degC$ as seen 
in 
Figure~\ref{fig:dataSummary}.

It can be clearly seen in Figures~\subref{fig:E15results:c} that the calibrated 
discrepancy 
standard deviation $\sigma(\BParams^{\text{post}})$ varies similarly 
across all four experiments.  It is influenced by two factors: (1) the 
temperature
variance from the data series $\Bdata^{(s)}$ (\ie the individual
sensor recordings) and (2) model insufficiencies that hinder a fully
accurate reproduction of the data. The latter can be
reduced by changing the parametrization: an additional model parameter 
between approximately $200\!\degC$ and $400\!\degC$ 
could reduce the calibrated mismatch standard deviation for 
the temperature increase after the $100\!\degC$ plateau. This 
temperature range 
currently lacks an independent parameter.

\subsection{Time-dependent sensitivity analysis}
\label{sec:results:sensitivity}

Using the PCA+PCE surrogate model and the derivations from
Section~\ref{sec:sensitivity}, the constructed surrogate
model of the temperature evolution can be reused to compute the time-dependent 
Sobol' indices as measures for the individual
model parameters' importance across the simulation time.

\begin{figure}
	\centering
	\includegraphics[width=14cm]{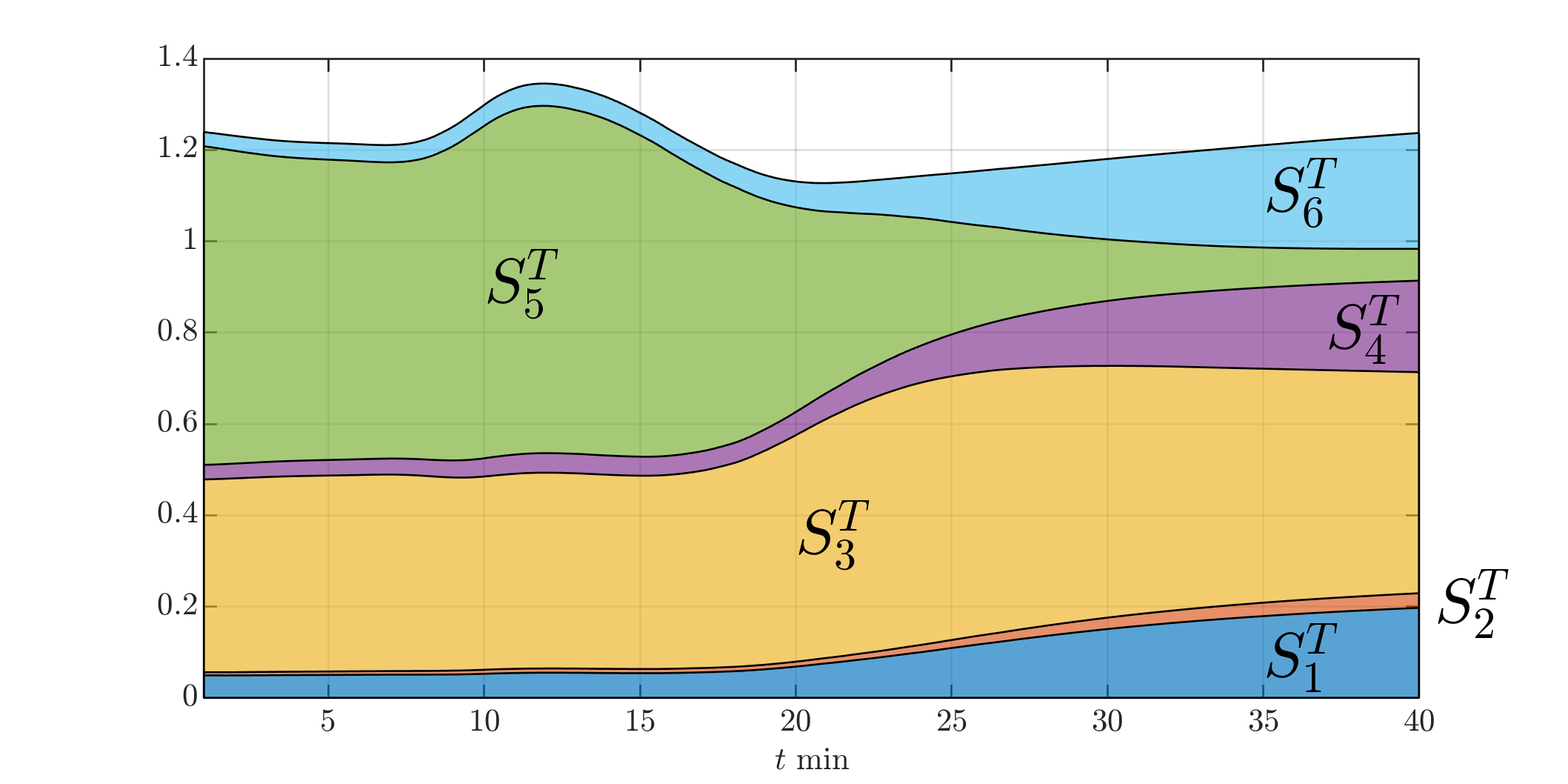}
	\caption{Time-dependent total Sobol' indices $S_i^{T}$ for the 
		surrogate
		model of Product A (E1).}%
	\label{fig:E16sensitivity}%
\end{figure}

The time-dependent Sobol' indices $S_i^T$ were computed for all four 
surrogate 
models constructed for the experimental setups E1-E4. They are 
displayed in Figures~\ref{fig:E16sensitivity} for Product A. In 
\ref{sec:additionalResults} the results of the remaining products are 
presented. The discussion in this section refers to them at times.

Generally there is no single parameter that clearly dominates the 
simulation output across the whole simulation time range. Instead, most 
parameters have a clear time range where they are important and other 
time ranges where their influence can be neglected. All parameters 
show a similar behaviour across the four models. 

At early simulation times before $t=15~\mathrm{min}$, the Sobol' 
indices 
$S_{3}^{T}$ and $S_{5}^{T}$ clearly dominate the sensitivity analysis
across all four models. These indices correspond to the parameters 
$X_{3}$ (low to mid temperature conductivity) and $X_{5}$ (specific 
heat 
capacity at $T=140\!\degC$). This is not surprising, because the system 
temperature is monotonously increasing and low temperature effects 
like these are expected to have a higher impact at earlier times. The 
Sobol' index $S_{3}^{T}$ carries on to influence the simulation 
at \modif{later times, while $S_{5}^{T}$ decreases in importance towards the 
	end of the simulation}.

Starting from $t=15~\mathrm{min}$, the Sobol' indices 
$S_{1}^{T}$, $S_{4}^{T}$ and $S_{6}^{T}$ become dominant. This can 
be explained similarly, since they influence the mid 
to high 
temperature behaviour. $X_{1}$ is the start of the second key 
process and thus heavily influences the high temperature behaviour of 
all temperature-dependent material properties. $X_{4}$ (high 
temperature conductivity) and $X_{6}$ (specific heat 
capacity at the second key process) are the high temperature material 
properties 
and as such are more important at later time steps of the simulation.

The second key process results in a temperature plateau (like the one at 
$100\!\degC$).
Because this process occurs at temperatures higher than the maximum 
temperature reached at the measurement location, the process has no 
direct effect there. 
The system does, however, reach higher temperatures at locations 
closer to the exposed surface and it is there that the second key 
process influences the system behaviour.
As can be seen from the results of the sensitivity analysis, the 
magnitude of the second key process ($X_6$) significantly influences 
the temperatures at the measurement location due to the locally lower 
thermal diffusivity. However, the
temperature where this peak in the thermal diffusivity appears ($X_2$), is not
so important.

\subsection{Validation of the calibration using V1-V2}
\label{sec:results:validation}

So far, all presented results were related to the experiments \emph{Test 
	1} 
(specimens E1-E4) described in Section~\ref{sec:expMod:experiments}. It 
was shown that the calibrated material properties for the examined 
materials can be used to conduct computer simulations that agree well 
with the same experimental observations that were used for calibration 
(Section~\ref{sec:results:calibration}). Before these calibrated material 
properties can be used to make predictions about the 
material behaviour in other experimental setups, validation experiments 
need to be carried out to judge the accuracy for the new intended use 
\citep{VVUQ:Oberkampf2010}.

In this work, the results from Test 2 (specimens V1 and V2) are 
used for this purpose (see Figure~\subref*{fig:dataSummaryV1} and 
\subref*{fig:dataSummaryV2}). These tests were conducted using two of 
the 
materials for which calibrated material properties were obtained 
(Product C (E3) and Product D (E4)). To validate the calibration 
results, the posterior predictive distribution of this setup needs to 
be computed. It is defined identically to Eq.~\eqref{eq:postPred} and 
samples from it can be drawn by
\begin{equation}
\label{eq:postPredValid}
\BData^{\text{postpred}} 
\sim\cn(\Bdata\vert\cm_{\textit{Test 
		2}}(\BparamsM),\errorCovPMult), 
\quad 
\text{where} \quad 
\BparamsM^{\text{post}}\sim\pi(\BparamsM\vert\Bdata_{\textit{Test 1}}),
\end{equation}

where the subscripts $_{\textit{Test 1}}$ and  $_{\textit{Test 2}}$ were 
introduced to distinguish between quantities and models belonging to the 
respective 
setups. In this sense, $\cm_{\textit{Test 
		2}}(\BParamsM)$ refers to the finite element model predicting 
the heat evolution in the two interfaces of Test 2 and 
$\Bdata_{\textit{Test 1}}$ are measurements from Test 1.

Using parameters calibrated in Test 1 for the 
predictions in Test 2 requires careful consideration of the 
implications. On the one hand, the model parameters $\BParamsM$ 
of the insulation products can be reused without further 
considerations (since this is the point of the validation). On the 
other 
hand, parameters related to the discrepancy model $\BParamsE$  cannot 
be 
transferred so easily. 

The discrepancy term captures measurement noise and model inadequacy 
and is assumed to follow a zero mean normal distribution 
(Eq.~\eqref{eq:additiveDiscrepancyRV}). It cannot be directly applied 
in drawing predictive quantities as neither the measurement noise nor 
the model inadequacy can be expected to be identical between the two 
setups. Therefore, simplifying assumptions about the discrepancy covariance 
matrix 
$\errorCovPMult$ have to be made. 

The only information available 
about the discrepancy covariance matrix are the 
calibration results of E1-E4 (Section~\ref{sec:results:calibration}). 
There it was parameterized as described in 
Eq.~\eqref{eq:covMatrixParam}. Across all posterior parameter 
distributions, the MAP estimator of the correlation length parameter 
$\theta=X_{14}$ was $\sim 30\,s$. The discrepancy standard deviation 
$\sigma(t)$, however, did not yield such a uniform result as can be 
clearly seen from Figures~\ref{fig:E15results} to \ref{fig:E18results}. 
To still parameterize the covariance matrix in the predictive draws 
from Eq.~\eqref{eq:postPredValid}, a conservative choice of 
$\theta=30~s$ 
and a constant 
$\sigma(t) = 10\!\degC$ is thus made. This corresponds to setting the 
discrepancy parameter vector to
\begin{equation}
\BparamsE=(10\!\degC, 0, 0, 0, 0, 0, 0, 30\,s)
\end{equation}

\begin{figure}
	\centering
	\subfloat[Model 
	predictions]{{\includegraphics[width=14cm]{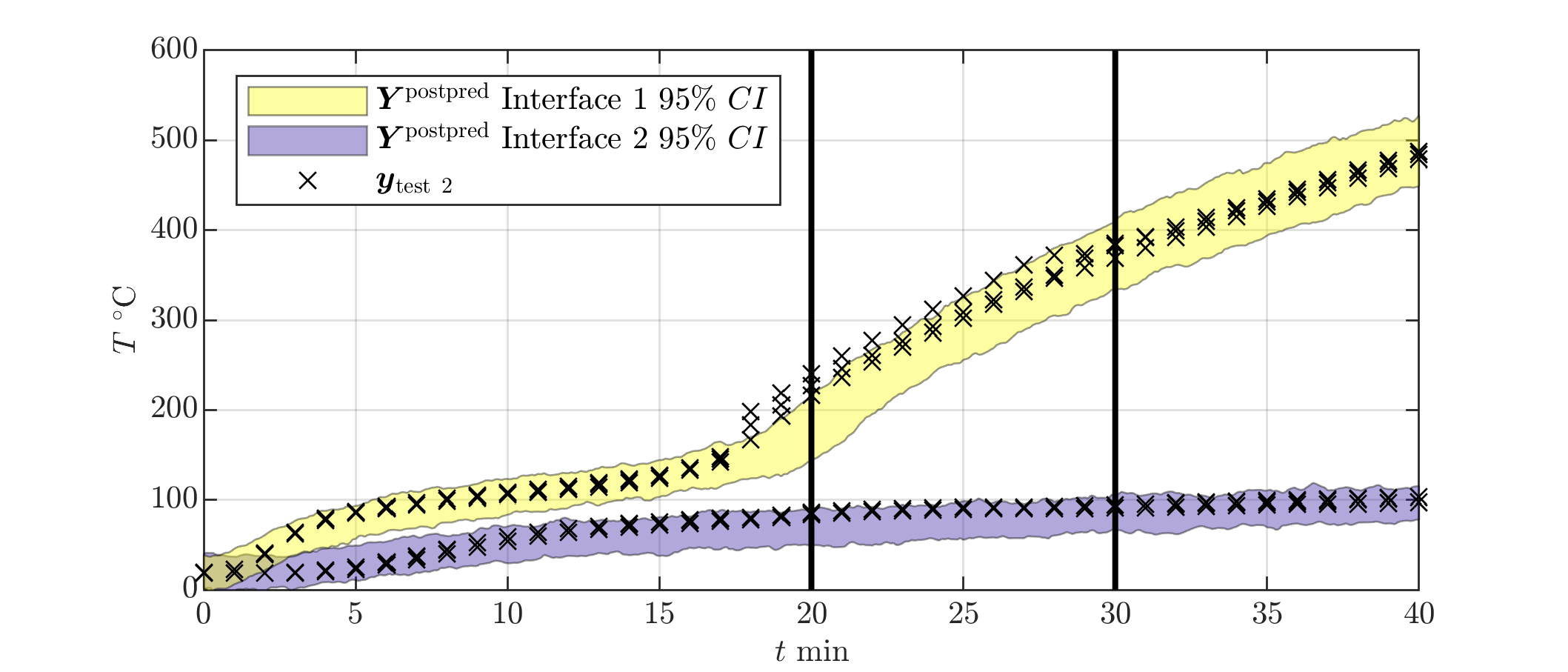}
			\label{fig:E13results:a}
	}}%
	\\
	\subfloat[Statistics of model predictions at two snapshot times]{{
			\begin{tabular}{r|cc|cc}
				\hline
				& \multicolumn{2}{c|}{$t = 20~\mathrm{min}$} & 
				\multicolumn{2}{c}{$t 
					= 30~\mathrm{min}$}\\
				($\degC$)& Interface 1 & Interface 2 & Interface 1 & Interface 
				2\\
				\hline
				& $83.4$ & $240.6$& $90.8$ & $382.8$\\
				$y_t^{(s)}$& $85.8$ & $227.1$ & $94.5$ & $369.0$\\
				& $85.7$ & $216.6$ 	& $94.2$ & $385.6$\\
				\hline
				$\hat{\mu}$ & $70.1$ & $185.4$ & $84.0$ & $370.7$\\
				$\hat{\sigma}$ & $10.7$ & $18.1$ & $10.2$ & $20.3$\\
				\hline
			\end{tabular}
			\label{fig:E13results:b}
	}}%
	\caption{Validation of the calibrated material properties of 
		Product C and Product D using 
		measurements from Test 2 (V1) including the empirical mean $\hat{\mu}$ 
		and 
		the 
		empirical standard deviation $\hat{\sigma}$.}%
	\label{fig:E13results}%
\end{figure}

\begin{figure}
	\centering
	\subfloat[Model 
	predictions]{{\includegraphics[width=14cm]{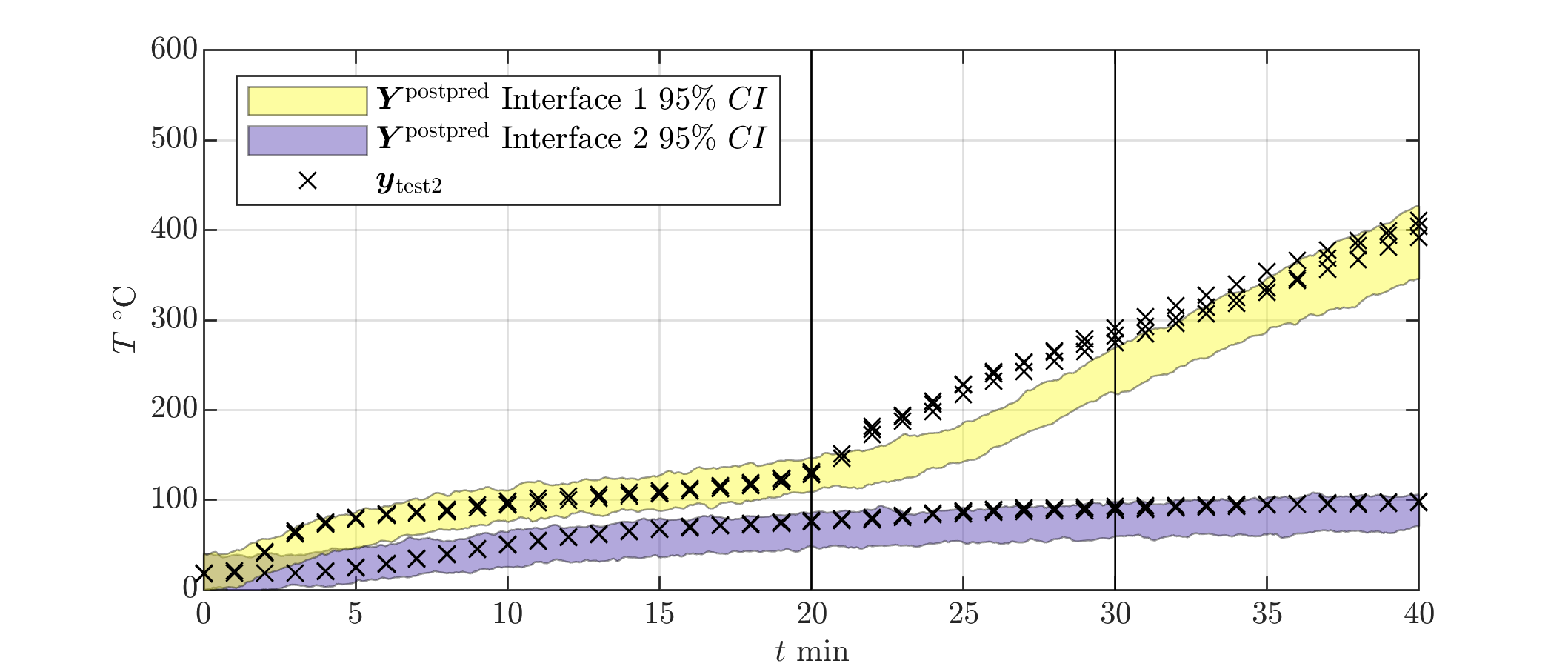}
	}}%
	\\
	\subfloat[Statistics of model predictions at two snapshot times]{{
			\begin{tabular}{r|cc|cc}
				\hline
				& \multicolumn{2}{c|}{$t = 20~\mathrm{min}$} & 
				\multicolumn{2}{c}{$t 
					= 30~\mathrm{min}$}\\
				($\degC$)& Interface 1 & Interface 2 & Interface 1 & Interface 
				2\\
				\hline
				& $75.3$ & $129.0$& $88.9$ & $283.0$\\
				$y_t^{(s)}$& $76.2$ & $131.5$ & $92.7$ & $291.6$\\
				& $77.2$ & $127.9$ 	& $90.4$ & $274.8$\\
				\hline
				$\hat{\mu}$ & $65.9$ & $127.8$ & $76.6$ & $243.2$\\
				$\hat{\sigma}$ & $9.9$ & $9.7$ & $9.7$ & $12.9$\\
				\hline
			\end{tabular}
	}}%
	\caption{Validation of the calibrated material properties of 
		Product D using 
		measurements from Test 2 (V2) including the empirical mean $\hat{\mu}$ 
		and 
		the 
		empirical standard deviation $\hat{\sigma}$.}%
	\label{fig:E14results}%
\end{figure}

Figures~\ref{fig:E13results} and \ref{fig:E14results} show the 
resulting confidence intervals from the posterior predictive distribution 
$\bm{Y}^{\text{postpred}}$ defined in 
Eq.~\eqref{eq:postPredValid} for interface 1 and 2 of Test 2. To 
show the agreement of the predictions with the measurements, 
$\Bdata_{\textit{Test 2}}$ are displayed as well. 
The figures contain one time-temperature plot in 
Figures~\subref{fig:E13results:a} and summary statistics of the model 
predictions in Figure~\subref{fig:E13results:b} at snapshot times 
$t=20~\mathrm{min}$ and 
$t=30~\mathrm{min}$ respectively. 

Generally the simulations agree remarkably well 
with the experimental observations, but there are also time intervals, 
where small differences between the predicted and observed 
temperature evolutions are visible. The predictions at large time instants, 
which are of interest in practice, appear excellent.

\section{Summary and conclusion}
\label{sec:conclusion}
In this paper, a procedure to calibrate temperature-dependent effective 
material properties of fire insulation panels was presented. Available 
experimental temperature measurements were modelled using a 1D finite 
element heat transfer model. Because the associated material properties 
vary with temperature, they were parameterized using a 
set of model parameters. The actual 
calibration was then carried out using the well-known Bayesian 
inference 
framework. The necessary sampling from the posterior distribution was 
conducted with the advanced AIES (Affine invariant 
ensemble sampler) MCMC algorithm. In an effort to reduce 
the computational burden from the 
required repeated finite element simulations, a surrogate 
model of the heat transfer problem was constructed by combining 
polynomial chaos expansions (PCE) with the principal component analysis 
(PCA) technique. This surrogate 
model offered the possibility to additionally conduct a sensitivity 
analysis using the time-dependent Sobol' indices at no additional 
computational cost. Finally, the calibration was validated using a 
secondary set of experiments.

The proposed approach is superior to the previously used brute-force 
calibration approach because it
automates the process, clearly defines the discrepancy and explicitly
considers the uncertainties present in the model and measurements.
Accordingly, it does not only deliver a single best fit property, but returns 
the full multivariate distribution of the calibrated
properties and allows the computation of confidence intervals. Furthermore, the 
used Bayesian framework is a natural way 
to 
update information 
through the use of conditional random variables. It is well suited 
for engineering problems, where often expert knowledge is available 
that thereby 
can be directly integrated into the calibration procedure. \modif{However, it 
	needs to be considered that depending on the application less general 
	calibration procedures might be more suitable \citep{Mottershead2011, 
		Bayesian:Patelli2017}.}

A valuable side 
product of the presented surrogate modelling technique is the 
\emph{free} computation of the time-dependent Sobol' indices. This 
sensitivity analysis offers valuable insight into the time-dependent effect of 
the used parametrization.

Finally, it is worth emphasizing that the proposed method, which combines 
surrogate modelling (PCE) with dimensionality reduction (PCA), an advanced MCMC 
algorithm (AIES) and global sensitivity analysis is general and can be applied 
to any calibration problem involving complex computer codes. All the algorithms 
used are available in the UQLab uncertainty quantification software 
\citep{MarelliUQLab2014}, especially the recently developed Bayesian inversion 
module \citep{UQdoc_12_113}. 

\section*{Acknowledgement}
The authors gratefully acknowledge the European COST Action FP1404 on the fire 
safe use of bio-based building products. Additionally, we would like to thank 
Prof. Alar Just and his team, as well as Dr. Joseph Nagel, who participated in 
the preliminary discussions about the case study definitions. The PhD thesis of 
the first author is 
supported 
by ETH grant \#44 17-1.

\appendix
\section{Surrogate model approximation error}
\label{sec:SurrogateError}
The following approximation error $\tilde{\eta}$ was originally derived 
in \citet{BlatmanIcossar2013}. The surrogate model from 
Eq.~\eqref{eq:surrogate} has a total 
$L_2$-approximation error that can be written by denoting the $2$-norm 
as $\norm{\cdot}_2$ by
\begin{align}
\varepsilon &= \expe{\norm{\BModelOut-\BModelOut^{\mathrm{PCA+PCE}}}^2_2}\\
&= \expe{\norm{\left(\BModelOut-\BModelOut^{\mathrm{PCA}}\right) + 
		\left(\BModelOut^{\mathrm{PCA}}-\BModelOut^{\mathrm{PCA+PCE}}\right)}^2_2}.
\end{align} 

Through the Cauchy-Schwarz inequality this error is bounded by
\begin{align}
\varepsilon &\le \left(
\sqrt{\expe{\norm{\BModelOut-\BModelOut^{\mathrm{PCA}}}^2_2}} + 
\sqrt{\expe{\norm{\BModelOut^{\mathrm{PCA}}-\BModelOut^{\mathrm{PCA+PCE}}}^2_2}}\right)^2\\
&\eqdef \left(\sqrt{\varepsilon_{\mathrm{PCA}}} + 
\sqrt{\varepsilon_{\mathrm{PCE}}}\right)^2.
\end{align} 

\begin{description}
	\item[PCA error $\varepsilon_{\mathrm{PCA}}$:] this incorporates the 
	error 
	from estimating the mean $\meanRespEst$ and covariance matrix 
	$\covRespEst$ of the response as well as the dimensionality 
	reduction error from leaving out $N-N'$ dimensions. The former is 
	neglected in this paper and for the latter the 
	sum of the discarded eigenvalues $\lambda_p$ can be directly used:
	\begin{equation}
	\varepsilon_{\mathrm{PCA}} \approx 
	\tilde{\varepsilon}_{\mathrm{PCA}} = \sum_{p=N'+1}^N\lambda_p.
	\end{equation}
	\item[PCE error $\varepsilon_{\mathrm{PCE}}$:] this is the error of 
	the 
	polynomial chaos approximation. It can be estimated as the sum of 
	the individual LOO errors for the $N'$ scalar-valued principal 
	component PCE's:
	\begin{equation}
	\varepsilon_{\mathrm{PCE}} \approx  
	\tilde{\varepsilon}_{\mathrm{PCE}} = 
	\sum_{p=1}^{N'} 
	\varepsilon_{p,\mathrm{LOO}}.
	\end{equation}
\end{description}

For practicality a relative error measure $\eta$ is preferred. This can 
be obtained by dividing the estimator of the absolute 
error bound
by an estimator of $\expe{\norm{\BModelOut}_2^2}$, such as the trace of 
its estimated covariance matrix $\mathrm{Tr}(\covRespEst)$:
\begin{equation}
\label{eq:errorEst}
\tilde{\eta} = 
\frac{\left(\sqrt{\tilde{\varepsilon}_{\mathrm{PCA}}} + 
	\sqrt{\tilde{\varepsilon}_{\mathrm{PCE}}}\right)^2}{\mathrm{Tr}(\covRespEst)}.
\end{equation}

\begin{figure}[H]%
	\centering
	\subfloat[Relative error $\tilde{\eta}$ of 
	surrogate for increasing experimental design size $K$ and included 
	principal 
	components 
	$N'$.]{{\includegraphics[width=0.45\textwidth]{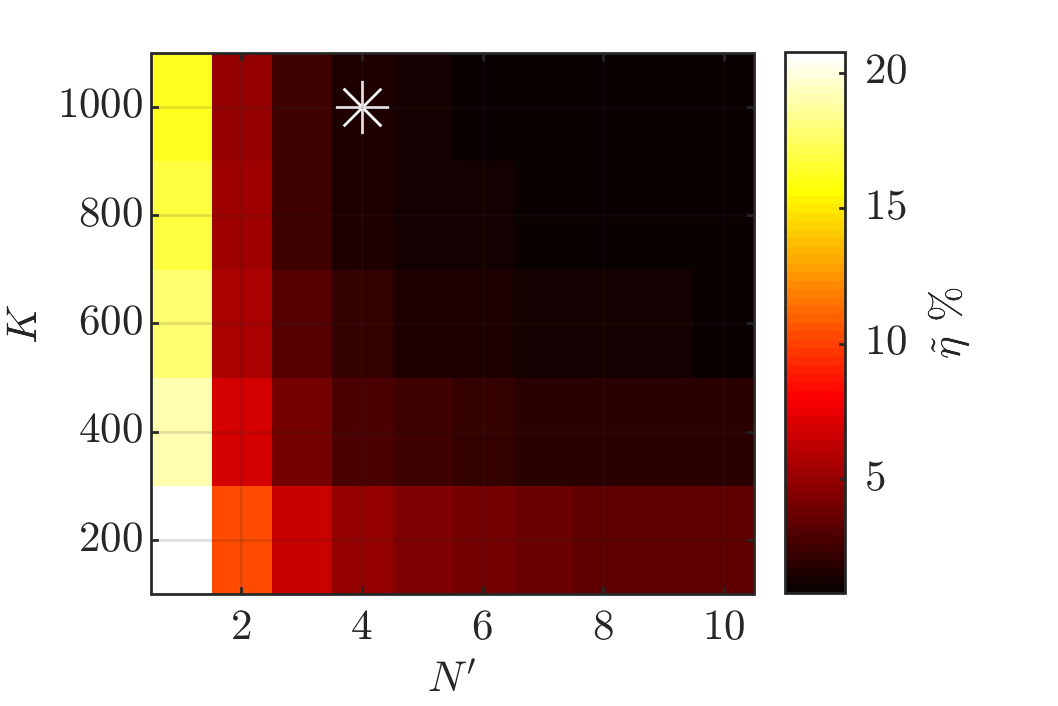}
	}}
	\hfill
	\subfloat[Comparison of forward model with 
	surrogate 
	model.]{{\includegraphics[width=0.45\textwidth]{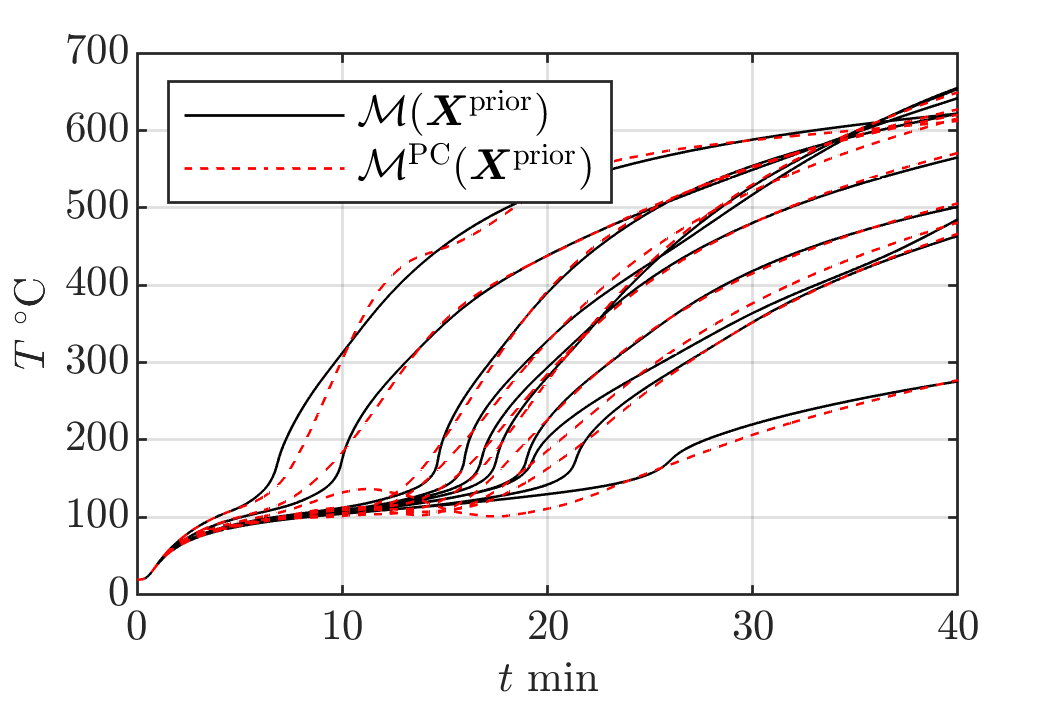}
	}}%
	\caption{Convergence diagnostics for the surrogate model for setup E1, 
		with $K=1{,}000$ and $N'=4$.}%
	\label{fig:conv}%
\end{figure}

Figure~\ref{fig:conv} shows the resulting error estimate for a set of 
experimental designs and included principal components using the setup E1. 
Additionally, the 
actual and surrogate model are run for a set of parameters and plotted for 
comparison.

\clearpage
\section{Derivation of PCA-based Sobol' indices}
\label{sec:PCASobolDeriv}

This appendix contains the derivations for PCA-based Sobol' indices. To 
simplify the derivations, we introduce the notation $\BParams_{\sim i} = 
(X_1,\dots,X_{i-1},X_{i+1},\dots,X_{M})\trans$ to denote the random vector that 
contains all \emph{but} the $i$-th 
random variable $X_i$.  In the 
following derivations, the subscript in the expectation and variance operators
$\mathbb{E}_{X_i}$ and $\mathrm{Var}_{X_i}$ denotes the variable(s) with 
respect to which expectation and variance are computed, \ie 
$\expcm{\cdot}{X_i}\eqdef\int (\cdot) \pi_i(x_i)\di{x_i}$.

The total Sobol' index $S_{i,t}^{T}$ for the $t$-th
element of a vector valued model 
output $\BModelOut = (Y_1,\dots,Y_N)\trans$ is defined as
\begin{equation}
\label{eq:totalSobol}
S_{i,t}^{T} = \frac{\expcm{\varcm{Y_t}{X_i}}{\BParams_{\sim i}}}{\varc{Y_t}} = 
1 - 
\frac{\varcm{\expcm{Y_t}{X_i}}{\BParams_{\sim i}}}{\varc{Y_t}}.
\end{equation}

This can be used to write an expression for the variance of the 
expectation $\varcm{\expcm{Y_t}{X_i}}{\BParams_{\sim i}}$ of the
$t$-th model output as
\begin{align}
\varcm{\expcm{Y_t}{X_i}}{\BParams_{\sim i}}
&= \expcm{\left(\expcm{Y_t}{X_i}\right)^2}{\BParams_{\sim i}} - 
\left(\expcm{\expcm{Y_t}{X_i}}{\BParams_{\sim i}}\right)^2\\
&= \expcm{\left(\expcm{Y_t}{X_i}\right)^2}{\BParams_{\sim i}} - 
\left(\expcm{Y_t}{\BParams}\right)^2.
\end{align}

By introducing the expression for the surrogate model from
Eq.~\eqref{eq:surrogateIndiv} and the fact that the expectation of the
$t$-th response is
$\expcm{Y_t}{\BParams}\approx\mu_{Y_t}$, the following is
obtained:
\begin{align}
\varcm{\expcm{Y_t}{X_i}}{\BParams_{\sim i}}
&= \expcm{\left(\expcm{\mu_{Y_t} + \ve{\phi}^{\mathrm{row}}_{t}
		\mat{A}\trans\PolyBasisVec(\BParams)}{X_i}\right)^2}{\BParams_{\sim i}}
- 
\mu_{Y_t}^2\\
&= \expcm{\left(\mu_{Y_t} + \ve{\phi}^{\mathrm{row}}_{t}
	\mat{A}\trans\expcm{\PolyBasisVec(\BParams)}{X_i}\right)^2}{\BParams_{\sim 
		i}}
- 
\mu_{Y_t}^2\\
\begin{split}
&= \mathbb{E}_{\BParams_{\sim 
		i}}\big[\mu_{Y_t}^2 + 
2\mu_{Y_t}\ve{\phi}^{\mathrm{row}}_{t}
\mat{A}\trans\expcm{\PolyBasisVec(\BParams)}{X_i}\\
&\qquad+ 
\left(\ve{\phi}^{\mathrm{row}}_{t}
\mat{A}\trans\expcm{\PolyBasisVec(\BParams)}{X_i}\right)^2\big] - 
\mu_{Y_t}^2.
\end{split}
\end{align}

Because the expectation of all principal components vanishes 
($\mat{A}\trans\expc{\PolyBasisVec(\BParams)} = 
\expc{\ve{Z}} =\ve{0}$),
one can write
\begin{equation}
\varcm{\expcm{Y_t}{X_i}}{\BParams_{\sim i}}
= \expcm{\left(\ve{\phi}^{\mathrm{row}}_{t}
	\mat{A}\trans\expcm{\PolyBasisVec(\BParams)}{X_i}\right)^2}{\BParams_{\sim 
		i}}.
\end{equation}

By switching to the summation notation, this can also be written as
\begin{equation}
\varcm{\expcm{Y_t}{X_i}}{\BParams_{\sim i}}  =
\expcm{\left(\sum_{\ve{\alpha}\in\ca^{\star}}\sum_{p=1}^{N'}\phi_{pt} 
	\PolyCoeffApproxPC\expcm{\PolyBasis(\BParams)}{X_i}\right)^2}{\BParams_{\sim
		i}}.
\end{equation}

Because the inner sum is only over the coefficients
$\phi_{pt}\PolyCoeffApproxPC$, this can be further simplified by
substituting
$c_{\alpha}\eqdef\sum_{p=1}^{N'}\phi_{pt}\PolyCoeffApproxPC$ to obtain:
\begin{align}
\varcm{\expcm{Y_t}{X_i}}{\BParams_{\sim i}}  &=
\expcm{\left(\sum_{\ve{\alpha}\in\ca^{\star}}c_{\alpha}\expcm{\PolyBasis(\BParams)}{X_i}\right)^2}{\BParams_{\sim
		i}}\\
&=
\expcm{\left(\sum_{\ve{\alpha}\in\ca^{\star}}c_{\alpha}\expcm{\PolyBasis(\BParams)}{X_i}\right)
	\left(\sum_{\ve{\beta}\in\ca^{\star}}c_{\beta}\expcm{\PolyBasisB(\BParams)}{X_i}\right)}{\BParams_{\sim
		i	}}\\
&=
\sum_{\ve{\alpha,\beta}\in\ca^{\star}}c_{\alpha}c_{\beta}\expcm{\expcm{\PolyBasis(\BParams)}{X_i}\expcm{\PolyBasisB(\BParams)}{X_i}}{\BParams_{\sim
		i}}.
\end{align}

Due to the orthonormality of the polynomial basis 
$\{\PolyBasis\}_{\ve{\alpha}\in\ca^{\star}}$, the 
conditional expectation in this equation can be expressed analytically 
as
%
%
%
\begin{multline}
\expcm{\expcm{\PolyBasis(\BParams)}{X_i}\expcm{\PolyBasisB(\BParams)}{X_i}}{\BParams_{\sim
		i}} = \delta_{\ve{\alpha},\ve{\beta}, i}\\
\text{with} ~\delta_{\ve{\alpha},\ve{\beta}, i} = 
\begin{cases}
1, \quad \mathrm{if}~ \text{$\ve{\alpha}=\ve{\beta}$ and
	$\alpha_i = 0$},\\
0, \quad \mathrm{otherwise}.
\end{cases}
\end{multline}

Therefore, the variance of the conditional expectation from
Eq.~\eqref{eq:totalSobol} becomes
\begin{equation}
\varcm{\expcm{Y_t}{X_i}}{\BParams_{\sim i}} =
\sum_{\ve{\alpha}\in\ca^{\star}_{i=0}}\left(\sum_{p=1}^{N'}\phi_{pt}\PolyCoeffApproxPC\right)^2,
\end{equation}
where $\ca^{\star}_{i=0}=\{\ve{\alpha}\in\ca^{\star}:\alpha_{i} = 0\}$ is the 
subset that contains only those
polynomials $\PolyBasis$ with $\alpha_{i} = 0$. For completeness the total 
variance in the 
denominator of Eq.~\eqref{eq:totalSobol} reads:
\begin{equation}
\varc{Y_t} =
\sum_{\ve{\alpha}\in\ca^{\star}}\left(\sum_{p=1}^{N'}\phi_{pt}\PolyCoeffApproxPC\right)^2.
\end{equation}
The total PCA-based index for the $t$-th component of the output vector 
$\BModelOut$
is thus obtained by plugging these results into
Eq.~\eqref{eq:totalSobol}:
\begin{equation}
S_{i,t}^{T} = 1 -
\frac{\sum_{\ve{\alpha}\in\ca^{\star}_{i=0}}\left(\sum_{p=1}^{N'}\phi_{pt} 
	\PolyCoeffApproxPC\right)^2}{\sum_{\ve{\alpha}\in\ca^{\star}}\left(\sum_{p=1}^{N'}\phi_{pt}
	\PolyCoeffApproxPC\right)^2}.
\end{equation}

\clearpage
\section{Additional results}
\label{sec:additionalResults}
This section presents the calibration and sensitivity analysis 
results for Product B, Product C and Product D.

\begin{table}
	\caption{Posterior statistics for the calibration with Product B 
		(E2). The 
		values are computed from the available posterior sample and include the 
		MAP 
		estimate, the 
		empirical mean $\hat{\mu}$, the empirical
		$95\%$ confidence interval, the 
		empirical standard deviation $\hat{\sigma}$, and the empirical 
		coefficient 
		of 
		variation $\mathrm{c.o.v.}\eqdef \hat{\sigma}/\hat{\mu}$. The prior 
		statistics are shown in 
		Table~\ref{tab:priorDist}.}
	\label{tab:postStatE1}
	\centering
	\resizebox{\textwidth}{!}{
		\begin{tabular}{rccccc}
			\hline
			& MAP & $\hat{\mu}$ & $95\%$ conf. interval & $\hat{\sigma}$ & 
			c.o.v. \\
			\hline
			$X_{1}$ & $5.38 \cdot 10^{2}$ & $5.05 \cdot 10^{2}$ & $[4.25 \cdot 
			10^{2}, 5.79 \cdot 10^{2}]$ & $4.25 \cdot 10^{1}$ & $8.41 \cdot 
			10^{-2}$\\$X_{2}$ & $3.97 \cdot 10^{-1}$ & $4.02 \cdot 10^{-1}$ & 
			$[3.37 \cdot 10^{-1}, 4.75 \cdot 10^{-1}]$ & $3.65 \cdot 10^{-2}$ & 
			$9.07 \cdot 10^{-2}$\\
			$X_{3}$ & $0.224$ & $0.226$ & $[0.217, 0.239]$ & $5.39 \cdot 
			10^{-3}$ & $2.38 \cdot 10^{-2}$\\
			$X_{4}$ & $0.107$ & $0.115$ & $[0.1, 0.138]$ & $1.12 
			\cdot 10^{-2}$ & $9.74 \cdot 10^{-2}$\\
			$X_{5}$ & $5.12 \cdot 10^{4}$ & 
			$5.16 \cdot 10^{4}$ & $[4.93 \cdot 10^{4}, 5.37 \cdot 10^{4}]$ & 
			$1.14 
			\cdot 10^{3}$ & $2.21 \cdot 10^{-2}$\\$X_{6}$ & $1.08 \cdot 10^{3}$ 
			& 
			$1.08 \cdot 10^{3}$ & $[1.00 \cdot 10^{3}, 1.30 \cdot 10^{3}]$ & 
			$8.20 
			\cdot 10^{1}$ & $7.56 \cdot 10^{-2}$\\$X_{7}$ & $8.80 $ & 
			$8.83 $ & $[8.30 , 9.40 ]$ & $2.83 
			\cdot 10^{-1}$ & $3.21 \cdot 10^{-2}$\\$X_{8}$ & $2.02 $ & 
			$2.04 $ & $[2.30 , 1.78 ]$ & $1.95 
			\cdot 10^{-1}$ & $4.28 \cdot 10^{-2}$\\$X_{9}$ & $1.17 $ & 
			$1.15 $ & $[1.39 , 0.91 ]$ & $1.81 
			\cdot 10^{-1}$ & $0.70 \cdot 10^{-1}$\\$X_{10}$ & $5.17 \cdot 
			10^{-1}$ 
			& $5.34 \cdot 10^{-1}$ & $[0.75 , 3.39 \cdot 10^{-1}]$ & 
			$1.56 \cdot 10^{-1}$ & $1.31 \cdot 10^{-1}$\\$X_{11}$ & $4.27$ & 
			$4.29 
			$ & $[3.93 , 4.67]$ & $1.93 \cdot 10^{-1}$ & $4.50 \cdot 
			10^{-2}$\\$X_{12}$ & 
			$1.89 $ & $1.90 $ & $[2.14 , 1.68 
			]$ & $1.78 \cdot 10^{-1}$ & $4.16 \cdot 10^{-2}$\\$X_{13}$ 
			& $4.04 \cdot 10^{-1}$ & $4.22 \cdot 10^{-1}$ & $[2.04 \cdot 
			10^{-1}, 
			6.41 \cdot 10^{-1}]$ & $1.11 \cdot 10^{-1}$ & $2.63 \cdot 
			10^{-1}$\\$X_{14}$ & $2.64 \cdot 10^{1}$ & $2.64 \cdot 10^{1}$ & 
			$[2.56 
			\cdot 10^{1}, 2.71 \cdot 10^{1}]$ & $3.81 \cdot 10^{-1}$ & $1.44 
			\cdot 
			10^{-2}$\\
			\hline
		\end{tabular}
	}
\end{table}

\begin{figure}
	\centering
	\includegraphics[width=14cm]{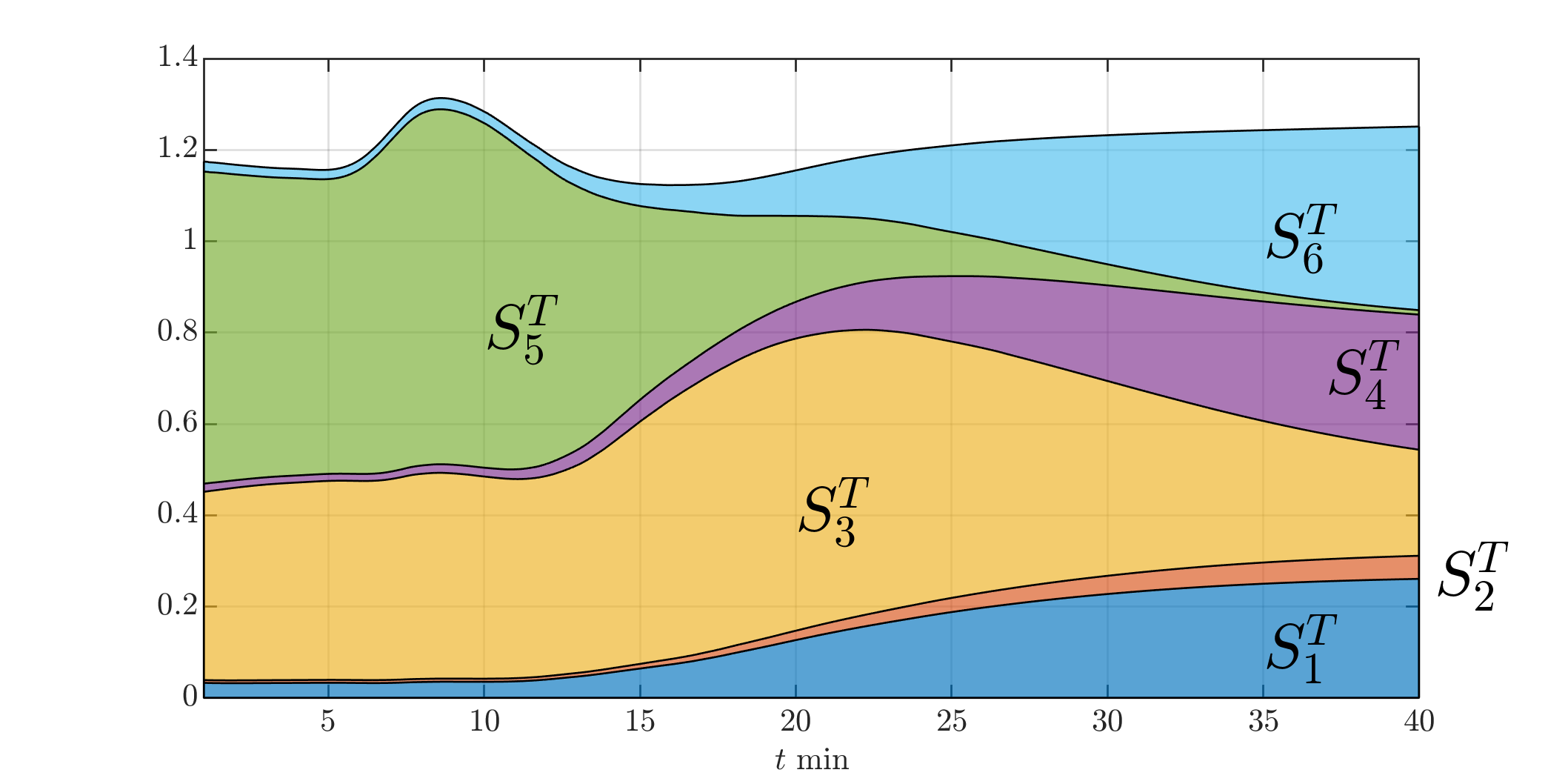}
	\caption{Time-dependent total Sobol' indices $S_i^{T}$ for the 
		surrogate
		model of Product B (E2).}%
	\label{fig:E15sensitivity}%
\end{figure}

\begin{figure}
	\centering
	\subfloat[Realizations of the conductivity 
	$\lambda(\BParams)$]{{\includegraphics[width=7cm]{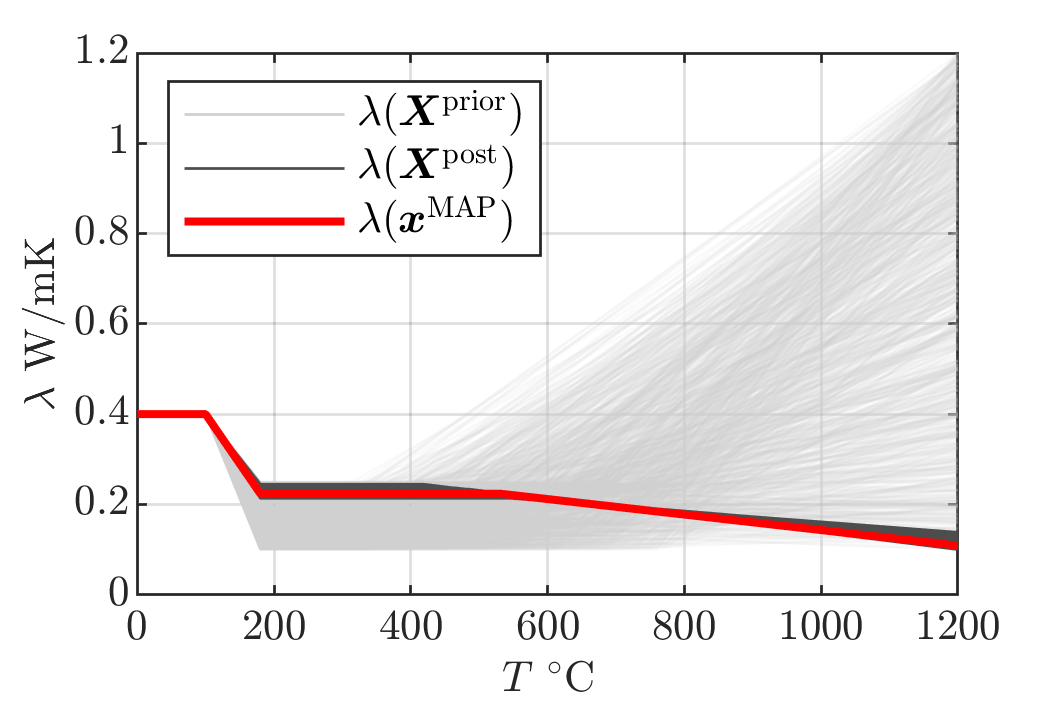}
			\label{fig:E15results:a}
	}}%
	\subfloat[Realizations of the heat capacity 
	$c(\BParams)$]{{\includegraphics[width=7cm]{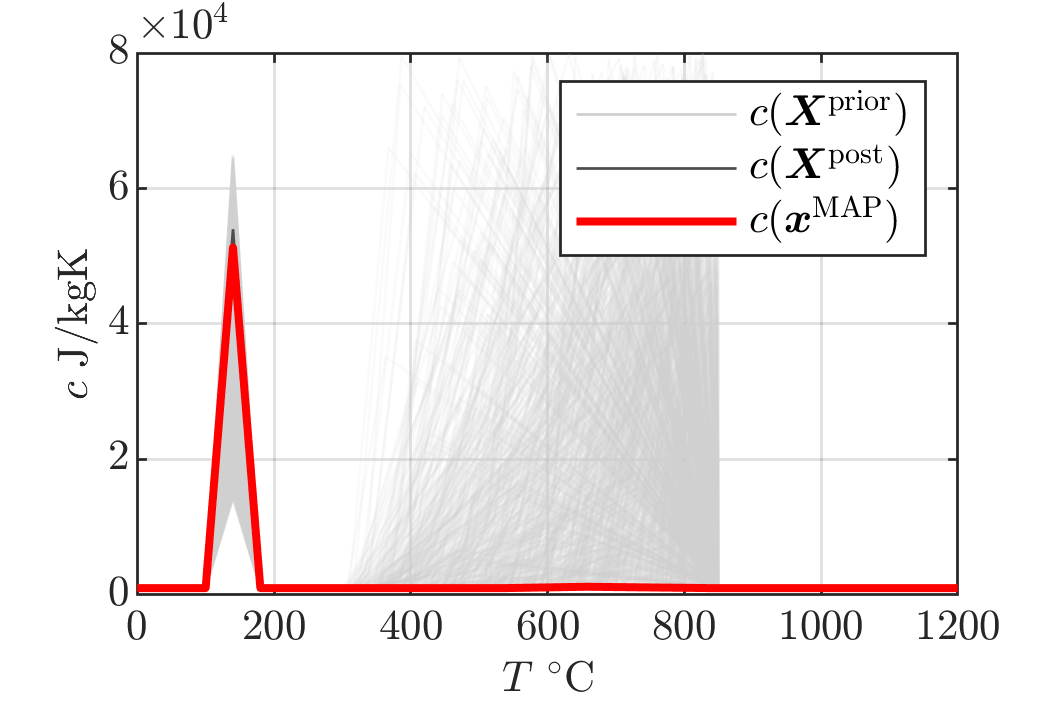}
			\label{fig:E15results:b} }}%
	\\
	\subfloat[Model predictions and calibrated discrepancy standard deviation]{{
			\begin{minipage}{\linewidth}
				\includegraphics[width=14cm,trim={0 0.5cm 0 
					0},clip]{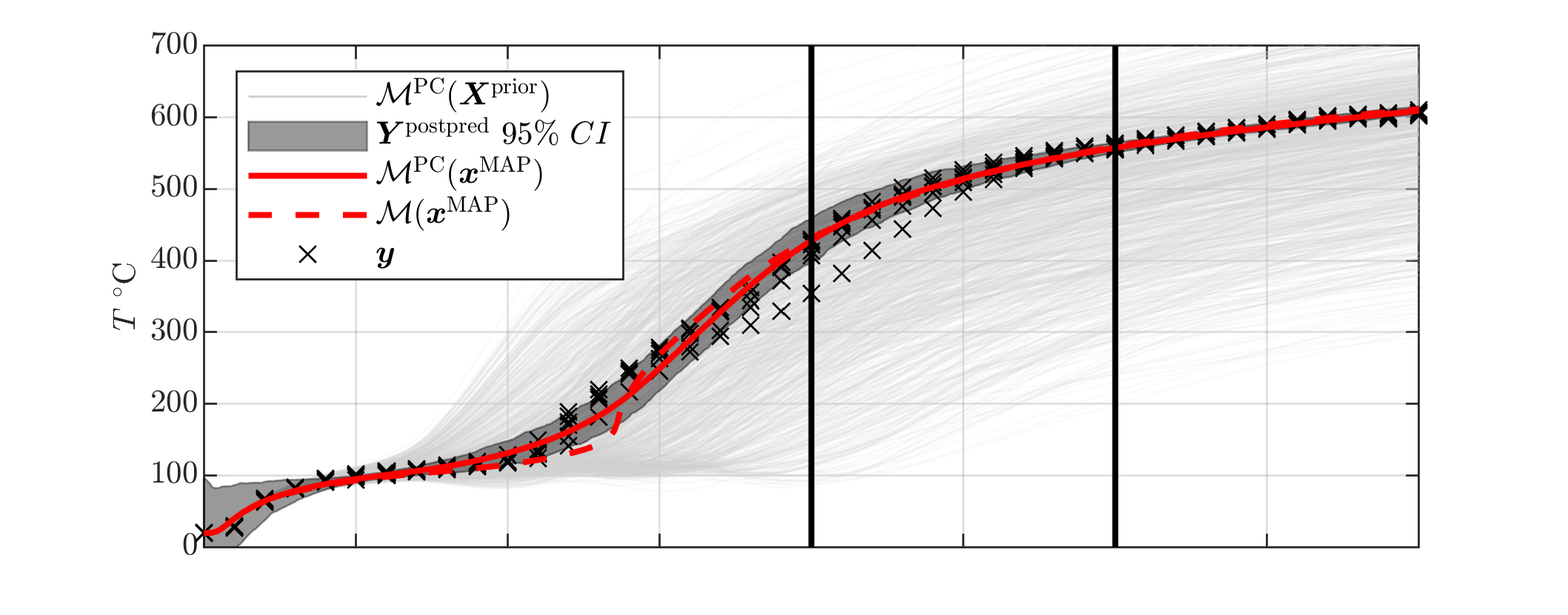}\\
				\includegraphics[width=14cm,trim={0 0 0 
					0.1},clip]{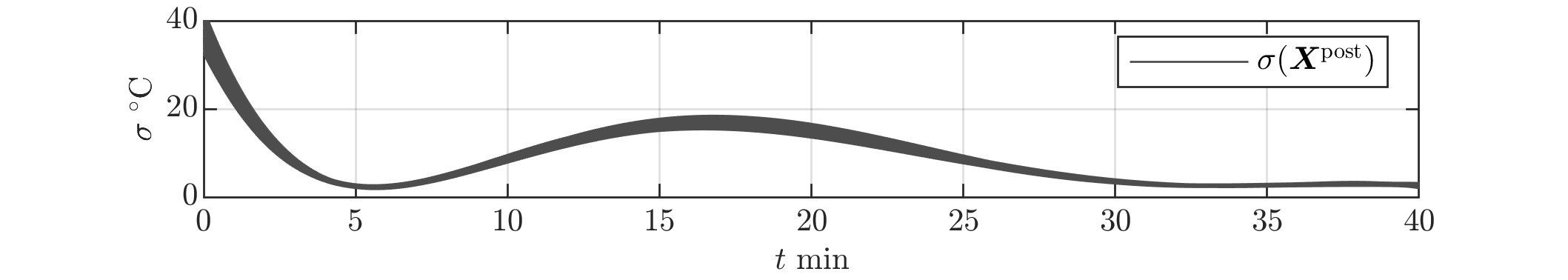} 
			\end{minipage}
			\label{fig:E15results:c}
	}}%
	\\
	\subfloat[Model predictions at 
	$t=20~\mathrm{min}$]{{\includegraphics[width=7cm]{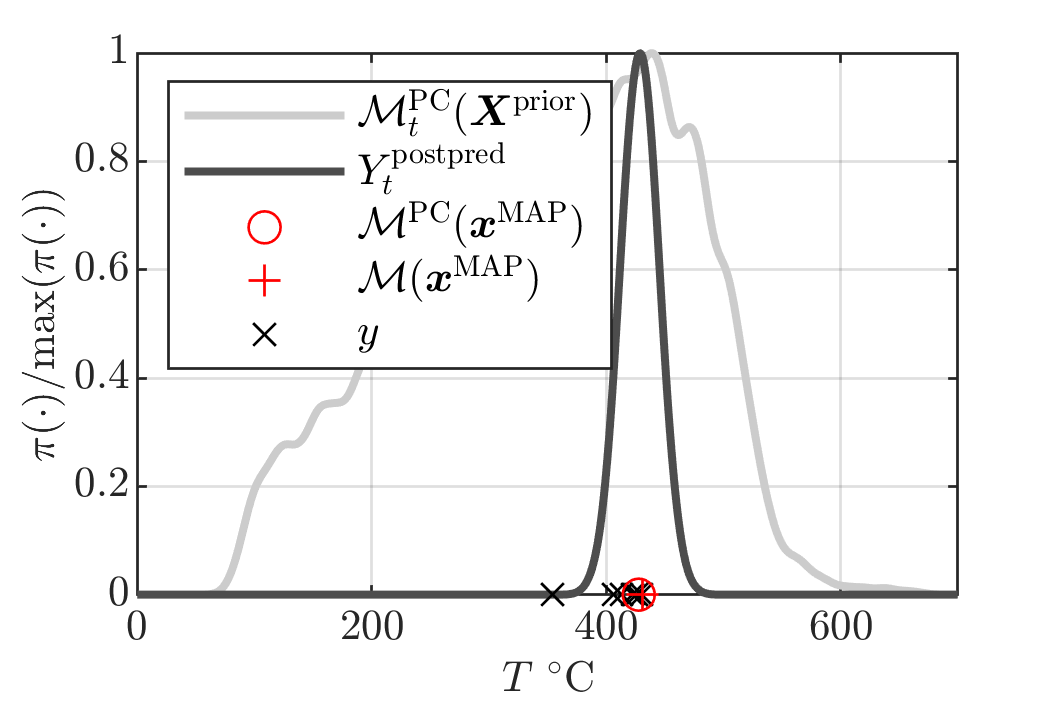}
			\label{fig:E15results:d}
	}}%
	\subfloat[Model predictions at 
	$t=30~\mathrm{min}$]{{\includegraphics[width=7cm]{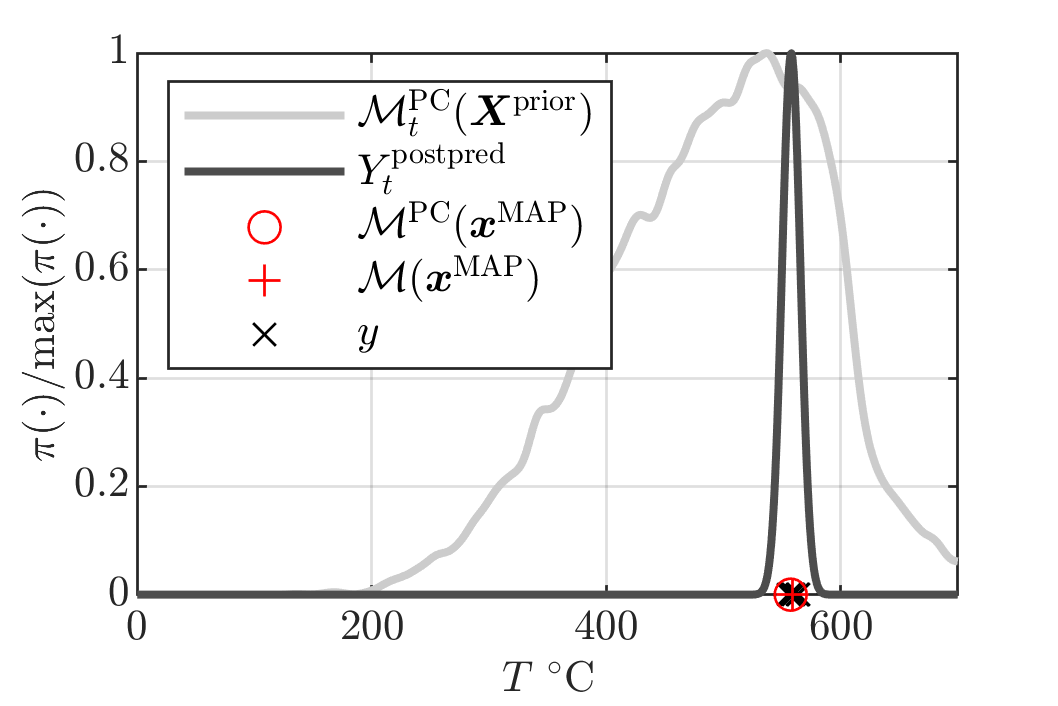}
			\label{fig:E15results:e}
	}}%
	\caption{Calibration results for Product B 9.5~mm insulation
		(E2), 
		experiments conducted by \citet{test:Gyproc2016}.}%
	\label{fig:E15results}%
\end{figure}


\begin{table}
	\caption{Posterior statistics for the calibration with Product C 
		(E3). The 
		values are computed from the available posterior sample and include the 
		MAP 
		estimate, the 
		empirical mean $\hat{\mu}$, the empirical
		$95\%$ confidence interval, the 
		empirical standard deviation $\hat{\sigma}$, and the empirical 
		coefficient 
		of 
		variation $\mathrm{c.o.v.}\eqdef \hat{\sigma}/\hat{\mu}$. The prior 
		statistics are shown in 
		Table~\ref{tab:priorDist}.}
	\label{tab:postStatE3}
	\centering
	\resizebox{\textwidth}{!}{
		\begin{tabular}{rccccc}
			\hline
			& MAP & $\hat{\mu}$ & $95\%$ conf. interval & $\hat{\sigma}$ & 
			c.o.v. \\
			\hline
			$X_{1}$ & $5.50 \cdot 10^{2}$ & $5.63 \cdot 10^{2}$ & $[4.63 \cdot 
			10^{2}, 7.10 \cdot 10^{2}]$ & $6.25 \cdot 10^{1}$ & $1.11 \cdot 
			10^{-1}$\\$X_{2}$ & $8.36 \cdot 10^{-1}$ & $7.93 \cdot 10^{-1}$ & 
			$[4.24 \cdot 10^{-1}, 9.95 \cdot 10^{-1}]$ & $1.60 \cdot 10^{-1}$ & 
			$2.01 \cdot 10^{-1}$\\
			$X_{3}$ & $0.165$ & $0.160$ & $[0.133, 0.184]$ & $1.28 \cdot 
			10^{-2}$ & $7.98 \cdot 10^{-2}$\\
			$X_{4}$ & $0.648$ & $0.701$ & $[0.539, 0.93]$ & $0.106 
			$ & $1.51 \cdot 10^{-1}$\\$X_{5}$ & $2.16 \cdot 10^{4}$ & 
			$2.06 \cdot 10^{4}$ & $[1.64 \cdot 10^{4}, 2.49 \cdot 10^{4}]$ & 
			$2.18 
			\cdot 10^{3}$ & $1.06 \cdot 10^{-1}$\\$X_{6}$ & $1.05 \cdot 10^{4}$ 
			& 
			$1.12 \cdot 10^{4}$ & $[3.99 \cdot 10^{3}, 2.65 \cdot 10^{4}]$ & 
			$4.99 
			\cdot 10^{3}$ & $4.44 \cdot 10^{-1}$\\$X_{7}$ & $1.08 \cdot 10^{1}$ 
			& 
			$1.08 \cdot 10^{1}$ & $[9.95 , 1.17 \cdot 10^{1}]$ & $4.53 
			\cdot 10^{-1}$ & $4.18 \cdot 10^{-2}$\\$X_{8}$ & $1.43 $ & 
			$1.39 $ & $[1.03 , 1.72 ]$ & $1.80 
			\cdot 10^{-1}$ & $1.30 \cdot 10^{-1}$\\$X_{9}$ & $1.32 $ & 
			$1.31 $ & $[1.56 , 1.06 ]$ & $1.82 
			\cdot 10^{-1}$ & $6.20 \cdot 10^{-2}$\\$X_{10}$ & $1.62 $ & 
			$1.62 $ & $[1.90 , 1.36 ]$ & $2.00 
			\cdot 10^{-1}$ & $5.51 \cdot 10^{-2}$\\$X_{11}$ & $1.58 $ & 
			$1.65 $ & $[1.37 , 1.94 ]$ & $1.53 
			\cdot 10^{-1}$ & $9.29 \cdot 10^{-2}$\\$X_{12}$ & $2.15 $ & 
			$2.24 $ & $[1.89 , 2.63 ]$ & $1.94 
			\cdot 10^{-1}$ & $8.64 \cdot 10^{-2}$\\$X_{13}$ & $1.25 $ & 
			$1.24 $ & $[1.43 , 1.05 ]$ & $1.44 
			\cdot 10^{-1}$ & $5.19 \cdot 10^{-2}$\\$X_{14}$ & $3.06 \cdot 
			10^{1}$ & 
			$3.06 \cdot 10^{1}$ & $[2.94 \cdot 10^{1}, 3.18 \cdot 10^{1}]$ & 
			$5.93 
			\cdot 10^{-1}$ & $1.94 \cdot 10^{-2}$\\
			\hline
		\end{tabular}
	}
\end{table}

\begin{figure}
	\centering
	\includegraphics[width=14cm]{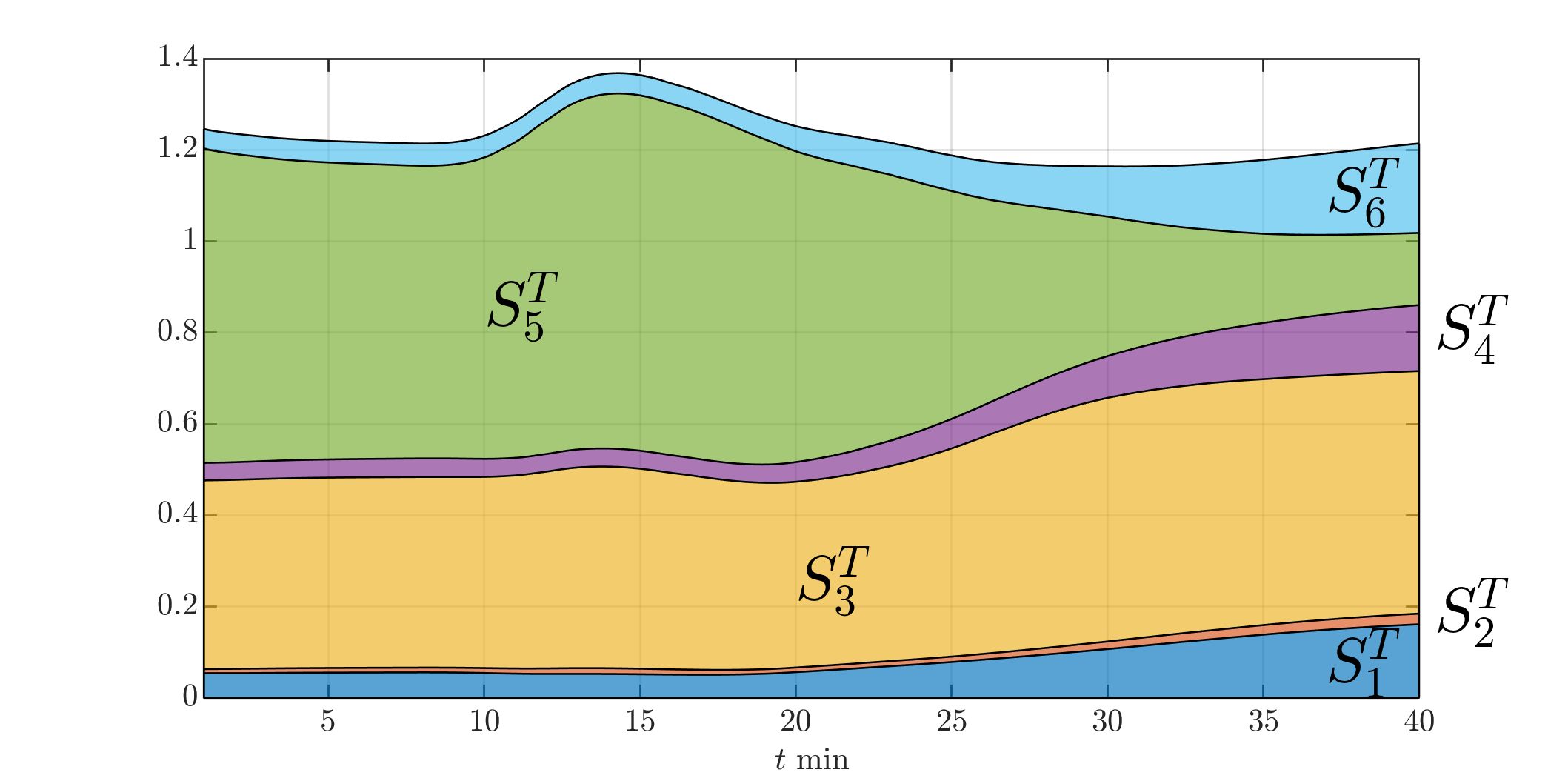}
	\caption{Time-dependent total Sobol' indices $S_i^{T}$ for the 
		surrogate
		model of Product C (E3).}%
	\label{fig:E17sensitivity}%
\end{figure}

\begin{figure}
	\centering
	\subfloat[Realizations of the conductivity 
	$\lambda(\BParams)$]{{\includegraphics[width=7cm]{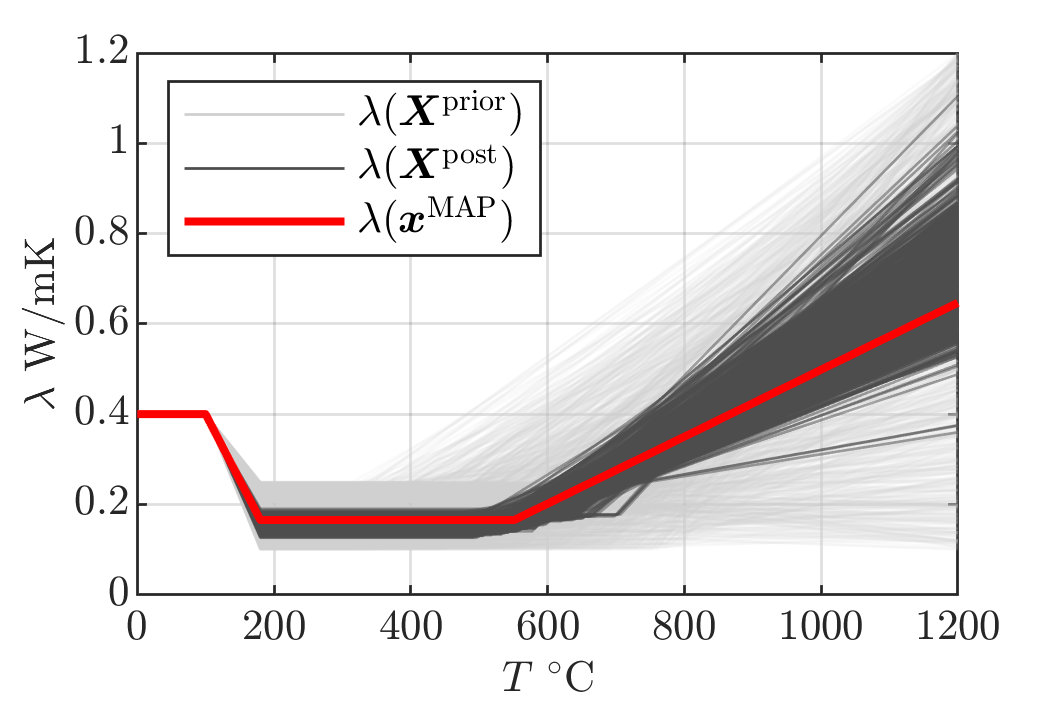}
	}}%
	\subfloat[Realizations of the heat capacity 
	$c(\BParams)$]{{\includegraphics[width=7cm]{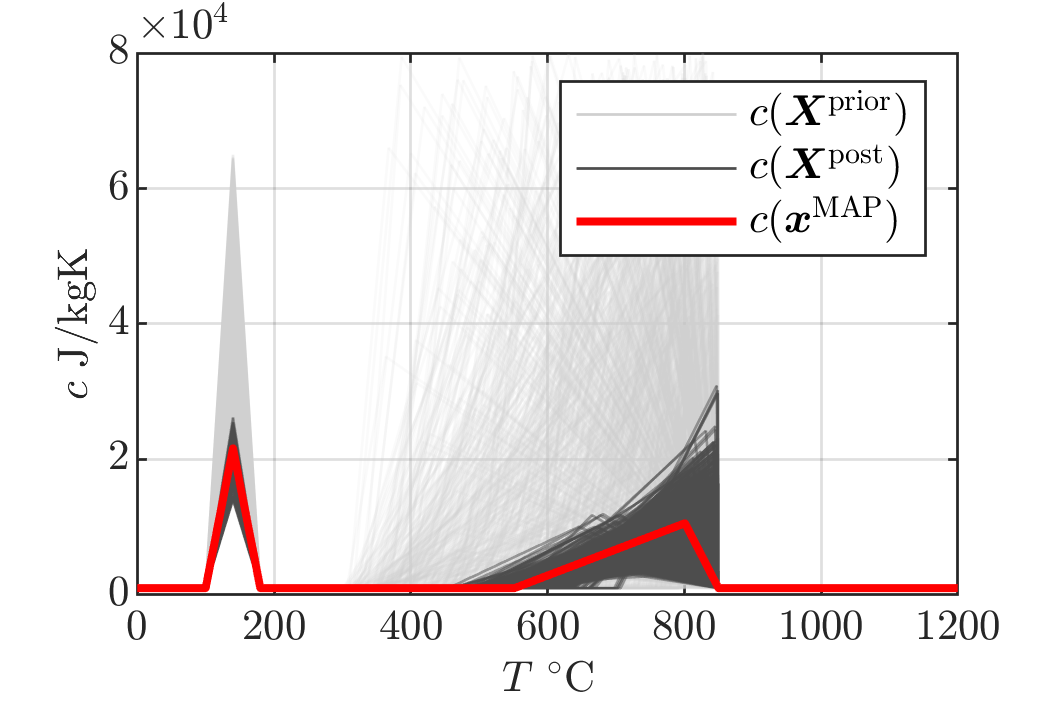}
	}}%
	\\
	\subfloat[Model predictions and calibrated discrepancy standard deviation]{{
			\begin{minipage}{\linewidth}
				\includegraphics[width=14cm,trim={0 0.5cm 0 
					0},clip]{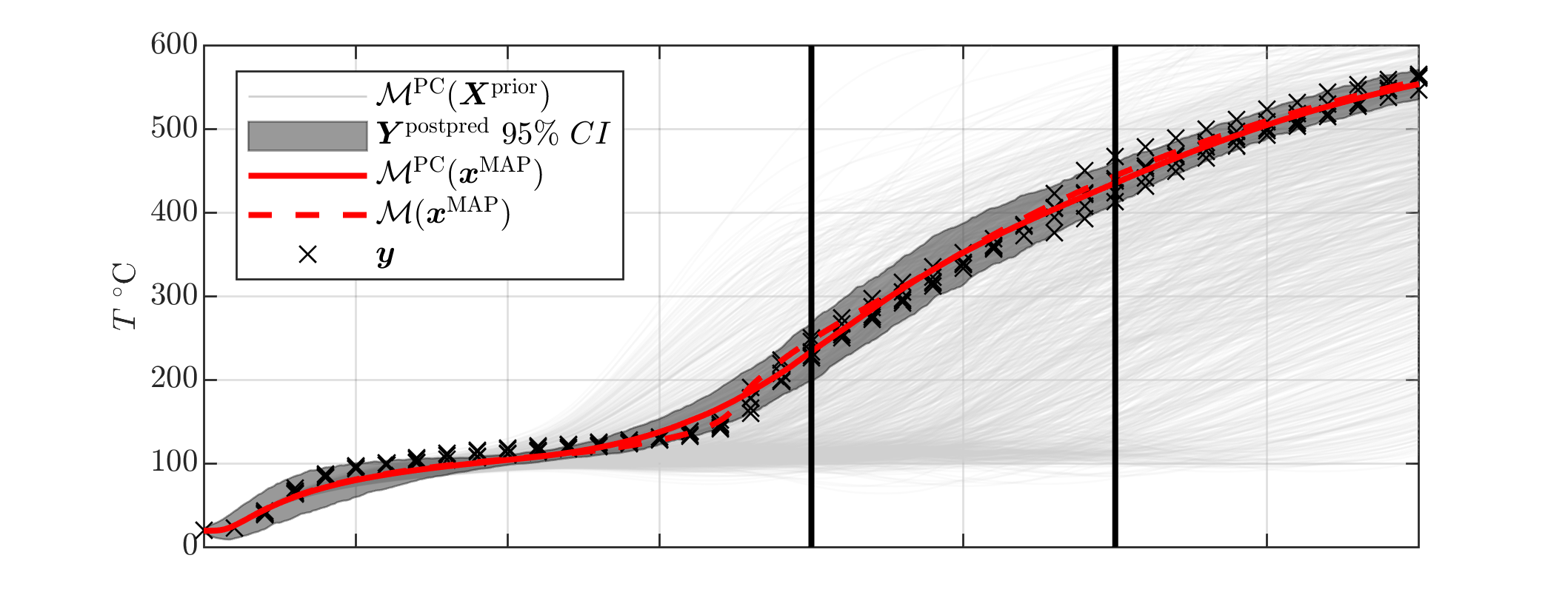}\\
				\includegraphics[width=14cm,trim={0 0 0 
					0.1},clip]{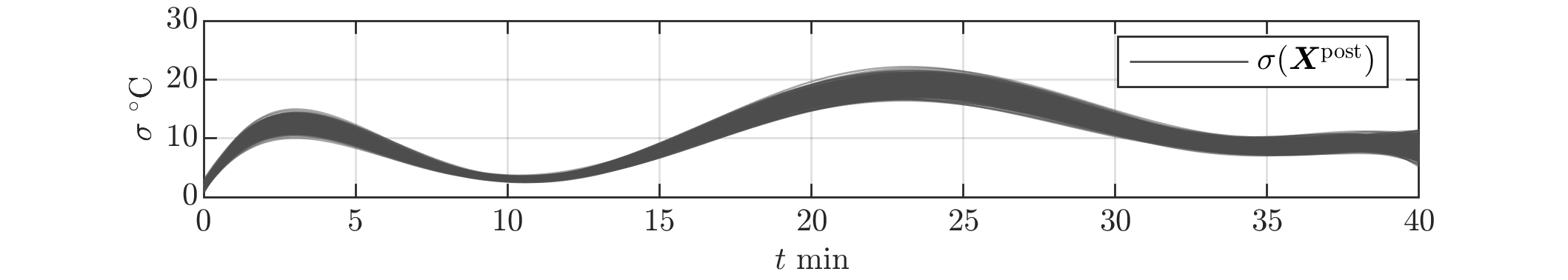} 
			\end{minipage}
	}}%
	\\
	\subfloat[Model predictions at 
	$t=20~\mathrm{min}$]{{\includegraphics[width=7cm]{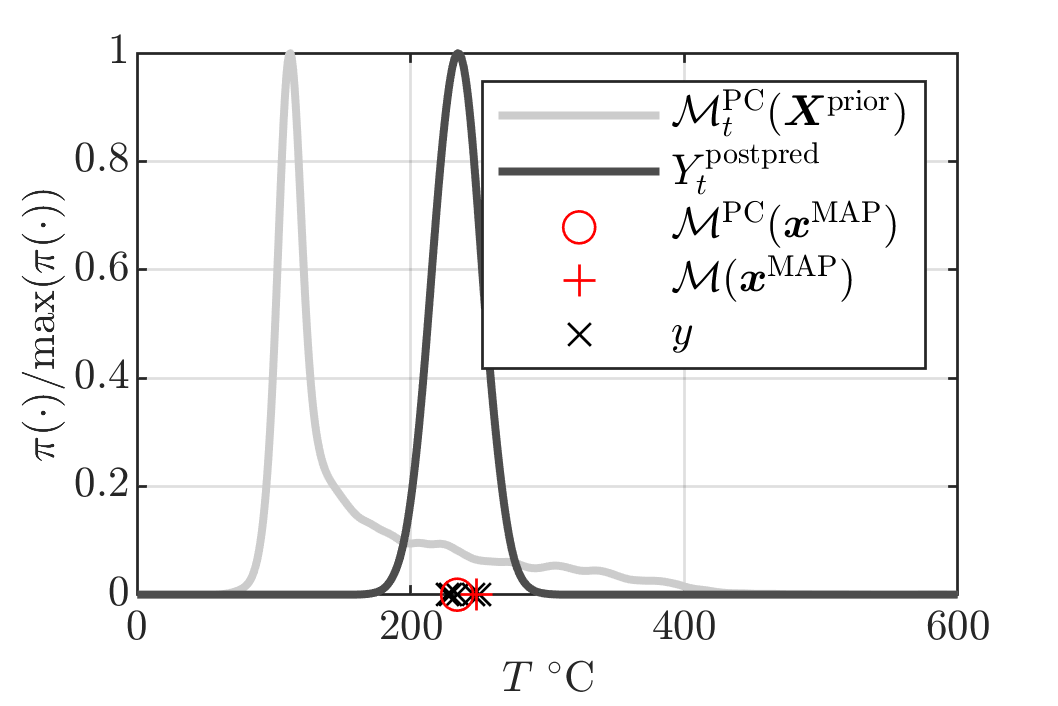}
			
	}}%
	\subfloat[Model predictions at 
	$t=30~\mathrm{min}$]{{\includegraphics[width=7cm]{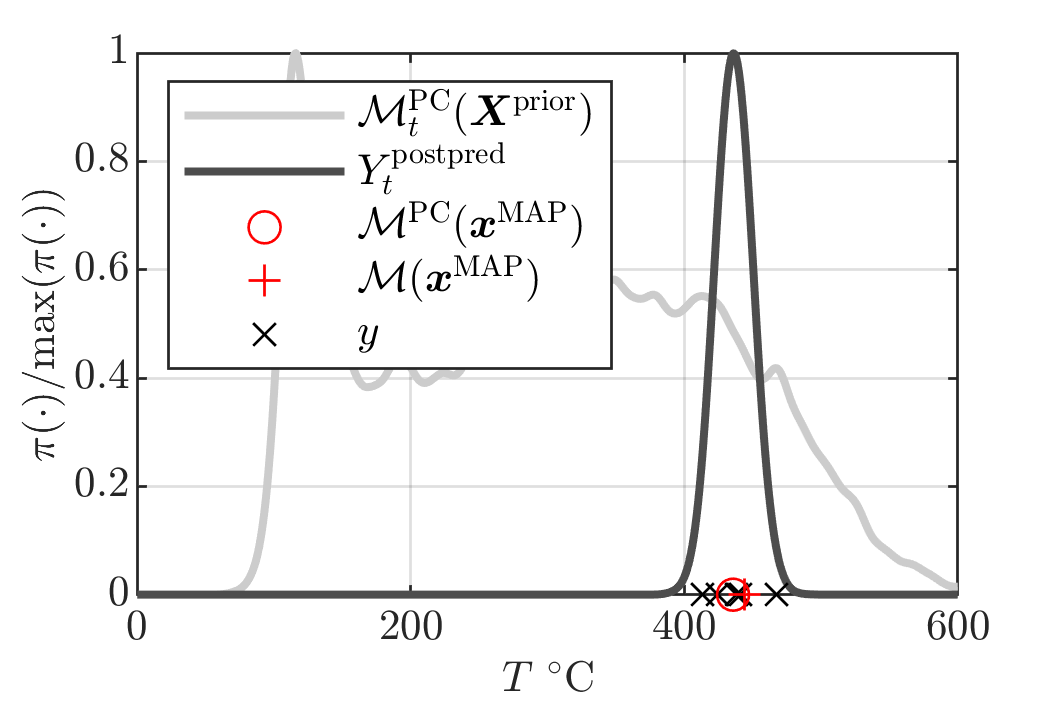}
			
	}}%
	\caption{Calibration results for Product C 12.5~mm insulation
		(E3), 
		experiments conducted by \citet{test:Gyproc2016}.}%
	\label{fig:E17results}%
\end{figure}


\begin{table}
	\caption{Posterior statistics for the calibration with Product D 
		(E4). The 
		values are computed from the available posterior sample and include the 
		MAP 
		estimate, the 
		empirical mean $\hat{\mu}$, the empirical
		$95\%$ confidence interval, the 
		empirical standard deviation $\hat{\sigma}$, and the empirical 
		coefficient 
		of 
		variation $\mathrm{c.o.v.}\eqdef \hat{\sigma}/\hat{\mu}$. The prior 
		statistics are shown in 
		Table~\ref{tab:priorDist}.}
	\label{tab:postStatE4}
	\centering
	\resizebox{\textwidth}{!}{
		\begin{tabular}{rccccc}
			\hline
			& MAP & $\hat{\mu}$ & $95\%$ conf. interval & $\hat{\sigma}$ & 
			c.o.v. \\
			\hline
			$X_{1}$ & $7.07 \cdot 10^{2}$ & $6.90 \cdot 10^{2}$ & $[6.41 \cdot 
			10^{2}, 7.43 \cdot 10^{2}]$ & $2.62 \cdot 10^{1}$ & $3.79 \cdot 
			10^{-2}$\\$X_{2}$ & $9.82 \cdot 10^{-1}$ & $8.07 \cdot 10^{-1}$ & 
			$[5.42 \cdot 10^{-1}, 9.95 \cdot 10^{-1}]$ & $1.25 \cdot 10^{-1}$ & 
			$1.55 \cdot 10^{-1}$\\
			$X_{3}$ & $0.157$ & $0.155$ & $[0.14, 0.173]$ & $9.21 \cdot 
			10^{-3}$ & $5.93 \cdot 10^{-2}$\\
			$X_{4}$ & $1.16$ & $1.06$ & $[0.843, 1.2]$ & $9.75 
			\cdot 10^{-2}$ & $9.16 \cdot 10^{-2}$\\$X_{5}$ & $2.04 \cdot 
			10^{4}$ & 
			$2.08 \cdot 10^{4}$ & $[1.81 \cdot 10^{4}, 2.42 \cdot 10^{4}]$ & 
			$1.70 
			\cdot 10^{3}$ & $8.16 \cdot 10^{-2}$\\$X_{6}$ & $5.18 \cdot 10^{3}$ 
			& 
			$4.91 \cdot 10^{3}$ & $[1.60 \cdot 10^{3}, 8.17 \cdot 10^{3}]$ & 
			$1.64 
			\cdot 10^{3}$ & $3.34 \cdot 10^{-1}$\\$X_{7}$ & $8.15 $ & 
			$8.39 $ & $[7.77 , 8.99 ]$ & $3.28 
			\cdot 10^{-1}$ & $3.91 \cdot 10^{-2}$\\$X_{8}$ & $2.88 $ & 
			$2.97 $ & $[2.61 , 3.33 ]$ & $1.86 
			\cdot 10^{-1}$ & $6.28 \cdot 10^{-2}$\\$X_{9}$ & $6.76 \cdot 
			10^{-1}$ & 
			$7.09 \cdot 10^{-1}$ & $[4.65 \cdot 10^{-1}, 9.73 \cdot 10^{-1}]$ & 
			$1.31 \cdot 10^{-1}$ & $1.84 \cdot 10^{-1}$\\$X_{10}$ & $1.67$ & 
			$1.69 
			$ & $[1.90 , 1.49]$ & $1.57 \cdot 10^{-1}$ & $4.14 \cdot 
			10^{-2}$\\$X_{11}$ & 
			$0.76 $ & $0.75 $ & $[0.91 , 5.75 
			\cdot 10^{-1}]$ & $1.25 \cdot 10^{-1}$ & $0.75 \cdot 
			10^{-1}$\\$X_{12}$ 
			& $1.38 $ & $1.41 $ & $[1.21 , 1.64 
			]$ & $1.09 \cdot 10^{-1}$ & $7.75 \cdot 10^{-2}$\\$X_{13}$ 
			& $5.10 \cdot 10^{-1}$ & $5.35 \cdot 10^{-1}$ & $[6.55 \cdot 
			10^{-1}, 
			4.15 \cdot 10^{-1}]$ & $9.24 \cdot 10^{-2}$ & $0.77 \cdot 
			10^{-1}$\\$X_{14}$ & $2.96 \cdot 10^{1}$ & $2.97 \cdot 10^{1}$ & 
			$[2.88 
			\cdot 10^{1}, 3.07 \cdot 10^{1}]$ & $4.73 \cdot 10^{-1}$ & $1.59 
			\cdot 
			10^{-2}$\\
			\hline
		\end{tabular}
	}
\end{table}

\begin{figure}
	\centering
	\includegraphics[width=14cm]{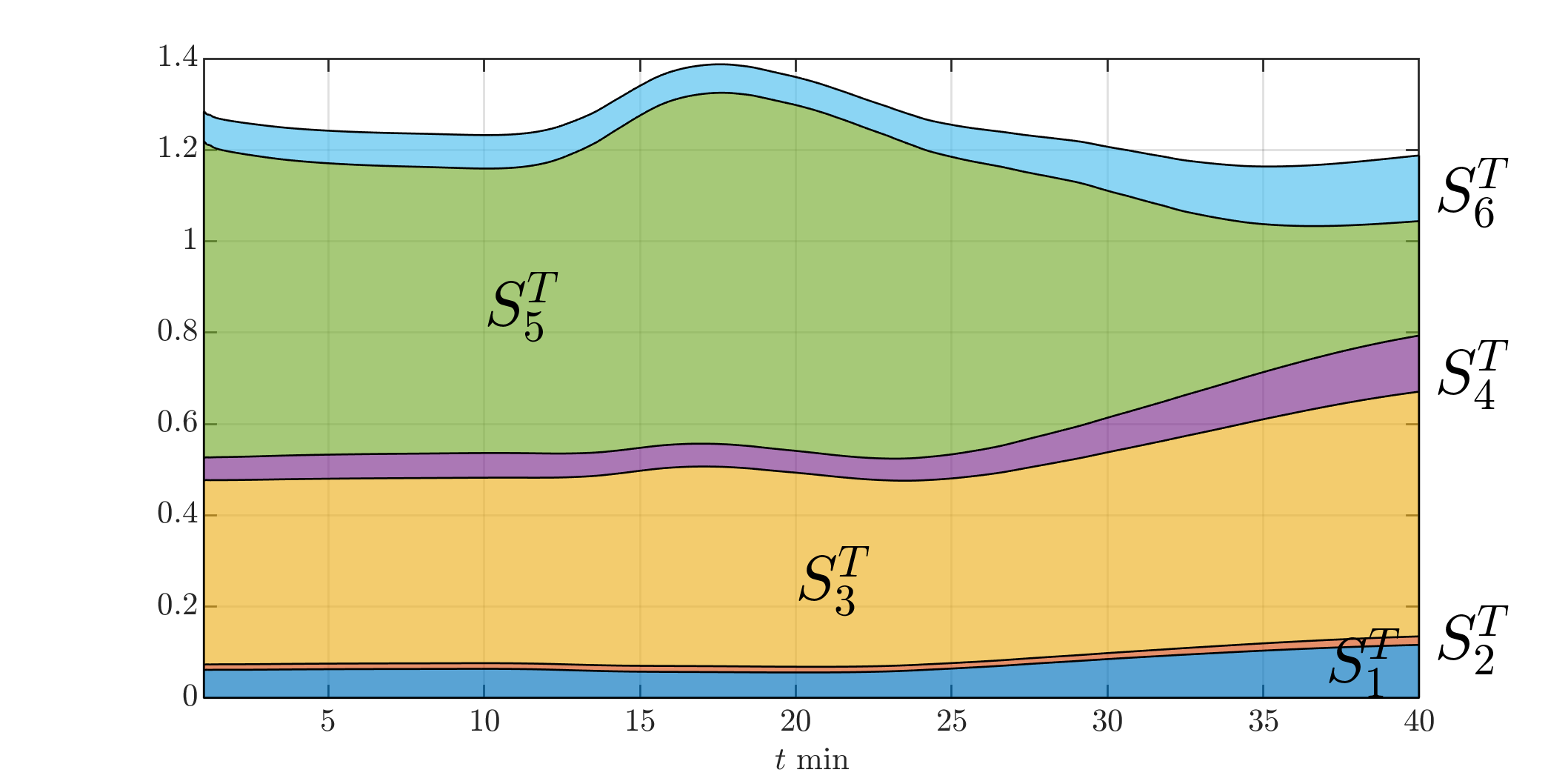}
	\caption{Time-dependent total Sobol' indices $S_i^{T}$ for the 
		surrogate
		model of Product D (E4).}%
	\label{fig:E18sensitivity}%
\end{figure}

\begin{figure}
	\centering
	\subfloat[Realizations of the conductivity 
	$\lambda(\BParams)$]{{\includegraphics[width=7cm]{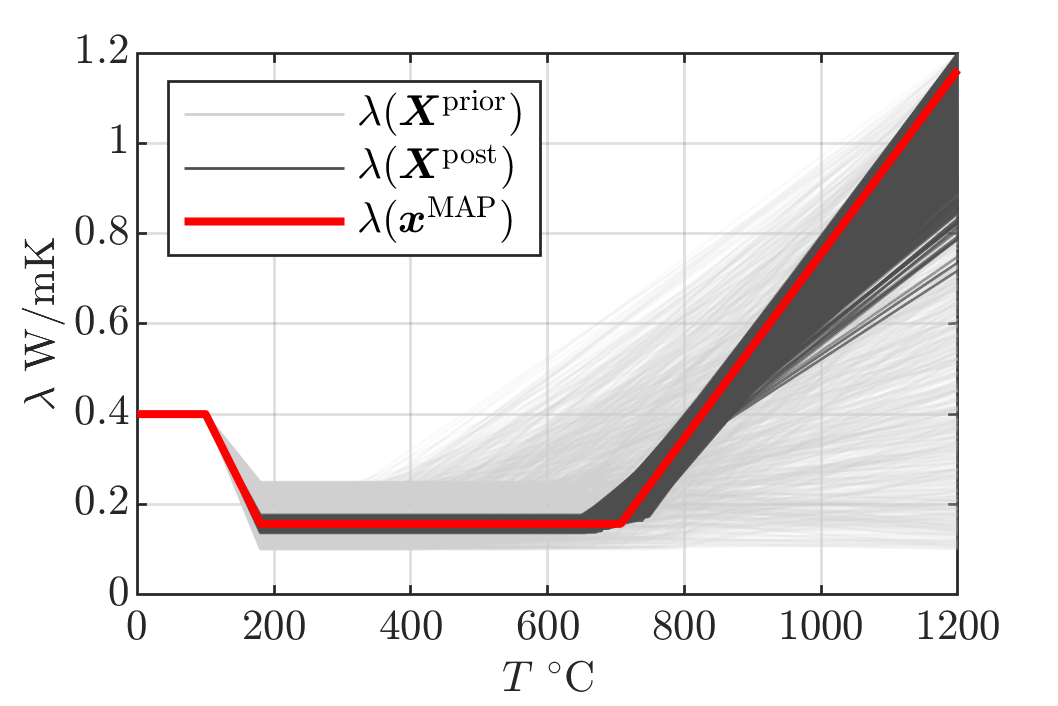}
	}}%
	\subfloat[Realizations of the heat capacity 
	$c(\BParams)$]{{\includegraphics[width=7cm]{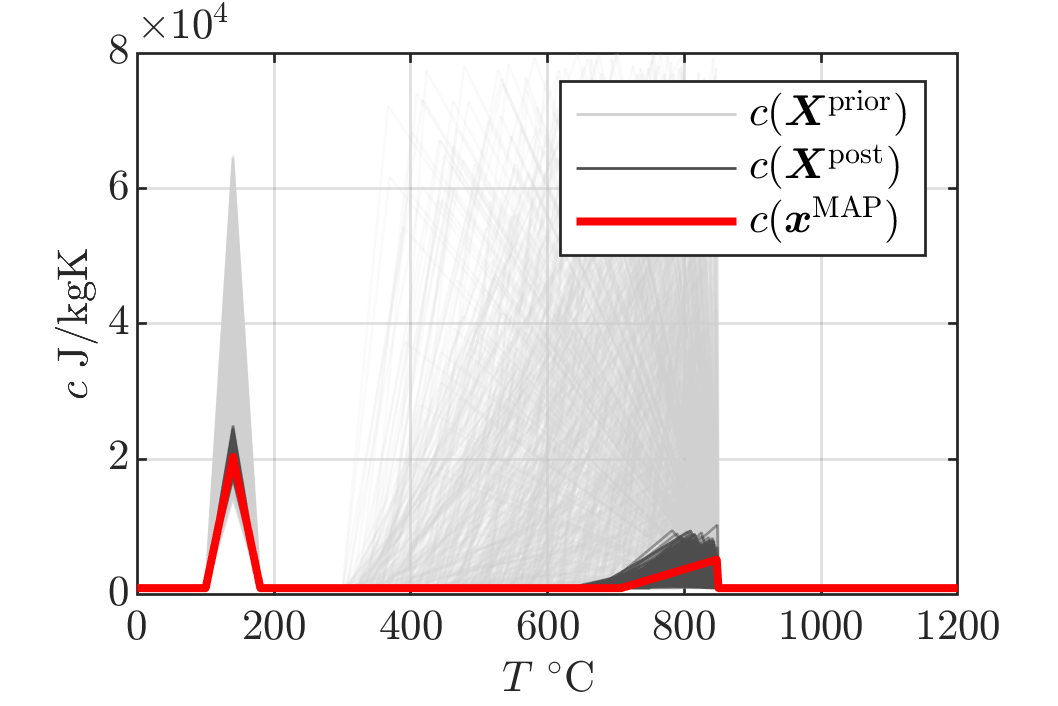}
	}}%
	\\
	\subfloat[Model predictions and calibrated discrepancy standard deviation]{{
			\begin{minipage}{\linewidth}
				\includegraphics[width=14cm,trim={0 0.5cm 0 
					0},clip]{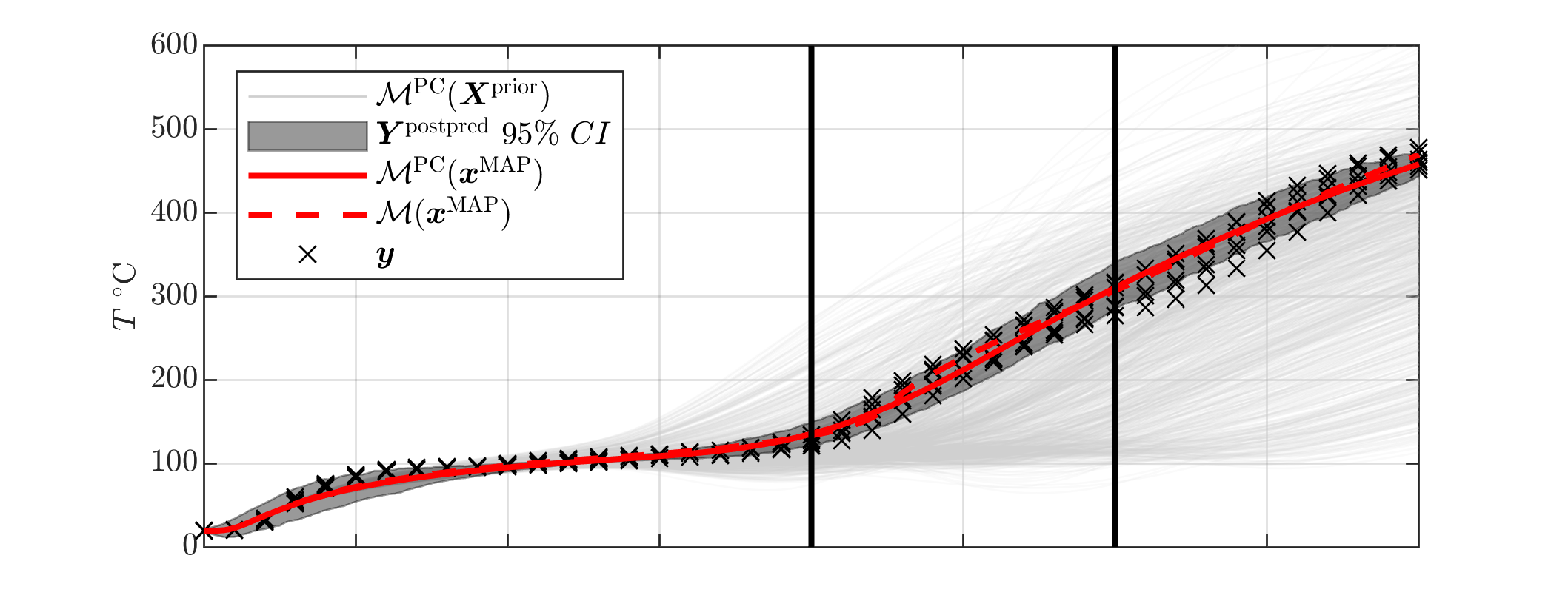}\\
				\includegraphics[width=14cm,trim={0 0 0 
					0.1},clip]{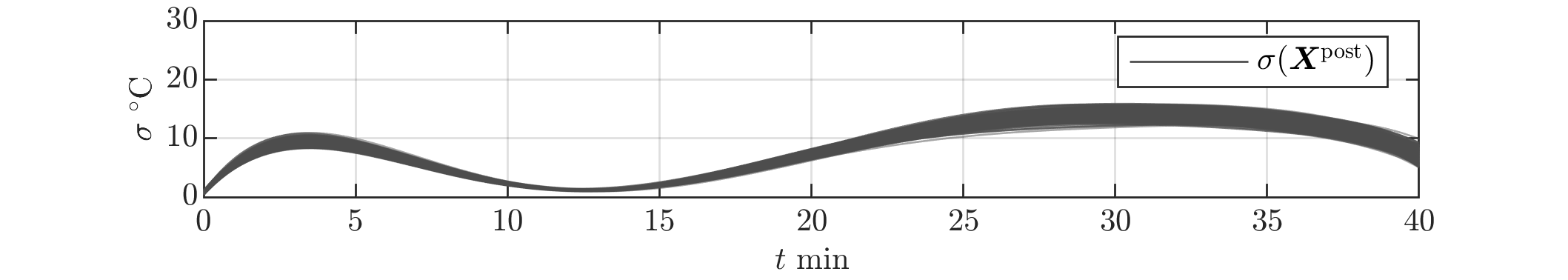} 
			\end{minipage}
	}}%
	\\
	\subfloat[Model predictions at 
	$t=20~\mathrm{min}$]{{\includegraphics[width=7cm]{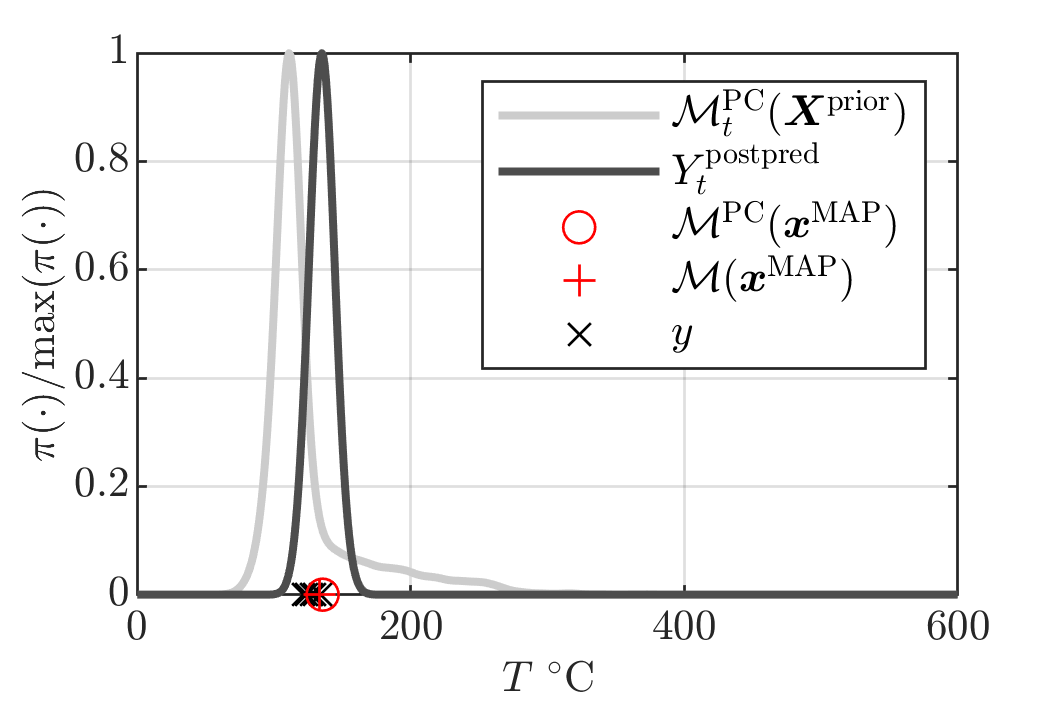}
			
	}}%
	\subfloat[Model predictions at 
	$t=30~\mathrm{min}$]{{\includegraphics[width=7cm]{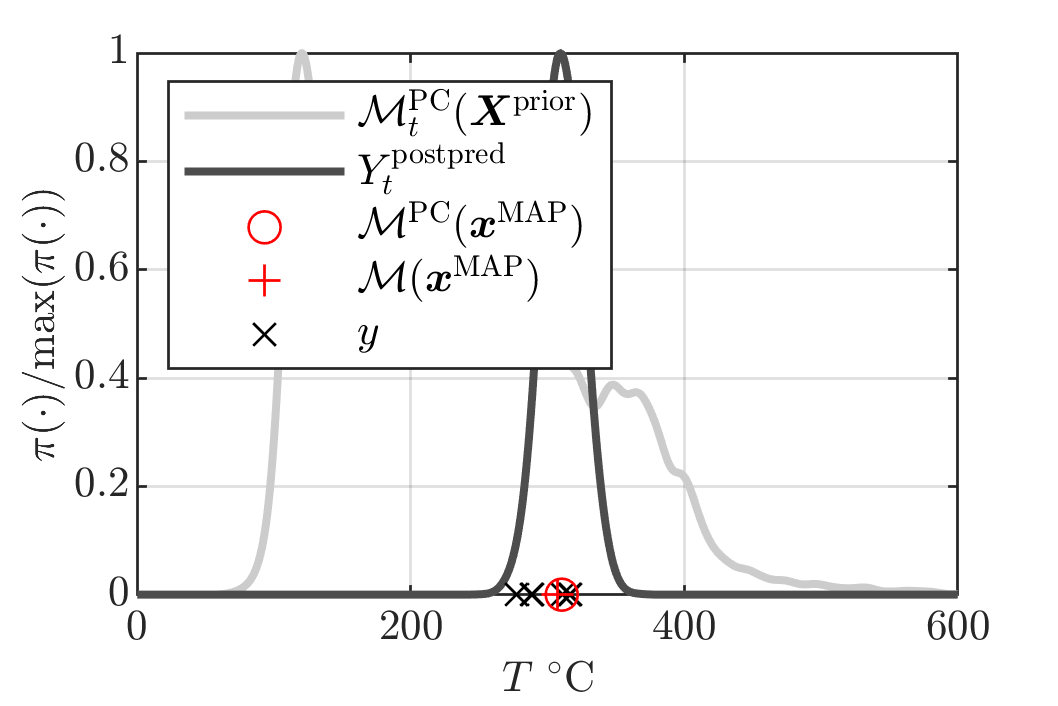}
	}}%
	\caption{Calibration results for Product D 15~mm insulation
		(E4), 
		experiments conducted by \citet{test:Gyproc2016}.}%
	\label{fig:E18results}%
\end{figure}

\bibliographystyle{chicago} 
\bibliography{fireCal}
\end{document}